\documentclass[10pt,journal,compsoc]{IEEEtran}

\usepackage{times}  
\usepackage{helvet}  
\usepackage{courier}  
\usepackage{url}  
\usepackage{graphicx}  

\usepackage[namelimits]{amsmath} 
\usepackage{amssymb}             
\usepackage{amsfonts}            
\usepackage{mathrsfs}            
\usepackage{algorithm}
\usepackage{algorithmic}
\usepackage{makecell,rotating,multirow,diagbox}
\usepackage[normalem]{ulem}
\usepackage{subfigure}
\usepackage{bm}
\begin{document}
\title{Deep Cross-Modal Hashing with Hashing \\Functions and Unified Hash Codes Jointly Learning}

\author{Rong-Cheng~Tu, Xian-Ling~Mao, Bing~Ma, Yong~Hu, Tan~Yan, Wei~Wei and Heyan Huang
\IEEEcompsocitemizethanks{\IEEEcompsocthanksitem R. Tu, X. Mao, B. Ma, Y. Hu, T. Y and H. Huang are with the Department of Computer Science and Technology, Beijing Institute of Technology, Beijing 100081, Chain.\protect\\
E-mail: \{tu\_rc, maoxl, 3220180838, huyong, yantan, hhy63\}@bit.edu.cn.
\IEEEcompsocthanksitem W. Wei is with School of Computer Science, Huazhong University of Science and Technology, Wuhan 430074, Chain.\protect\\
E-mail: weiw@hust.edu.cn}
}

%
%

\markboth{Journal of \LaTeX\ Class Files,~Vol.~14, No.~8, August~2015}%
{Shell \MakeLowercase{\textit{et al.}}: Bare Demo of IEEEtran.cls for IEEE Journals}
\IEEEtitleabstractindextext{%
\begin{abstract}
	Due to their high retrieval efficiency and low storage cost, cross-modal hashing methods have attracted considerable attention. Generally, compared with shallow cross-modal hashing methods, deep cross-modal hashing methods can achieve a more satisfactory performance by integrating feature learning and hash codes optimizing into a same framework. However, most existing deep cross-modal hashing methods either cannot learn a unified hash code for the two correlated data-points of different modalities in a database instance or cannot guide the learning of unified hash codes by the feedback of hashing function learning procedure, to enhance the retrieval accuracy. To address the issues above, in this paper, we propose a novel end-to-end Deep Cross-Modal Hashing with Hashing Functions and Unified Hash Codes Jointly Learning (DCHUC). Specifically, by an iterative optimization algorithm, DCHUC jointly learns unified hash codes for image-text pairs in a database and a pair of hash functions for unseen query image-text pairs. With the iterative optimization algorithm, the learned unified hash codes can be used to guide the hashing function learning procedure; Meanwhile, the learned hashing functions can feedback to guide the unified hash codes optimizing procedure. Extensive experiments on three public datasets demonstrate that the proposed method outperforms the state-of-the-art cross-modal hashing methods.
\end{abstract}
	}

\maketitle


\IEEEdisplaynontitleabstractindextext

%
\IEEEpeerreviewmaketitle

\section{Introduction}
\IEEEPARstart{W}{ith} a tremendous amount of  multimedia data being generated on the Internet everyday such as texts, images and so on, similarity-preserving hashing methods \cite{cao2017hashnet,liu2018fast,zhang2016collaborative,bronstein2010data,jin2014density,huang2017unsupervised,ghasedi2018unsupervised,wang2017supervised,qiu2017deep} have been extensively studied for large-scale multimedia search due to their high retrieval efficiency and low storage cost. Because the corresponding data of different modalities may have semantic correlations, it is essential to support cross-modal retrieval that returns relevant results of one modality when querying another modality, e.g., retrieving images with text queries. Hence, cross-modal hashing methods \cite{liu2017cross,erin2017cross,feng2014cross,deng2016discriminative,wang2017adversarial,yang2017pairwise} get more and more attention.

Roughly speaking, cross-modal hashing methods can be divided into shallow cross-modal hashing methods \cite{liu2014discrete,liu2018fast,liu2012supervised,liu2017cross,luo2018sdmch,jiang2019discrete} and deep cross-modal hashing methods \cite{cao2016correlation,cao2016deep,wang2017adversarial,li2018self,erin2017cross,shen2017deep}. Shallow cross-modal hashing methods mainly use hand-crafted features to learn projections for mapping each example into a binary code. The feature extraction procedure in shallow cross-modal hashing methods is independent of the hash codes learning procedure. It means that the shallow cross-modal hashing methods may not achieve satisfactory performance in real applications, because the hand-crafted features might not be optimally suitable for hash codes optimizing procedure. Compared with shallow cross-modal hashing methods, deep cross-modal hashing methods can integrate feature learning and hash codes learning into a same framework, and capture non-linear correlations among cross-modal instances more effectively to get better performance, where each instance contains two correlated data-points of different modalities like image-text pairs. 

However, the existing deep cross-modal hashing methods either cannot learn a unified hash code for the two correlated data-points of different modalities in a database instance or cannot guide the learning of unified hash codes by the feedback of hashing function learning procedure, to enhance the retrieval accuracy. 
First, most deep cross-modal hashing methods assume that there are different hash codes for the two correlated data-points of different modalities in a database instance, and then try to decrease the gap between two hash codes through optimizing certain pre-defined loss functions. Thus, they just learn the similar hash codes for two correlated data-points of different modalities in a same instance, and cannot obtain unified hash codes. However, the unified hash code schema has been proved that it can enhance the retrieval accuracy \cite{rasiwasia2010new,zhou2014latent,ding2014collective}. Second, as far as we know, until now there is only one deep cross-modal hashing method that can learn unified hash codes \cite{erin2017cross}. The method is a two step framework. It first learns unified hash codes for instances in a database, and then utilizes the learned unified hash codes to learn modal-specific hashing function. It means the deep hashing method cannot guide the learning of unified hash codes by the feedback of hashing function learning procedure.

To address the issues above, in this paper, we propose a novel Deep Cross-Modal Hashing with Hashing Functions and Unified Hash Codes Jointly Learning, called DCHUC. DCHUC can jointly learn unified hash codes for database instances and modal-specific hashing functions for unseen query points in an end-to-end framework. 
More specifically, by minimising the objective function, DCHUC uses a four-step iterative scheme to optimize the unified hash codes of the database instances and the hash codes of query data-points generated by the learned hashing networks. With the iterative optimization algorithm, the learned unified hash codes can guide the hashing functions learning procedure; Meanwhile, the learned hashing function can feedback to guide the unified hash codes optimizing procedure. Moreover, the objective function consists of a hashing loss and a classification loss. The hashing loss is used to make the learned hash codes can preserve both inter-modal and intra-modal similarity, and the classification loss can be used to make the learned hashing codes preserve more discriminative semantic information.

In addition, because the training phase of deep models is typically time-consuming, so it is hard to use all instances in a large-scale database to train hashing model. Inspired by ADSH \cite{jiang2018asymmetric}, we use an asymmetric scheme to reduce the training time complexity to O(mn). Specially, we samples $m$ anchors instances from $n$ database instances ($m \ll n$) to approximate query datasets, and constructs an asymmetric affinity to supervise hashing functions learning for unseen query instances and unified hash codes optimizing for instances in a database. 

To summarize, the main contributions of DCHUC are outlined as follows:
\begin{itemize}
	\item To the best of our knowledge,  DCHUC is the first deep method that can jointly learn unified hash codes for database instances and hashing functions for unseen query points in an end-to-end framework. By using the end-to-end framework, our method can get the high-quality hash codes to improve the retrieval accuracy.
	\item By treating the query instances and database instances in an asymmetric way, DCHUC can use the whole set of database instances in training phase to generate higher-quality hash codes even if the size of a database is large.  
	\item Experiments on three large-scale datasets show that DCHUC can outperform the state-of-the-art cross-modal hashing baselines in real applications.  
\end{itemize}

\begin{figure*}[tb]
	\centering
	\includegraphics[width=\textwidth]{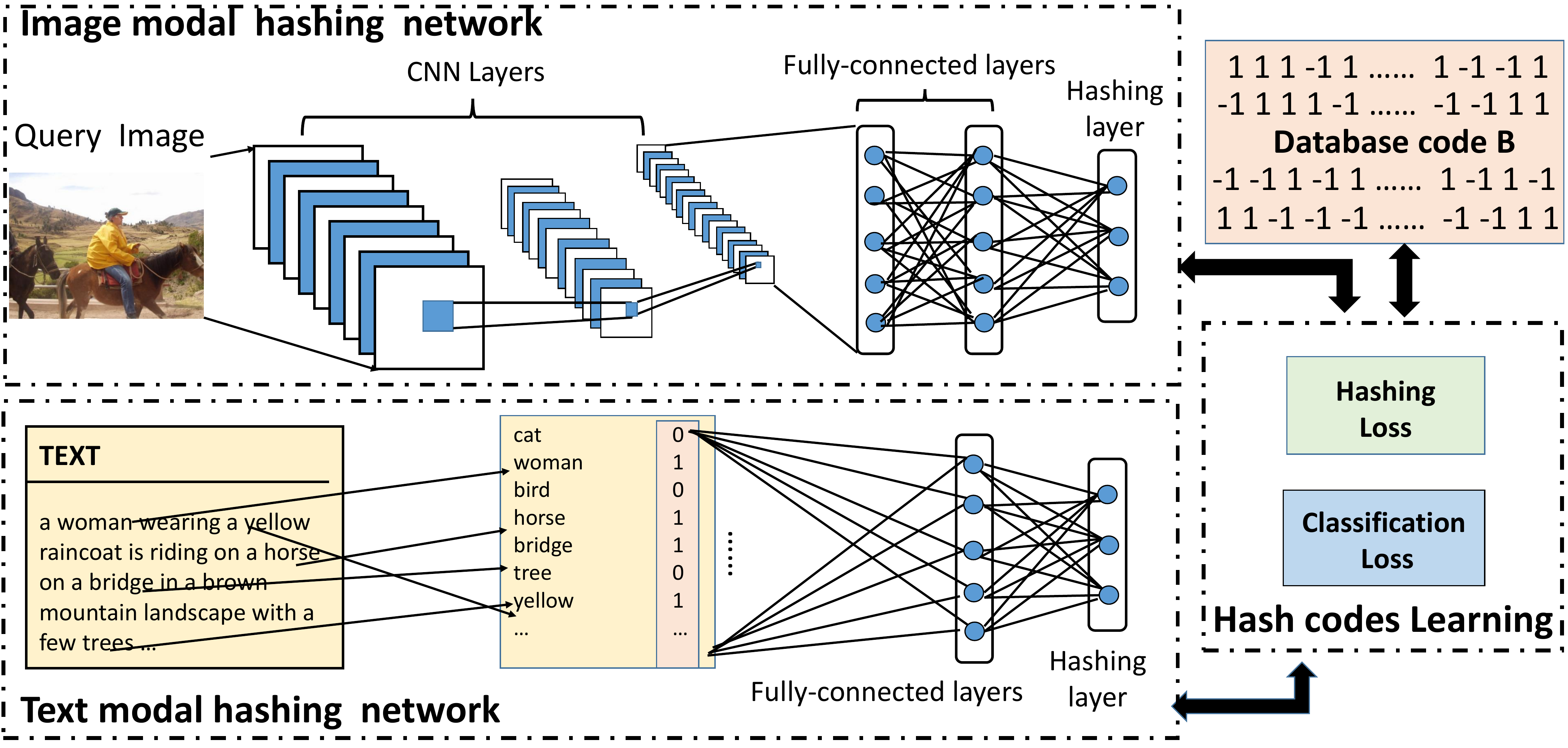}
	\caption{The DCHUC learning framework. It contains three parts: image modal hashing network, text modal hashing network and hash code optimizing. The hash codes of image modal query data-points can be generated by the image modal hashing network with an element-wise function $sign(\cdot)$, and the hash codes of text modal data-points can be generated text modal hashing network with an element-wise function $sign(\cdot)$. In the hash codes optimizing part, a four-step iterative scheme is used to optimize hash codes for both database instances and query instances by minimising the hashing loss and the classification loss}
	\label{fig_architecture}
\end{figure*} 
\section{Related Work}
In this section, we briefly review the related works of cross-modal hashing methods, including shallow cross-modal hashing methods and deep cross-modal hashing methods.
\subsection{Shallow Cross-Modal Hashing Methods}
Shallow cross-modal hashing methods \cite{bronstein2010data,zhang2014large,kumar2011learning,zhou2014latent,luo2018sdmch,jiang2019discrete} mainly use hand-crafted features to learn a single pair of linear or non-linear projections to map each example into a binary vector. The representative methods in this category include Cross Modality Similarity Sensitive Hashing (CMSSH) \cite{bronstein2010data}, Semantic Correlation Maximization (SCM) \cite{zhang2014large}, Cross View Hashing (CVH) \cite{kumar2011learning},Latent Semantic Sparse Hashing (LSSH) \cite{zhou2014latent}, Collective Matrix Factorization Hashing (CMFH) \cite{ding2014collective}, Semantics Preserving Hashing (SePH) \cite{lin2015semantics}, Supervised Discrete Manifold-embedded Cross-Modal Hashing (SDMCH) \cite{luo2018sdmch}, Discrete Latent Factor hashing (DLFH) \cite{jiang2019discrete} and Discrete Cross-modal Hashing (DCH) \cite{xu2017learning}. CMSSH is a supervised hashing methods, which designs a cross-modal hashing method by preserving the intra-class similarity via eigen-decomposition and boosting. SCM utilizes label information to learn a modality-specific transformation, and preserves the maximal correlation between modalities. CVH presents an unsupervised cross-modal spectral hashing method so that the cross-modality similarity is also preserved in the learned hash functions. LSSH utilizes sparse coding and matrix factorization in the common space to obtain a unified binary by a latent space learning method. CMFH learns a unified binary hash code by performing matrix factorization with latent factor model in the training stage. SePH generates a unified binary hash code by constructing an affinity matrix in a probability distribution while at the same time minimizing the Kullback-Leibler divergence. SDMCH generates binary hash codes by exploiting the non-linear manifold structure of data and constructing the correlations among heterogeneous multiple modalities with semantic information. DLFH directly learns the binary hash codes without continuous relaxation by a discrete latent factor model. DCH jointly learns the unified binary codes and the modality-specific hash functions under the classification framework with discrete optimization algorithm.

Despite of significant progress in this category has been achieved, the performance of hand-crafted feature based methods are still unsatisfactory in many real-world applications. Because the feature extraction procedure is independent of the hash-code learning procedure in hand-crafted feature based methods, which means that the hand-crafted features might not be optimally suitable for the hash codes optimizing procedure.
\subsection{Deep Cross-Modal Hashing Methods}
Recently, deep cross-modal hashing methods \cite{cao2016deep,jiang2017deep,cao2016correlation,wang2017adversarial,li2018self,erin2017cross} have been proposed to achieve promising performance due to the powerful arbitrary non-linear representation of deep neural network. For example, deep visual-semantic hashing (DVSH) \cite{cao2016deep} learns a visual semantic fusion network with cosine hinge loss to generate the binary codes and learns modality-specific deep networks to obtain hashing functions. However, DVSH can only be used for some special cross-modal cases where one of the modalities have to be temporal dynamics. Deep cross-modal hashing (DCMH) \cite{jiang2017deep} utilized a negative log-likelihood loss to generate cross-modal similarity preserving hash codes by an end-to-end deep learning framework. Correlation Autoencoder Hashing (CAH) \cite{cao2016correlation} learns hashing functions by designing an auto-encoder architecture to jointly maximize the feature and semantic correlation between different modalities. Adversarial cross-modal retrieval (ACMR) \cite{wang2017adversarial} utilizes a classification manner with adversarial learning approach to discriminate between different modalities and generate binary hash codes. Self-supervised adversarial hashing (SSAH) \cite{li2018self} generates binary hash codes by utilizing two adversarial networks to jointly model different modalities and capture their semantic relevance under the supervision of the learned semantic feature. Cross-modal deep variational hashing (CMDVH) \cite{erin2017cross} uses a two step framework. In the first step the method learns unified hash code for image-text pair in a database, and utilize the learned unified hash codes to learn hashing functions in the second step. Thus, for CMDVH, the learned hashing function in the second stage cannot give feedback to guide unified hash codes optimizing.

Typically, deep cross-modal hashing methods can outperform shallow hashing methods in terms of retrieval accuracy. However, most of existing deep cross-modal hashing methods cannot bridge the modality gap well to generate unified hash codes for image-text pairs in a database. Although CMDVH can generates unified binary codes for points of modalities, its hashing function learning procedure cannot feedback to guide the unified hash codes optimizing. Hence, CMDVH cannot get the optimal unified hash codes to bridge the modality gap well. Furthermore, please note that, although DCH can jointly learn unified hash codes for instances in a database and hashing functions for query instances, it is a shallow hashing method. Its feature extraction procedure is independent of the hash codes learning procedure, and DCH need use all the database instances to lean hashing functions which means it is hard to reconstruct DCH to a deep architecture. Thus, we propose a novel deep hashing method that can learn the unified hash codes for instances in a database and hashing functions for query instances in an end-to-end framework. 
\section{Our Method}
\subsection{Problem Definition}
Assume that we have $n$ training instances in a database, and each instance has two modal data points. Without loss of generality, we use image-text databases for illustration in this paper, which means that each instance in the database has both a data point of text modality and a data point of image modality. We use $\boldsymbol{O} = \{\boldsymbol{o}_i\}^n_{i = 1}$ to denote a cross-modal dataset with n instances, and $\boldsymbol{o}_i = (\boldsymbol{x}_i, \boldsymbol{y}_i, \boldsymbol{l}_i)$, where $\boldsymbol{x}_i$ and $\boldsymbol{y}_i$ denote the original image and text points in the $i^{th}$ instance $\boldsymbol{o}_i$, respectively. $\boldsymbol{l}_i = [\boldsymbol{l}_{i1},\boldsymbol{l}_{i2},\dots,\boldsymbol{l}_{ic} ]^T$ is the label annotation assigned to $\boldsymbol{o}_i$, where c is the class number. If $\boldsymbol{o}_i$ belongs to the $j^{th}$ class $\boldsymbol{l}_{ij} = 1$, otherwise $\boldsymbol{l}_{ij} = 0$. Furthermore, a pairwise similarity matrix $\boldsymbol{S} \in \{-1, +1\}^{n \times n}$ is used to describe the semantic similarities between two instances. If $\boldsymbol{S}_{ij} = 1$, it means that $\boldsymbol{o}_i$ is semantically similar to $\boldsymbol{o}_j$, otherwise $\boldsymbol{S}_{ij} = -1$. Specifically, if two instances $\boldsymbol{o}_i$ and $\boldsymbol{o}_j$ are annotated by multiple labels, we define $\boldsymbol{S}_{ij} = 1$ when $\boldsymbol{o}_i$ and $\boldsymbol{o}_j$ share as least one label, otherwise $\boldsymbol{S}_{ij} = -1$.

Given the above database $\boldsymbol{O}$ and similarity information $\boldsymbol{S}$, the goal of DCHUC is to learn the similarity-preserving hash codes $\boldsymbol{B} = \{\boldsymbol{b}_i\}^n_{i = 1} \in \{-1, +1\}^{n \times k}$ for instances in the database, where $k$ is the length of each binary code and $\boldsymbol{b}_i$ denotes the learned hash code for the instance $\boldsymbol{o}_i$, i.e., a unified hash code for the image-text pair $\boldsymbol{x}_i$ and $\boldsymbol{y}_i$. Meanwhile, the Hamming distance between $\boldsymbol{b}_i$ and $\boldsymbol{b}_j$ should be as small as possible when $\boldsymbol{S}_{ij} = 1$. Otherwise, the Hamming distance should be as large as possible. Moreover, in order to generate a binary code for any unseen image modal query point $\boldsymbol{x_q}$ or text modal query point $\boldsymbol{y_q}$, DCHUC should learn two modal-specific hashing functions $\boldsymbol{b}_{\boldsymbol{x}_q} = \mathcal{\boldsymbol{F}}(\boldsymbol{x}_q) \in \{-1, +1\}^k$ and $\boldsymbol{b}_{\boldsymbol{y}_q} = \mathcal{\boldsymbol{P}}(\boldsymbol{y}_q) \in \{-1, +1\}^k$, respectively. In order to learn the two hash functions, we sample a subset or use the whole set of  $\boldsymbol{O}$ as the query set $\boldsymbol{Q} = \boldsymbol{O}^{\Phi}$ for training, where $\boldsymbol{O}^{\Phi}$ denotes the query instances indexed by $\Phi$ from the database $\boldsymbol{O}$. Moreover, we use $\Upsilon = \{1, 2, \dots, n\}$ to denote the indices of all the database instances and $\Phi = \{i_1, i_2, \dots,i_m\} \subseteq \Upsilon$ to denote the indices of the $m$ sampled query instances, and $\boldsymbol{X}^\Phi$ and $\boldsymbol{Y}^\Phi$ denote image modal points and text modal points in query set $\boldsymbol{Q}$, respectively. Correspondingly, the similarity between query instances and database instances can be denoted as $\boldsymbol{S}^\Phi \in \{-1, +1\}^{m \times n}$, which is formed by the rows of $\boldsymbol{S}$ indexed by $\Phi$. In addition, in this paper, $sign(\cdot)$ is an element-wise sign function which returns $1$ if the element is positive and returns $-1$ otherwise. 
\subsection{Deep Cross-Modal Hashing with Hashing Functions and Unified Hash Codes Jointly Learning}
The model architecture for DCHUC is shown in Fig. \ref{fig_architecture}, which contains three parts: \textbf{image modal hashing network}, \textbf{text modal hashing network}  and \textbf{hash codes optimizing}.

For the image modal hashing network part, it contains a convolutional neural network (CNN) which is adapted from Alexnet \cite{krizhevsky2012imagenet}. The CNN component contains eight layers. The first seven layers are the same as those in Alexnet \cite{krizhevsky2012imagenet}. The eighth layer is a fully-connected layer with the output being the learned image features, which is named as hashing layer. The hashing layer contains $k$ units where $k$ is the length of hash codes. An activation function  $tanh(\cdot)$ is used to make the output features close to $"-1"\ or\ "+1"$. We use $\boldsymbol{v}_i = \mathcal{F}(\boldsymbol{x}_i;\boldsymbol{\Theta}) \in \mathcal{R}^k$ to denote the final output features of the image modal hashing network.

For the text modal hashing network part, a neural network containing two fully-connected layers is used to learn text modal features. We represent each text point $\boldsymbol{y}_i$ as a bag-of-words (BoW) vector, and use the BoW as the input of the two-fully-connected neural network. The first fully-connected layer has $10,240$ hidden units, and the activation function for the first fully-connect layer is RELU \cite{krizhevsky2012imagenet}. The second fully-connected layer is also named as hashing layer with $k$ nodes. Similar to the image feature learning part, a $tanh(\cdot)$ function is used as an activation function to make the output features close to $"-1"\ or\ "+1"$. We use $\boldsymbol{t}_i = \mathcal{P}(\boldsymbol{y}_i;\boldsymbol{\Psi}) \in \mathcal{R}^k$ to denote the final output features of the text modal hashing network.

For the hash codes optimizing part, it will optimize hash codes for both database instances and query instances the objective function whose details will be introduced in section \ref{obf}. More specially, with a four-step iterative scheme, the unified hash codes $\boldsymbol{B}$ for database instances will be learned directly and the modal-specific hashing functions can be learned by back-propagation algorithm which will be introduced in section \ref{optim} in detail. Furthermore, the hash codes for query instances are generated by the final output features of modal-specific hashing network with an element-wise function $sign(\cdot)$. Specifically, for an image modal query point $\boldsymbol{x_i}$, we can get its binary hash codes $\boldsymbol{h}_i = sign(\boldsymbol{v}_i)$; for a text modal query point $\boldsymbol{y_i}$, its binary hash codes can be generated by $\boldsymbol{g}_i = sign(\boldsymbol{t}_i)$.

\subsection{Objective Function}
\label{obf}
The goal of DCHUC is to map instances in the database and the unseen query data-points into a semantic similarity-preserving Hamming space where the hash codes of data-points from the same categories should be similar no mater which modalities they belong to, and the hash codes of data-points from different categories should be dissimilar. In the following, we present more details about the objective function of our CMDAH.

In order to bridge the gap across different modalities well, we first assume the image point $\boldsymbol{x}_i$ and text point $\boldsymbol{y}_i$ for any instance $\boldsymbol{o}_i$ in a database share the same hash code $\boldsymbol{b}_i$, i.e., learn a unified hash code $\boldsymbol{b}_i$ for an image-text pair $\boldsymbol{x}_i$ and $\boldsymbol{y}_i$. Thus, the hash code $\boldsymbol{b}_i$ can preserve the image modal information and text modal information at the same time. Moreover, in order to make the learned hash codes of instances in the database and the hash codes of query data-points generated by the learned hashing functions can preserve the semantic similarity, one common way is to minimize the Frobenius norm loss between the semantic similarities and inner product of binary code pairs. Therefore, the hashing loss can be defined as follow:
\begin{equation} 
\begin{aligned}
\min\limits_{\boldsymbol{B}, \boldsymbol{H}, \boldsymbol{G}} \mathcal{L}_h &=\left\|\boldsymbol{H}\boldsymbol{B}^T - k\boldsymbol{S}^\Phi \right\|_F^2 + \left\|\boldsymbol{G}\boldsymbol{B}^T - k\boldsymbol{S}^\Phi\right\|_F^2 \\
&\ \ \ \  + \mu\left\|\boldsymbol{H}\boldsymbol{G}^T - k\boldsymbol{S}^\Phi_\Phi \right\|_F^2\\
&s.t.\ \ \ \boldsymbol{B} = [\boldsymbol{b}_1, \boldsymbol{b}_2, \cdots, \boldsymbol{b}_n]^T\in \{-1, +1\}^{n\times k}, \\
&\ \ \ \ \ \boldsymbol{H} = sign(\boldsymbol{V}) \in \{-1, +1\}^{m \times k},\\
&\ \ \ \ \ \boldsymbol{G} = sign(\boldsymbol{T}) \in \{-1, +1\}^{m \times k}
\end{aligned} \label{l1}
\end{equation}
where $\mu$ is a hype-parameter, $\boldsymbol{B}\in \{-1, +1\}^{n\times k}$ denotes the unified binary hash codes for $n$ database instances; $\boldsymbol{S}^\Phi_\Phi$ denotes the columns of $\boldsymbol{S}^\Phi$ indexed by $\Phi$; $\boldsymbol{H} \in \{-1, +1\}^{m \times k}$ denotes the binary hash codes for $m$ images modal query data-points, and $\boldsymbol{G} \in \{-1, +1\}^{m \times k}$ denotes the binary hash codes for $m$ text modal query data-points; $\boldsymbol{V} = [\boldsymbol{v}_{i_1}, \boldsymbol{v}_{i_2}, \cdots, \boldsymbol{v}_{i_m}]^T$ is the output of images modal hashing network for image query set $\boldsymbol{X}^\Phi$, and $\boldsymbol{T} = [\boldsymbol{t}_{i_1}, \boldsymbol{t}_{i_2}, \cdots, \boldsymbol{t}_{i_m}]^T$ is the output of text modal hashing network for text query set $\boldsymbol{Y}^\Phi$. 

Furthermore, in order to make the learned hashing codes preserve more discriminative semantic information, we expect the learned hashing codes can be ideal for classification too. Then the classification loss function can be defined as follow:
\begin{equation} 
\begin{aligned}
\min\limits_{\boldsymbol{B}, \boldsymbol{H}, \boldsymbol{G}, \boldsymbol{W}} \mathcal{L}_c &= \alpha(\left\|\boldsymbol{H}\boldsymbol{W} - \boldsymbol{L}^\Phi \right\|_F^2 + \left\|\boldsymbol{G}\boldsymbol{W} - \boldsymbol{L}^\Phi \right\|_F^2) \\
&+ \beta\left\|\boldsymbol{B}\boldsymbol{W} - \boldsymbol{L}\right\|_F^2  + \eta\left\|\boldsymbol{W} \right\|_F^2\\
&s.t.\ \ \boldsymbol{B} = [\boldsymbol{b}_1, \boldsymbol{b}_2, \cdots, \boldsymbol{b}_n]^T\in \{-1, +1\}^{n\times k}, \\
&\ \ \ \ \ \ \boldsymbol{H} = sign(\boldsymbol{V}) \in \{-1, +1\}^{m \times k},\\
&\ \ \ \ \ \ \boldsymbol{G} = sign(\boldsymbol{T}) \in \{-1, +1\}^{m \times k}.
\end{aligned} \label{l2}
\end{equation}
where $\boldsymbol{L} = [\boldsymbol{l}_1, \boldsymbol{l}_2, \cdots, \boldsymbol{l}_n]^T\in \{0, 1\}^{n\times c}$ is the label matrix of instances in the database $\boldsymbol{O}$, and $\boldsymbol{L}^{\Phi} \in {0, 1}^{m\times c}$ denotes the label matrix of query instances indexed by $\Phi$ from the label matrix $\boldsymbol{L}$. $\boldsymbol{W} = [\boldsymbol{w}_1, \boldsymbol{w}_2,\cdots, \boldsymbol{w}_c] \in \mathcal{R}^{k\times c}$ and $\boldsymbol{w}_j \in \mathcal{R}^{k\times 1}$ is the classification projected vector of the class $j$.

Thus, our objective hashing function can be defined as follow:
\begin{equation} 
\begin{aligned}
&\ \ \ \ \ \ \ \ \min\limits_{\boldsymbol{B}, \boldsymbol{H}, \boldsymbol{G}, \boldsymbol{W}} \mathcal{L} = \mathcal{L}_h + \mathcal{L}_c\\
&s.t.\ \ \boldsymbol{B} = [\boldsymbol{b}_1, \boldsymbol{b}_2, \cdots, \boldsymbol{b}_n]^T\in \{-1, +1\}^{n\times k}, \\
&\ \ \ \ \ \ \boldsymbol{H} = sign(\boldsymbol{V}) \in \{-1, +1\}^{m \times k},\\
&\ \ \ \ \ \ \boldsymbol{G} = sign(\boldsymbol{T}) \in \{-1, +1\}^{m \times k}.
\end{aligned} \label{ln}
\end{equation}

However, it is hard to learn functions $\boldsymbol{V} = \mathcal{F}(\boldsymbol{X}^\Phi;\boldsymbol{\Theta})^T$ and $\ \boldsymbol{T} = \mathcal{P}(\boldsymbol{Y}^\Phi;\boldsymbol{\Psi})^T$ due to the derivation of $sign(\cdot)$ function is $0$. Moreover, considering the query set is sampled from the whole database, the hash codes generated by the learned hashing function should be the same with the directly learned hash codes, i.e., if an instance $\boldsymbol{o}_i$ in the database is sampled as query instance, then the hash code $h_i$ for image modality data-point and $g_i$ for text modality data-point in $\boldsymbol{o}_i$ should be the same with $\boldsymbol{b}_i$. Thus, we can further reformulate Formula (\ref{ln}) as:
\begin{equation}
\begin{aligned}
\min\limits_{\boldsymbol{B}, \boldsymbol{V}, \boldsymbol{T}} \mathcal{L}&=\left\|\boldsymbol{V}\boldsymbol{B}^T - k\boldsymbol{S}^\Phi \right\|_F^2 + \left\|\boldsymbol{T}\boldsymbol{B}^T - k\boldsymbol{S}^\Phi\right\|_F^2 \\
&+ \mu\left\|\boldsymbol{V}\boldsymbol{T}^T - k\boldsymbol{S}^\Phi_\Phi \right\|_F^2
+ \beta\left\|\boldsymbol{B}\boldsymbol{W} - \boldsymbol{L}\right\|_F^2\\
&+ \alpha(\left\|\boldsymbol{V}\boldsymbol{W} - \boldsymbol{L}^\Phi \right\|_F^2 + \left\|\boldsymbol{T}\boldsymbol{W} - \boldsymbol{L}^\Phi \right\|_F^2) \\
&+ \eta\left\|\boldsymbol{W} \right\|_F^2 + \gamma\left\|\boldsymbol{B}^\Phi -0.5(\boldsymbol{V} + \boldsymbol{T})\right\|_F^2\\
&s.t.\ \ \ \boldsymbol{B}\in \{-1, +1\}^{n\times k}.
\end{aligned} \label{l}
\end{equation}
where $\alpha, \beta, \eta, \gamma, \mu$ are hyper-parameters, $\boldsymbol{B}^\Phi\in \{-1, +1\}^{m\times k} $ is formed by the rows of $\boldsymbol{B}$ indexed by $\Phi$.

\begin{algorithm}[t]
	\caption{Learning algorithm for DCHUC}
	\label{alg:A}
	\begin{algorithmic}[1]
		\REQUIRE
		Database instances $\boldsymbol{O} = \{\boldsymbol{X}, \boldsymbol{Y}, \boldsymbol{L}\}$,  the length of  hash codes $k$.
		\ENSURE 
		Database instances codes $\boldsymbol{B}$, image modal hashing network parameters $\boldsymbol{\Theta}$ and text modal hashing network parameters $\boldsymbol{\Psi}$.
		\STATE Initialize parameters:  $\boldsymbol{B}$, $\boldsymbol{\Theta}$, $\boldsymbol{\Psi}$, $\alpha$, $\eta$, $\gamma$. learning rate: $lr$, iteration number: $t_{out}$, $t_{in}$, the size of mini-batch $z=64$ (see Implementation Details).
		\STATE Utilize label $\boldsymbol{L}$ to generate similarity matrix $\boldsymbol{S}$.
		\REPEAT 
		\STATE Randomly generate index set $\Phi$ and sample $m$ instances $\boldsymbol{O}^\Phi = \{\boldsymbol{X}^\Phi, \boldsymbol{Y}^\Phi, \boldsymbol{L}^\Phi\}$ from database $\boldsymbol{O}$ as query set. Select $\boldsymbol{S}^\Phi$ from $\boldsymbol{S}$  
		\FOR{$iter = 1,2,\cdots, t_{in}$}
		\FOR{$iter\_batch = 1, 2, \cdots, m/z$}
		\STATE Randomly sample $z$ image points from $\boldsymbol{X}^\Phi$ as a mini-batch
		\STATE Update parameter $\Theta$ based on Formula (\ref{lv}) 
		\ENDFOR
		\FOR{$iter\_batch = 1, 2, \cdots, m/z$}
		\STATE Randomly sample $z$ image points from $\boldsymbol{Y}^\Phi$ as a mini-batch
		\STATE Update parameter $\Psi$ based on Formula (\ref{lt}) 
		\ENDFOR
		\ENDFOR
		\FOR{$iter\_bit = 1, 2, \cdots, k$}
		\STATE Update $\boldsymbol{B}_{*iter\_bit}$ based on Formula (\ref{l8}) 
		\ENDFOR
		\STATE Update $\boldsymbol{W}$ based on Formula (\ref{l9}) 
		\UNTIL Up to $t_{out}$
	\end{algorithmic}
\end{algorithm}
\subsection{Optimization}
\label{optim}
In order to optimize Formula (\ref{l}), we propose a four-step iterative scheme as shown below. More specifically, in each iteration we sample a query set from the database and then carry out our learning algorithm based on both the query set and database. The whole four-step learning algorithm for DCHUC is briefly outlined in Algorithm 1, and the detailed derivation steps will be introduced in the following content of this subsection.
\subsubsection{Learn $\boldsymbol{\Theta}$ with $\boldsymbol{\Psi}$, $\boldsymbol{B}$ and $\boldsymbol{W}$ fixed}
When $\boldsymbol{\Psi}$, $\boldsymbol{B}$ and $\boldsymbol{L}$ are fixed, we update the parameter $\boldsymbol{\Theta}$ of image hashing network by using a mini-batch stochastic gradient descent with back-propagation (BP) algorithm. More specifically, for each sampled image point $\boldsymbol{x}_i$ in $\boldsymbol{X}^\Phi$, we first compute the following gradient:
\begin{equation}
\begin{aligned}
\frac{\partial \mathcal{L}}{\partial \boldsymbol{v}_i} &= 2\sum\limits_{j = 1}^{n}[(\boldsymbol{v}_i^T\boldsymbol{b}_j - k\boldsymbol{S}_{ij}^\Phi)\boldsymbol{b}_j] + 2\mu\sum\limits_{j = 1}^{m}[(\boldsymbol{v}_i^T\boldsymbol{t}_j - k{\boldsymbol{S}^\Phi_\Phi}_{ij})\boldsymbol{t}_j]\\
&\ \ \ \ + 2\alpha\sum\limits_{j = 1}^{c}[(\boldsymbol{v}_i^T\boldsymbol{w}_j - \boldsymbol{L}_{ij}^\Phi)\boldsymbol{w}_j] + \gamma(\boldsymbol{v}_i + \boldsymbol{t}_i - 2\boldsymbol{b}_i)
\label{lv}
\end{aligned}
\end{equation}
Then we can compute $\frac{\partial \mathcal{L}}{\partial \boldsymbol{\Theta}}$ based on $\frac{\partial \mathcal{L}}{\partial \boldsymbol{v}_i}$ by using chain rule, and use BP to update the parameter $\boldsymbol{\Theta}$.
\subsubsection{Learn $\boldsymbol{\Psi}$ with $\boldsymbol{\Theta}$, $\boldsymbol{B}$ and $\boldsymbol{L}$ fixed}
When $\boldsymbol{\Theta}$ and $\boldsymbol{B}$ are fixed, we also update the parameter $\boldsymbol{\Psi}$ of text hashing network by using a mini-batch stochastic gradient descent with BP algorithm. More specifically, for each sampled text point $\boldsymbol{y}_i$ in $\boldsymbol{Y}^\Phi$, we first compute the following gradient:
\begin{equation}
\begin{aligned}
\frac{\partial \mathcal{L}}{\partial \boldsymbol{t}_i} &= 2\sum\limits_{j = 1}^{n}[(\boldsymbol{t}_i^T\boldsymbol{b}_j - k\boldsymbol{S}_{ij}^\Phi)\boldsymbol{b}_j] + 2\mu\sum\limits_{j = 1}^{m}[(\boldsymbol{t}_i^T\boldsymbol{v}_j - k{\boldsymbol{S}^\Phi_\Phi}_{ij})\boldsymbol{v}_j]\\
&\ \ \ \ + 2\alpha\sum\limits_{j = 1}^{c}[(\boldsymbol{t}_i^T\boldsymbol{w}_j - \boldsymbol{L}_{ij}^\Phi)\boldsymbol{w}_j] + \gamma(\boldsymbol{v}_i + \boldsymbol{t}_i - 2\boldsymbol{b}_i)
\label{lt}
\end{aligned}
\end{equation}
Then we can compute $\frac{\partial \mathcal{L}}{\partial \boldsymbol{\Psi}}$ based on $\frac{\partial \mathcal{L}}{\partial \boldsymbol{t}_i}$ by using chain rule, and use BP to update the parameter $\boldsymbol{\Psi}$.
\subsubsection{Learn $\boldsymbol{B}$ with $\boldsymbol{\Theta}$, $\boldsymbol{\Psi}$ and $\boldsymbol{W}$ fixed}
When $\boldsymbol{\Theta}$, $\boldsymbol{\Psi}$ and $\boldsymbol{W}$ are fixed, we can reformulate Formula (\ref{l}) as follows:
\begin{equation}
\begin{aligned}
\min\limits_{\boldsymbol{B}} \mathcal{L}&=\left\|\boldsymbol{V}\boldsymbol{B}^T - k\boldsymbol{S}^\Phi \right\|_F^2 + \left\|\boldsymbol{T}\boldsymbol{B}^T - k\boldsymbol{S}^\Phi\right\|_F^2\\
&\ \ \ \ + \beta \left\|\boldsymbol{B}\boldsymbol{W} - \boldsymbol{L} \right\|_F^2 + \gamma\left\|\boldsymbol{B}^\Phi -0.5 (\boldsymbol{V} + \boldsymbol{T}) \right\|_F^2 \\
&=\left\|\boldsymbol{V}\boldsymbol{B}^T\right\|^2_F - 2ktr(\boldsymbol{B}\boldsymbol{V}^T\boldsymbol{S}^\Phi) + \left\|\boldsymbol{T}\boldsymbol{B}^T\right\|^2_F \\
&\ \ \ \ - 2k tr(\boldsymbol{B}\boldsymbol{T}^T\boldsymbol{S}^\Phi) + \beta\left\|\boldsymbol{B}\boldsymbol{W}\right\|^2_F - 2\beta tr(\boldsymbol{B}\boldsymbol{W}\boldsymbol{L}^T)\\
&\ \ \ \ - \gamma tr(\boldsymbol{B}^\Phi (\boldsymbol{V}^T + \boldsymbol{T}^T)) + const\\
&s.t.\ \ \boldsymbol{B}\in \{-1, +1\}^{n\times k}\\
\end{aligned} 
\label{l5}
\end{equation}
where $const$ is a constant independent of $\boldsymbol{B}$ and $tr(\cdot)$ is the trace norm. For convenience of calculations, we can further reformulate Formula (\ref{l5}) as follows:
\begin{equation}
\begin{aligned}
\min\limits_{\boldsymbol{B}} \mathcal{L}&=\left\|\boldsymbol{V}\boldsymbol{B}^T\right\|^2_F + \left\|\boldsymbol{T}\boldsymbol{B}^T\right\|^F_2 + \beta\left\|\boldsymbol{B}\boldsymbol{W}\right\|^2_F - tr(\boldsymbol{B}(\gamma \mathring{\boldsymbol{V}}^T \\
&\ \ \ \ + \gamma\mathring{\boldsymbol{T}}^T + 2k\boldsymbol{V}^T\boldsymbol{S}^\Phi +2k \boldsymbol{T}^T\boldsymbol{S}^\Phi +2\beta \boldsymbol{W}\boldsymbol{L}^T)) \\
&\ \ \ \ + const\\
&= \left\|\boldsymbol{V}\boldsymbol{B}^T\right\|^2_F + \left\|\boldsymbol{T}\boldsymbol{B}^T\right\|^F_2 + \beta\left\|\boldsymbol{B}\boldsymbol{W}\right\|^2_F\\
&\ \ \ \ - tr(\boldsymbol{B}\boldsymbol{D}) + const\\
&s.t.\ \ \boldsymbol{B}\in \{-1, +1\}^{n\times k}\\
\end{aligned} 
\label{l6}
\end{equation}
where $ \boldsymbol{D} = \gamma \mathring{\boldsymbol{V}}^T + \gamma\mathring{\boldsymbol{T}}^T + 2k\boldsymbol{V}^T\boldsymbol{S}^\Phi +2k \boldsymbol{T}^T\boldsymbol{S}^\Phi +2\beta \boldsymbol{W}\boldsymbol{L}^T$; $\ \mathring{\boldsymbol{V}} = [\mathring{\boldsymbol{v}}_1, \mathring{\boldsymbol{v}}_2, \cdots, \mathring{\boldsymbol{v}}_n]^T$; $\ \mathring{\boldsymbol{T}} = [\mathring{\boldsymbol{t}}_1, \mathring{\boldsymbol{t}}_2, \cdots, \mathring{\boldsymbol{t}}_n]^T$, and $\mathring{\boldsymbol{v}}_i$, $\mathring{\boldsymbol{t}}_i$ are respectively defined as follows:
\begin{equation}
\mathring{\boldsymbol{v}}_i = \left\{
\begin{array}{lrc}
\boldsymbol{v}_i,   && if\ \ i \in \Phi,\\
\boldsymbol{0},    && if\ \ i \notin \Phi.
\end{array} \right.
\end{equation}
\begin{equation}
\mathring{\boldsymbol{t}}_i = \left\{
\begin{array}{lrc}
\boldsymbol{t}_i,   && if\ \ i \in \Phi,\\
\boldsymbol{0},    && if\ \ i \notin \Phi.
\end{array} \right.
\end{equation}

The above Formula (\ref{l6}) is NP hard. Inspired by SDH \cite{shen2015supervised}, the binary codes $\boldsymbol{B}$ can be learned by the discrete cyclic coordinate descent (DCC) method. It means that we directly learn hash codes $\boldsymbol{B}$ bit by bit. Specifically, we update one column of $\boldsymbol{B}$ with the other column fixed. We let $\boldsymbol{B}_{*i}$ denotes the $i^{th}$ column of $\boldsymbol{B}$, and $\widehat{\boldsymbol{B}}_i$ denotes the matrix of $\boldsymbol{B}$ without the column $\boldsymbol{B}_{*i}$;  Let $\boldsymbol{V}_{*i}$ denotes the $i^{th}$ column of $\boldsymbol{V}$, and $\widehat{\boldsymbol{V}}_i$ denotes the matrix of $\boldsymbol{V}$ without the column $\boldsymbol{V}_{*i}$;  Let $\boldsymbol{T}_{*i}$ denotes the $i^{th}$ column of $\boldsymbol{T}$, and $\widehat{\boldsymbol{T}}_i$ denotes the matrix of $\boldsymbol{T}$ without the column $\boldsymbol{T}_{*i}$; Let $\boldsymbol{W}_{i*}$ denotes the $i^{th}$ row of $\boldsymbol{W}$, and $\widetilde{\boldsymbol{W}}_i$ denotes the matrix of $\boldsymbol{W}$ without the row $\boldsymbol{W}_{i*}$; Let $\boldsymbol{D}_{i*}$ denotes the $i^{th}$ row of $\boldsymbol{D}$, and $\widetilde{\boldsymbol{D}}_i$ denotes the matrix of $\boldsymbol{D}$ without the row $\boldsymbol{D}_{i*}$. Then we can optimize $\boldsymbol{B}_{*i}$ by the following function:
\begin{equation}
\begin{aligned}
\min\limits_{\boldsymbol{B}_{*i}} \mathcal{L}&= \left\|\boldsymbol{V}\boldsymbol{B}^T\right\|^2_F + \left\|\boldsymbol{T}\boldsymbol{B}^T\right\|^F_2 + \beta\left\|\boldsymbol{B}\boldsymbol{W}\right\|^2_F\\
&\ \ \ \  - tr(\boldsymbol{B}\boldsymbol{D}) + const\\
&=tr(\boldsymbol{B}_{*i}(2\boldsymbol{V}_{*i}^T\widehat{\boldsymbol{V}}_i \widehat{\boldsymbol{B}}_i^T + 2\boldsymbol{T}_{*i}^T\widehat{\boldsymbol{T}}_i \widehat{\boldsymbol{B}}_i^T + 2\beta \boldsymbol{W}_{i*}\widetilde{\boldsymbol{W}}_i^T\widehat{\boldsymbol{B}}_i^T\\
&\ \ \ \  - \widetilde{\boldsymbol{D}}_{i*}) ) + const\\
&s.t.\ \ \ \boldsymbol{B}_{*i}\in \{-1, +1\}^{n\times k}\\
\end{aligned} 
\label{l7}
\end{equation}
Finally, we can get the optimal solution of Formula (\ref{l7}):
\begin{equation}
\boldsymbol{B}_{*i} = -sign(2\widehat{\boldsymbol{B}}_i \widehat{\boldsymbol{V}}_i^T \boldsymbol{V}_{*i} + 2\widehat{\boldsymbol{B}}_i \widehat{\boldsymbol{T}}_i^T \boldsymbol{T}_{*i} + 2\beta\widehat{\boldsymbol{B}}_i\widetilde{\boldsymbol{W}}_i\boldsymbol{W}_{i*}^T - \widetilde{\boldsymbol{D}}_{i*}^T)
\label{l8}
\end{equation}
then we can use Formula (\ref{l8}) to update $\boldsymbol{B}$ bit by bit.
\subsubsection{Learn $\boldsymbol{W}$ with $\boldsymbol{\Theta}$, $\boldsymbol{\Psi}$ and $\boldsymbol{B}$ fixed}
When $\boldsymbol{\Theta}$, $\boldsymbol{\Psi}$ and $\boldsymbol{B}$ are fixed, we can reformulate Formula (\ref{l}) as follows:
\begin{equation}
\begin{aligned}
\min\limits_{\boldsymbol{W}} \mathcal{L}&=\alpha(\left\|\boldsymbol{V}\boldsymbol{W} - \boldsymbol{L}^\Phi \right\|_F^2 + \left\|\boldsymbol{T}\boldsymbol{W} - \boldsymbol{L}^\Phi \right\|_F^2) \\
&+ \beta\left\|\boldsymbol{B}\boldsymbol{W} - \boldsymbol{L}\right\|_F^2  + \eta\left\|\boldsymbol{W} \right\|_F^2\\
\end{aligned}
\label{lw}
\end{equation}
For Formula (\ref{lw}), it is easy to solve W by the regularized least squares problem, which has a closed-form solution:
\begin{equation}
\boldsymbol{W}=(\alpha\boldsymbol{V}^T\boldsymbol{V}+ \alpha\boldsymbol{T}^T\boldsymbol{T} + \beta\boldsymbol{B}^T\boldsymbol{B}+ \eta\boldsymbol{I})^{-1}(\alpha \mathring{\boldsymbol{V}} + \alpha \mathring{\boldsymbol{T}} +\beta \boldsymbol{B})^T\boldsymbol{L}
\label{l9}
\end{equation}

\begin{table*}[t]
	\begin{minipage}{\textwidth}
		\setlength{\abovecaptionskip}{0pt}
		\setlength{\belowcaptionskip}{0pt}
		\caption{MAP. The best accuracy is shown in boldface and the second best accuracy is underlined. The baselines are based on Alexnet features}
		\label{table_map}
		\resizebox{\linewidth}{!}{
			\begin{tabular}{|l|l||l|l|l|l||l|l|l|l||l|l|l|l|}
				\hline
				\multirow{2}{*}{Task} & \multirow{2}{*}{Method} & \multicolumn{4}{c||}{MIRFLICKR-25K} & \multicolumn{4}{c||}{IAPR TC-12}   & \multicolumn{4}{c|}{NUS-WIDE}     \\ \cline{3-14}
				&                         & 16bits  & 32bits & 48bits & 64bits & 16bits & 32bits & 48bits & 64bits & 16bits & 32bits & 48bits & 64bits \\ \hline \hline
				\multirow{8}{*}{$T \rightarrow I$}  & CCA-ITQ                     &0.599    &0.587   &0.582   &0.578   &0.403   &0.399   &0.396   &0.390   &0.426   &0.415   &0.410   &0.401   \\ 
				& SCM                     &0.639    &0.612   &0.584   &0.592   &0.438   &0.423   &0.414   &0.398   &0.403   &0.371   &0.349   &0.328   \\ 
				& DCH                    &0.759    &0.780   &0.793   &0.794   &0.536   &0.559   &0.564   &0.582   &\underline{0.619}   &\underline{0.652}   &\underline{0.653}   &\underline{0.681}   \\ 
				& DLFH                    &0.769    &\underline{0.796}   &\underline{0.805}   &\underline{0.809}   &0.470   &0.498   &0.516   &0.555   &0.599   &0.608   &0.619   &0.630   \\ 
				& DCMH                    &0.763    &0.771   &0.771   &0.779   &0.511   &0.525   &0.527   &0.535   &0.629   &0.642   &0.652   &0.662   \\  
				& CMDVH                   &0.612    &0.610   &0.553   &0.600   &0.381   &0.383   &0.396   &0.381   &0.371   &0.359   &0.399   &0.424   \\  
				& SSAH                    &\underline{0.783}    &0.793   &0.800   &0.783   &\underline{0.538}   &\underline{0.566}   &\underline{0.580}   &\underline{0.586}   &0.613   &0.632   &0.635   &0.633   \\  
				& DCHUC                   &\textbf{0.850}    &\textbf{0.857}   &\textbf{0.853}   &\textbf{0.854}   &\textbf{0.615}   &\textbf{0.666}   &\textbf{0.681}   &\textbf{0.693}   &\textbf{0.698}   &\textbf{0.728}   &\textbf{0.742}  &\textbf{0.749}  \\ \hline \hline
				\multirow{8}{*}{ $I \rightarrow T$ }  & CCA-ITQ                     &0.593    &0.582   &0.577   &0.574   &0.312   &0.311   &0.310   &0.309  &0.424  &0.412   &0.398   &0.387   \\  
				& SCM                     &0.626    &0.595   &0.588   &0.578   &0.313   &0.310   &0.309   &0.308   &0.395   &0.368   &0.353   &0.335   \\  
				& DCH                    &0.748    &0.786   &\underline{0.799}   &\underline{0.805}   &0.486   &0.486   &0.496   &0.502   &0.648   &\underline{0.678}   &\underline{0.699}   &\underline{0.708}   \\ 
				& DLFH                    &0.719    &0.732   &0.742   &0.748   &0.417   &0.451   &0.484   &0.490   &0.558   &0.578   &0.591   &0.593   \\ 
				& DCMH                    &0.721    &0.733   &0.729   &0.742   &0.464   &0.485   &0.490   &0.498   &0.588   &0.607   &0.615   &0.632   \\  
				& CMDVH                   &0.611    &0.626   &0.553   &0.598   &0.376   &0.373   &0.365   &0.376   &0.370   &0.373   &0.414   &0.425   \\ 
				& SSAH                    &\underline{0.779}    &\underline{0.789}   &0.796   &0.794   &\underline{0.539}   &\underline{0.564}   &\underline{0.581}   &\underline{0.587}   &\underline{0.659}   &0.666   &0.679   &0.667  \\ 
				& DCHUC                   &\textbf{0.878}    &\textbf{0.882}   &\textbf{0.880}   &\textbf{0.881}   &\textbf{0.630}   &\textbf{0.677}   &\textbf{0.695}   &\textbf{0.701}   &\textbf{0.750}   &\textbf{0.771}   &\textbf{0.783}   &\textbf{0.791}  \\ \hline
		\end{tabular}}
		\vspace{0.4cm}
	\end{minipage}
	\begin{minipage}{\textwidth}
		\setlength{\abovecaptionskip}{0pt}
		\setlength{\belowcaptionskip}{0pt}
		\caption{Precision@1000. The best accuracy is shown in boldface and the second best accuracy is underlined. The baselines are based on Alexnet features}
		\label{table_precision}
		\resizebox{\linewidth}{!}{
			\begin{tabular}{|l|l||l|l|l|l||l|l|l|l||l|l|l|l|}
				\hline
				\multirow{2}{*}{Task} & \multirow{2}{*}{Method} & \multicolumn{4}{c||}{MIRFLICKR-25K}                                & \multicolumn{4}{c||}{IAPR TC-12}                                   & \multicolumn{4}{c|}{NUS-WIDE}                                     \\ \cline{3-14} 
				&                         & 16bits  & 32bits & 48bits & 64bits & 16bits & 32bits & 48bits & 64bits & 16bits & 32bits & 48bits & 64bits         \\ \hline \hline
				\multirow{8}{*}{$T \rightarrow I$}  & CCA-ITQ                     & 0.690          & 0.676          & 0.666          & 0.652          & 0.491          & 0.492          & 0.488          & 0.482          & 0.622          & 0.672          & 0.684          & 0.683          \\ 
				& SCM                     & 0.749          & 0.714          & 0.675          & 0.639          & 0.504          & 0.506          & 0.523          & 0.497          & 0.598          & 0.576          & 0.532          & 0.668          \\  
				& DCH                    & \underline{0.848}          & 0.848          & 0.843          & 0.852          & \underline{0.664}          & \underline{0.695}          & \underline{0.701}          & \underline{0.712}          & \textbf{0.808}          & \underline{0.819}          & \underline{0.808}          & \underline{0.815}          \\  
				& DLFH                    & 0.834          & \underline{0.857}          & \underline{0.865}          & \underline{0.870}          & 0.563          & 0.604          & 0.638          & 0.660          & 0.685          & 0.707          & 0.717          & 0.735          \\ 
				& DCMH                    & 0.815          & 0.824          & 0.834          & 0.835          & 0.596          & 0.610          & 0.613          & 0.626          & 0.694          & 0.710          & 0.721          & 0.731          \\ 
				& CMDVH                   & 0.613          & 0.636          & 0.545          & 0.601          & 0.396          & 0.410          & 0.403          & 0.396          & 0.340          & 0.293          & 0.408          & 0.417          \\ 
				& SSAH                    & 0.824          & 0.834          & 0.846          & 0.855          & 0.641          & 0.664          & 0.674          & 0.677          & 0.701          & 0.729          & 0.736          & 0.731          \\ 
				& DCHUC                   & \textbf{0.896} & \textbf{0.897} & \textbf{0.890} & \textbf{0.888} & \textbf{0.711} & \textbf{0.760} & \textbf{0.771} & \textbf{0.782} & \underline{0.799}          & \textbf{0.825} & \textbf{0.839} & \textbf{0.849} \\ \hline \hline
				\multirow{8}{*}{$I \rightarrow T$}  & CCA-ITQ                     & 0.666          & 0.656          & 0.649          & 0.635        & 0.401          & 0.341          & 0.302          & 0.302          & 0.607          & 0.657          & 0.667          & 0.666          \\  
				& SCM                     & 0.738          & 0.704          & 0.676          & 0.660          & 0.376          & 0.349          & 0.324          & 0.315          & 0.606          & 0.565          & 0.550          & 0.504          \\ 
				& DCH                    & \underline{0.844}          & \underline{0.866}          & \underline{0.860}          & \underline{0.868}          & 0.593          & 0.604          & 0.612          & 0.617          & \underline{0.813}          & \underline{0.829}          & \underline{0.822}          & \underline{0.817}          \\ 
				& DLFH                    & 0.800          & 0.817          & 0.824          & 0.825          & 0.480          & 0.536          & 0.584          & 0.596          & 0.646          & 0.682          & 0.703          & 0.698          \\ 
				& DCMH                    & 0.764          & 0.795          & 0.817          & 0.822          & 0.546          & 0.572          & 0.580          & 0.595          & 0.667          & 0.686          & 0.704          & 0.709          \\ 
				& CMDVH                   & 0.693          & 0.761          & 0.695          & 0.733          & 0.371          & 0.380          & 0.331          & 0.371          & 0.493          & 0.527          & 0.598          & 0.589             \\ 
				& SSAH                    & 0.840          & 0.854          & 0.859          & 0.863          & \underline{0.648}          & \underline{0.663}          & \underline{0.681}          & \underline{0.678}          & 0.738          & 0.749          & 0.765          & 0.749          \\ 
				& DCHUC                   & \textbf{0.917} & \textbf{0.918} & \textbf{0.912} & \textbf{0.911} & \textbf{0.724} & \textbf{0.766} & \textbf{0.781} & \textbf{0.783} & \textbf{0.845} & \textbf{0.859} & \textbf{0.872} & \textbf{0.881} \\ \hline
		\end{tabular}}
	\end{minipage}
\end{table*}
\subsection{Out-of-Sample Extension}
For any instance which is not in the retrieval set, we can obtain the hash code of its two modalities. In particular, given the image modality $\boldsymbol{x}_q$ in an instance $\boldsymbol{o}_q$, we can adopt forward propagation to generate the hash code as follows:
\begin{equation}
\boldsymbol{h}_q = sign(\mathcal{\boldsymbol{F}}(\boldsymbol{x}_i;\boldsymbol{\Theta}))
\end{equation}
Similarly, we can also use the text hashing network to generate the hash code of the instance $\boldsymbol{o}_q$ with only textual modality $\boldsymbol{y}_q$:
\begin{equation}
\boldsymbol{g}_q = sign(\mathcal{\boldsymbol{P}}(\boldsymbol{y}_i;\boldsymbol{\Psi}))
\end{equation}

\section{Experiments}
To evaluate the performance of DCHUC, we will carry out extensive experiments on three image-text datasets and compared it with seven state-of-the-art cross-modal hashing methods.

\subsection{Datasets}
Three datasets are used for evaluation, i.e., \textit{\textbf{MIRFLICKR-25K}} \cite{huiskes2008mir}, \textit{\textbf{IAPR TC-12}} \cite{escalante2010segmented} and \textit{\textbf{NUS-WIDE}} \cite{chua2009nus}, which are described below.

The \textit{\textbf{MIRFLICKR-25K}} dataset \cite{huiskes2008mir} contains 25,000 instances collected from Flickr website. Each image is labeled with several textual tags. Here, we follow the experimental protocols given in DCMH \cite{jiang2017deep}. In total, 20,015 data instances which have at least 20 textual tags have been selected for our experiment. The text modality for each instance is represented as a 1,386-dimensional bag-of-words (BoW) vector. Furthermore each instance is manually labeled with at least one of the 24 unique labels. For this dataset, we randomly sampled 2,000 instances as the test set, and the remaining as the database (retrieval set). Furthermore, the training phase of the existing deep cross-modal hashing methods are typically time-consuming, which makes them cannot efficiently work on large-scale datasets. Therefore, for deep methods, we randomly sample 10,000 instances from the retrieval set as the training set.

The \textit{\textbf{IAPR TC-12}} \cite{escalante2010segmented} consists of 20,000 instances which are annotated using 255 labels. After pruning the instance that is without any text information, a subset of 19999 image-text pairs are select for our experiment. The text modality for each instance is represented as a 2000-dimensional BoW vector. For this dataset, we randomly sampled 2,000 instances as test set, with the rest of the instances as retrieval set. We randomly select 10,000 instances from retrieval set for training deep cross-modal baselines.

The \textit{\textbf{NUS-WIDE}} dataset \cite{chua2009nus} contains 269,648 instances crawled from Flickr. Each image is associated with textual tags, and each instance is annotated with one or multiple labels from 81 concept labels. Only 195,834 image-text pairs that belong to the 21 most frequent concepts are selected for our experiment. The text modality for each instance is represented as a 1000-dimensional BoW vector. For this dataset, we randomly sampled 2,100 instance as test set, with the rest of the instances as retrieval set. Because the deep hashing baselines are very time-consuming for training, we randomly select 10,500 instances from database for training deep cross-modal baselines. 

For all the shallow cross-modal baselines, all the database are used for training. For all datasets, the image $\boldsymbol{x}_i$ and text $\boldsymbol{y}_j$ will be defined as a similar pair if $\boldsymbol{x}_i$ and $\boldsymbol{y}_j$ share at least one common label. Otherwise, they will be defined as a dissimilar pair.

\begin{figure*}[]
	\centering
	\subfigure[]{
		\begin{minipage}[t]{0.25\textwidth}
			\centering
			\includegraphics[width=\linewidth]{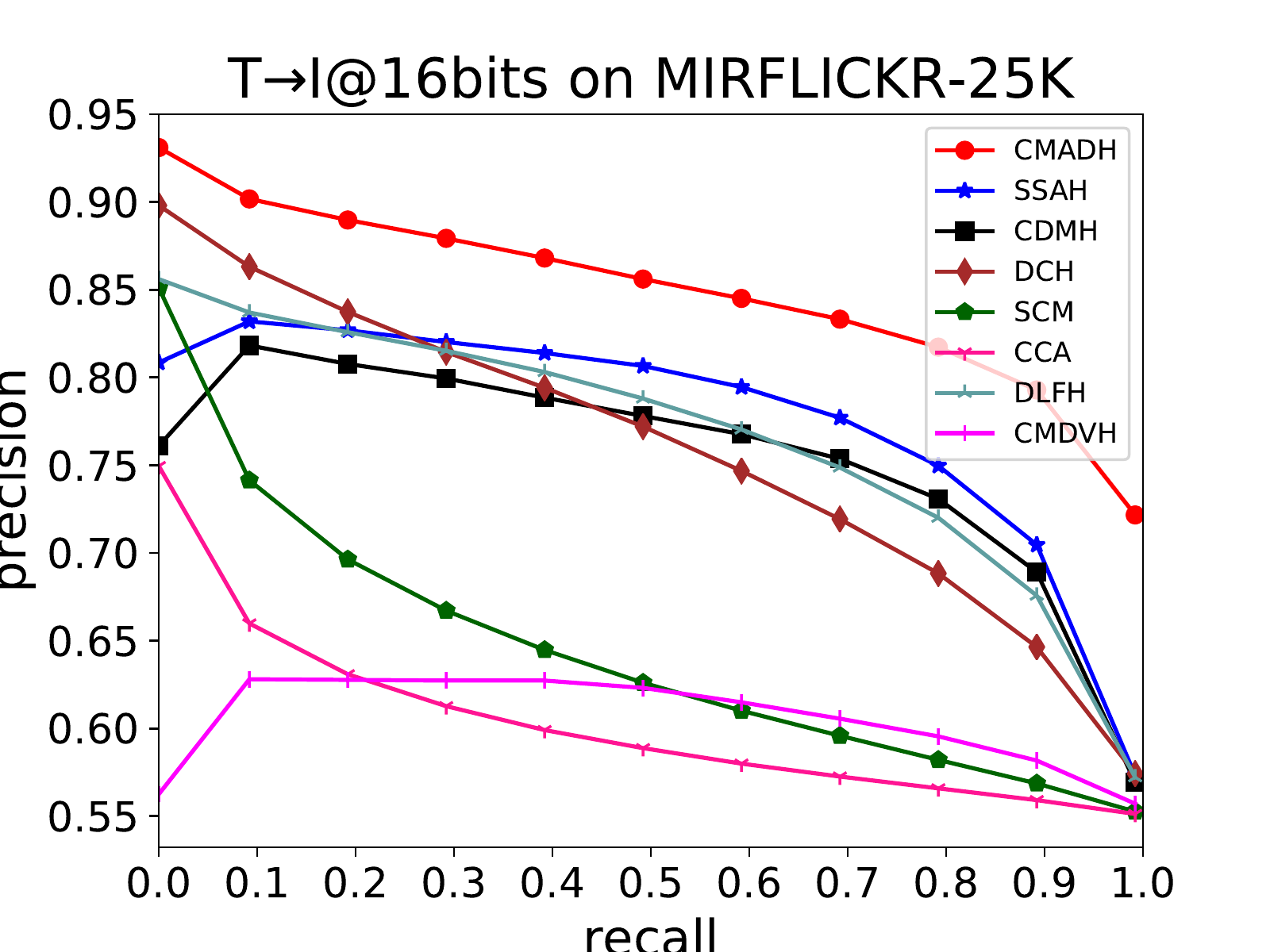}
		\end{minipage}%
	}%
	\subfigure[]{
		\begin{minipage}[t]{0.25\textwidth}
			\centering
			\includegraphics[width=\linewidth]{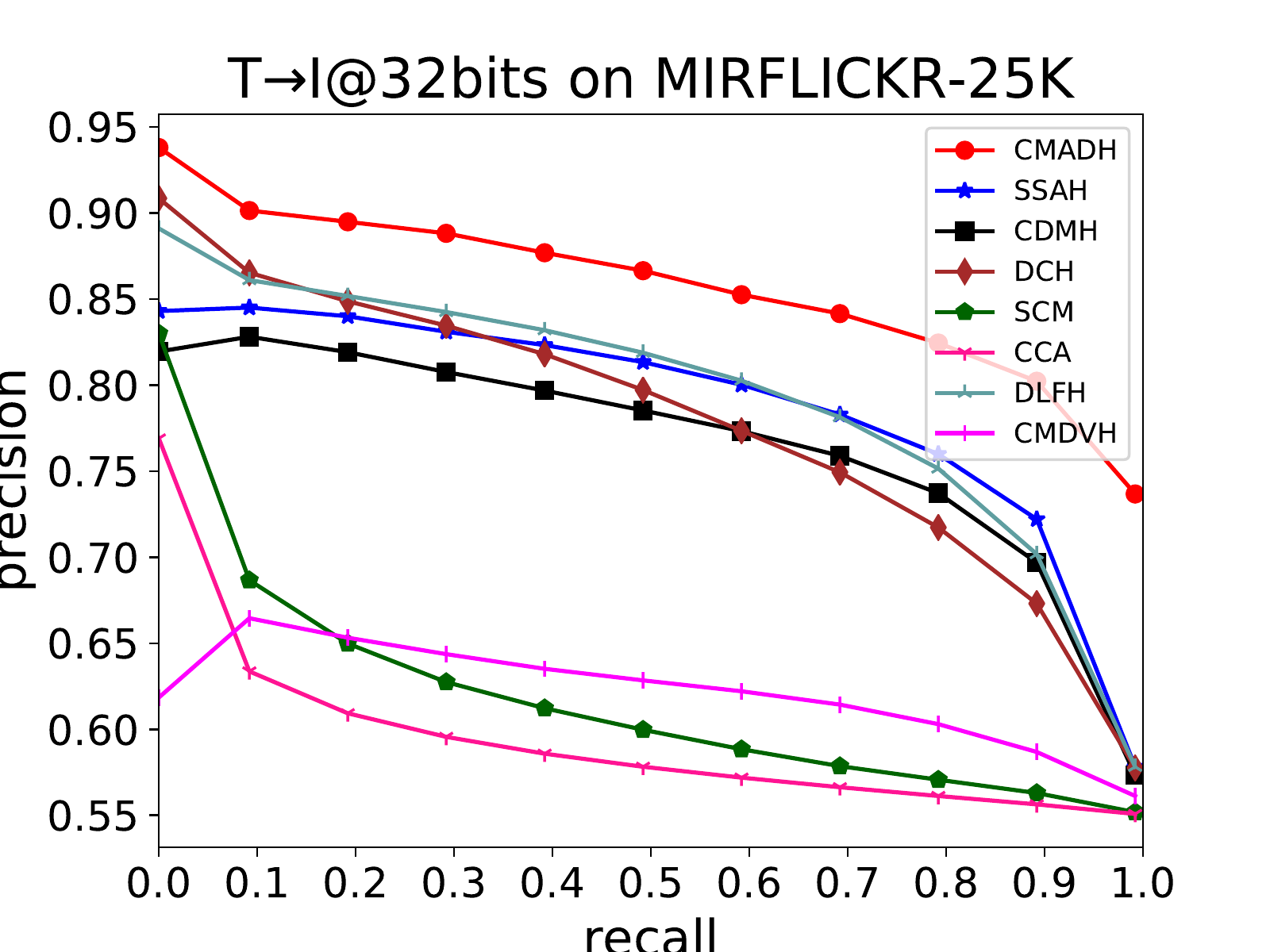}
		\end{minipage}%
	}%
	\subfigure[]{
		\begin{minipage}[t]{0.25\textwidth}
			\centering
			\includegraphics[width=\linewidth]{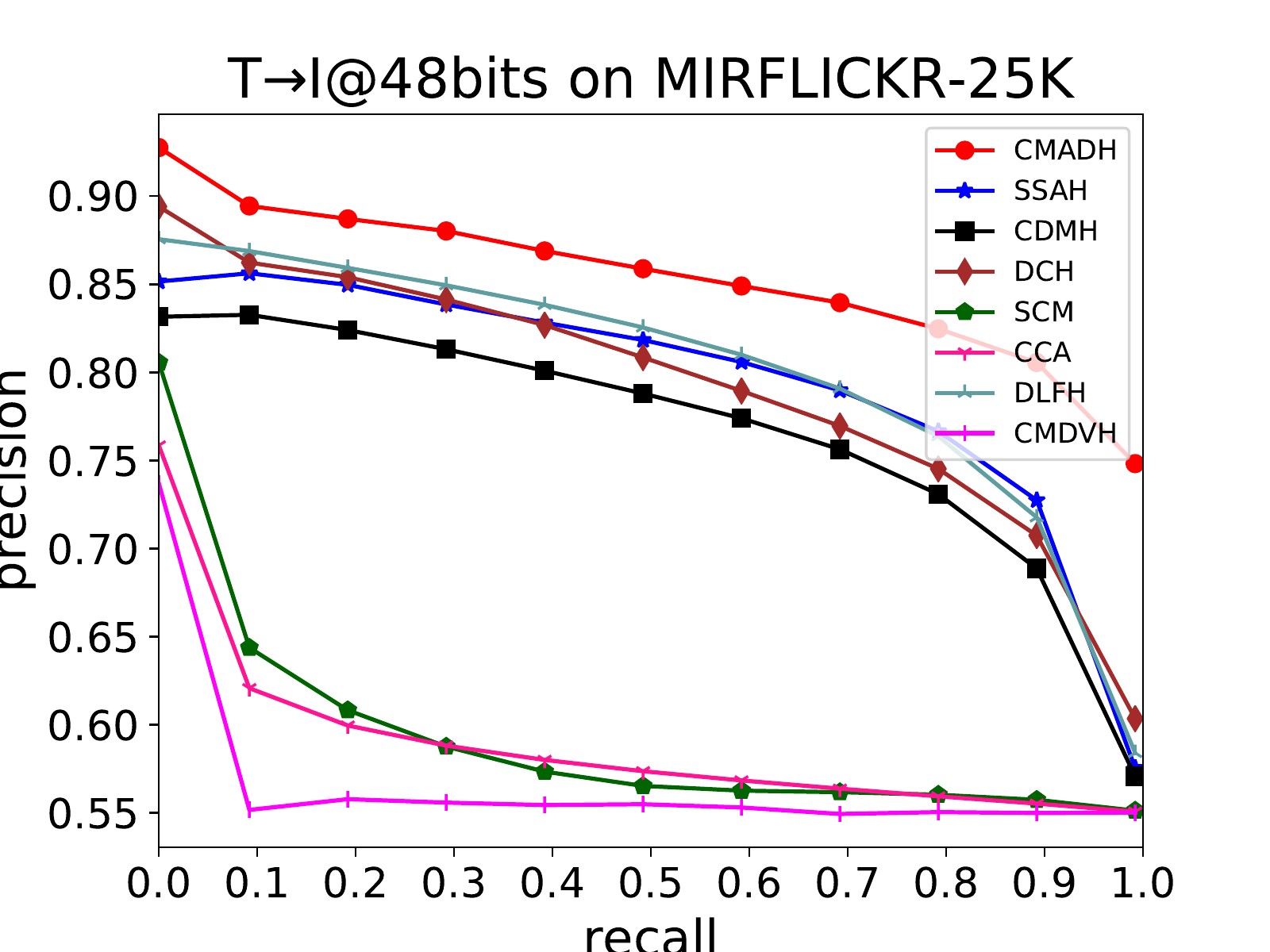}
		\end{minipage}%
	}%
	\subfigure[]{
		\begin{minipage}[t]{0.25\textwidth}
			\centering
			\includegraphics[width=\linewidth]{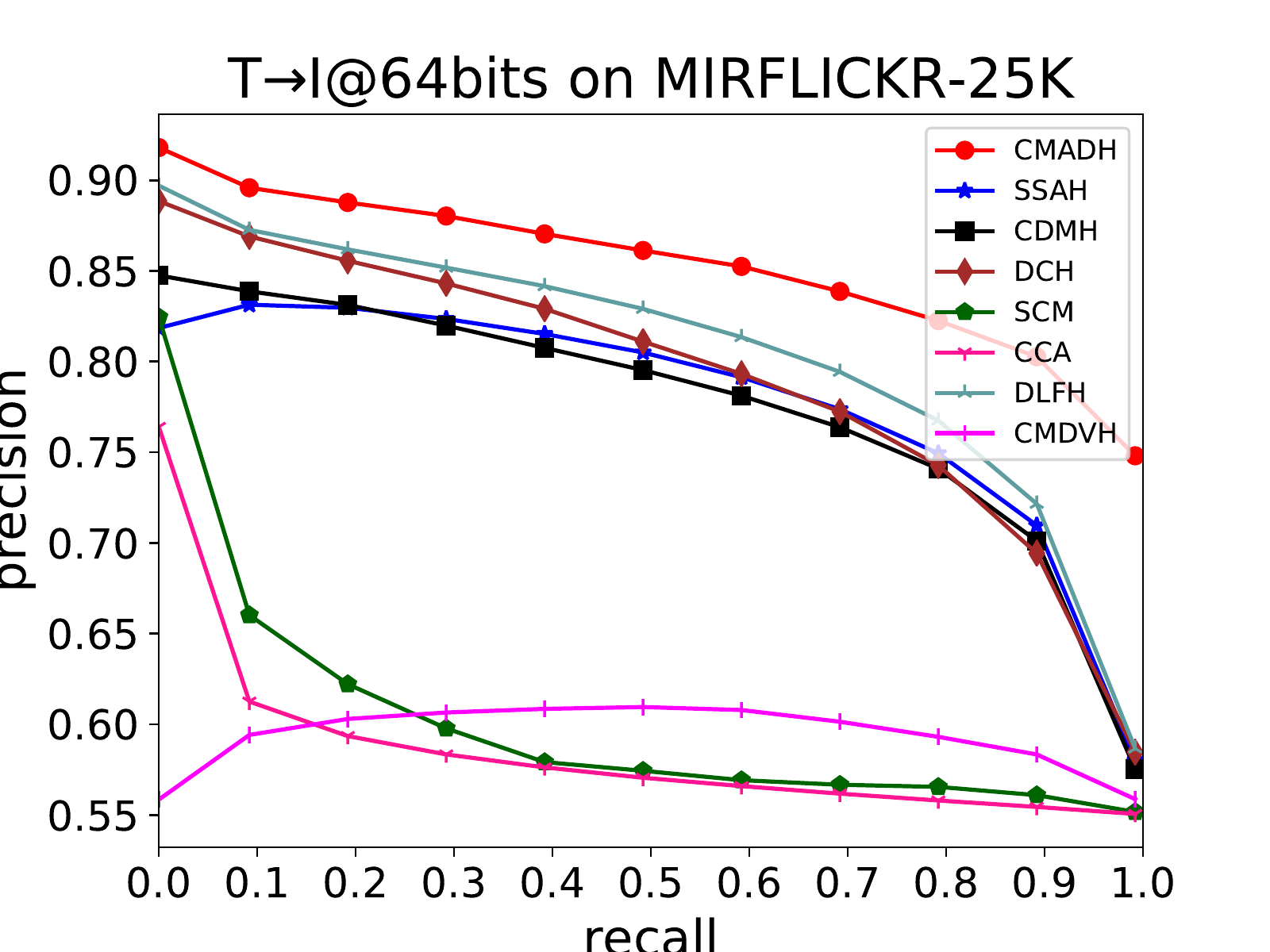}
		\end{minipage}%
	}%
	\quad
	\subfigure[]{
		\begin{minipage}[t]{0.25\textwidth}
			\centering
			\includegraphics[width=\linewidth]{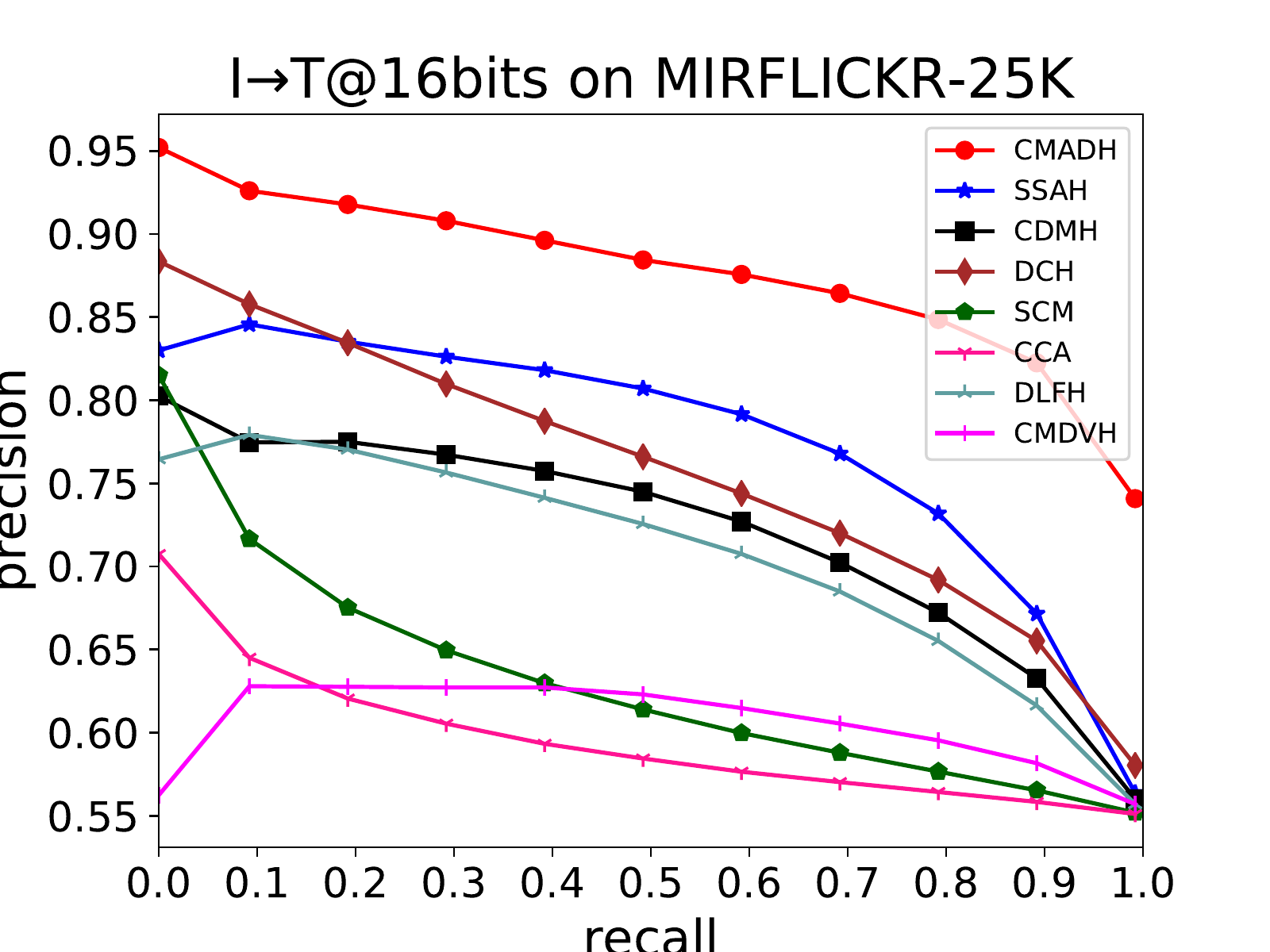}
		\end{minipage}%
	}%
	\subfigure[]{
		\begin{minipage}[t]{0.25\textwidth}
			\centering
			\includegraphics[width=\linewidth]{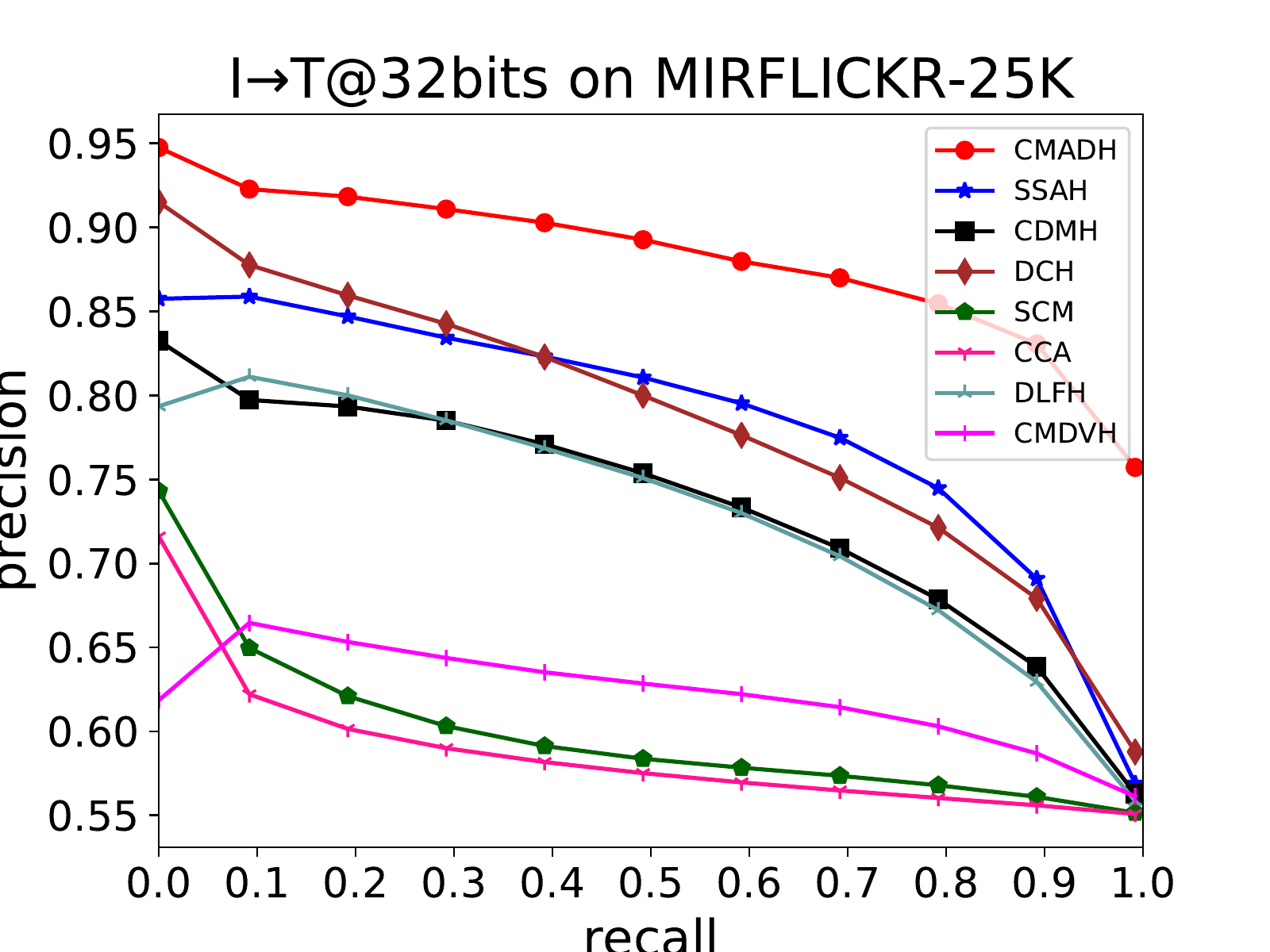}
		\end{minipage}%
	}%
	\subfigure[]{
		\begin{minipage}[t]{0.25\textwidth}
			\centering
			\includegraphics[width=\linewidth]{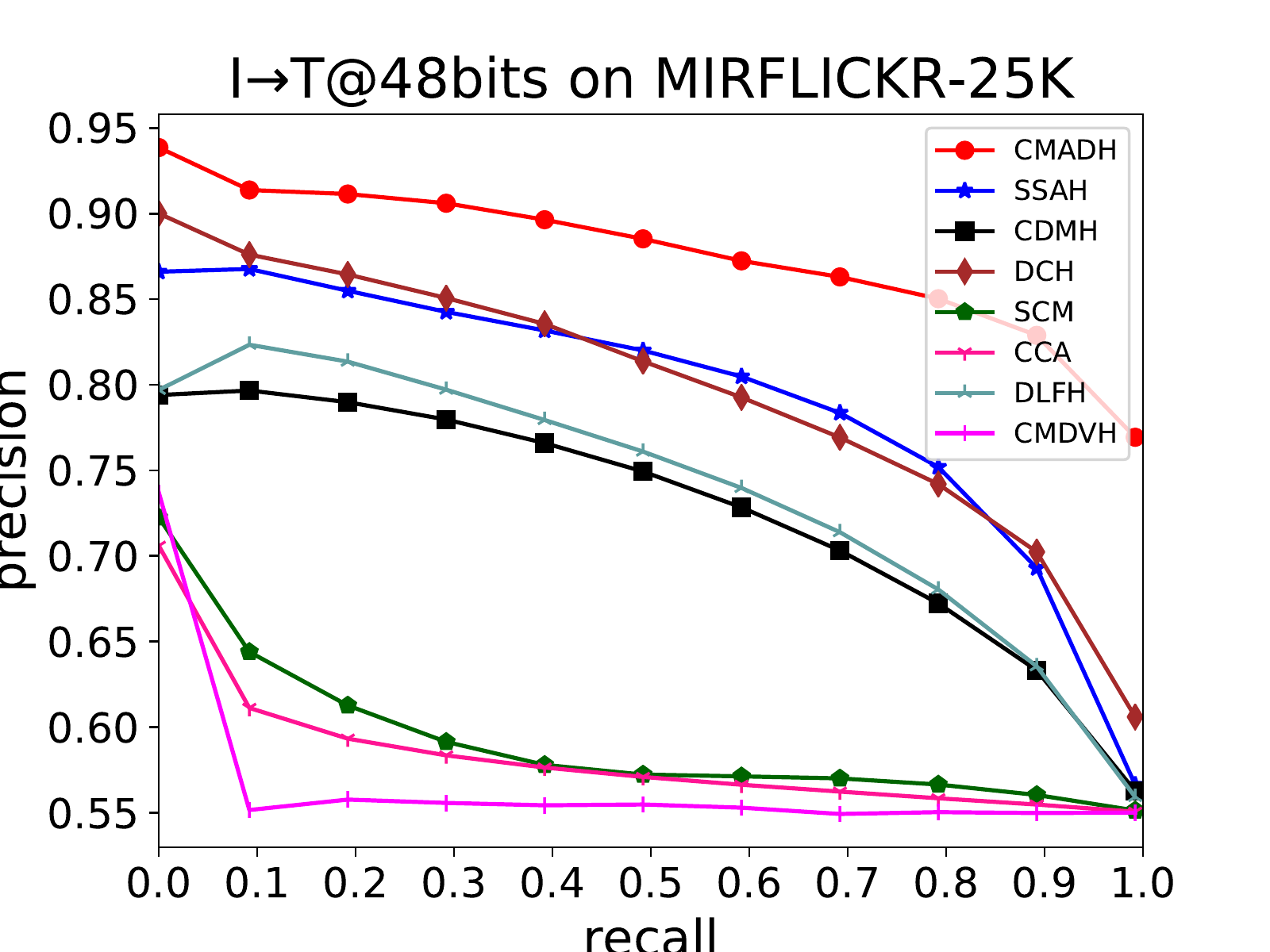}
		\end{minipage}%
	}%
	\subfigure[]{
		\begin{minipage}[t]{0.25\textwidth}
			\centering
			\includegraphics[width=\linewidth]{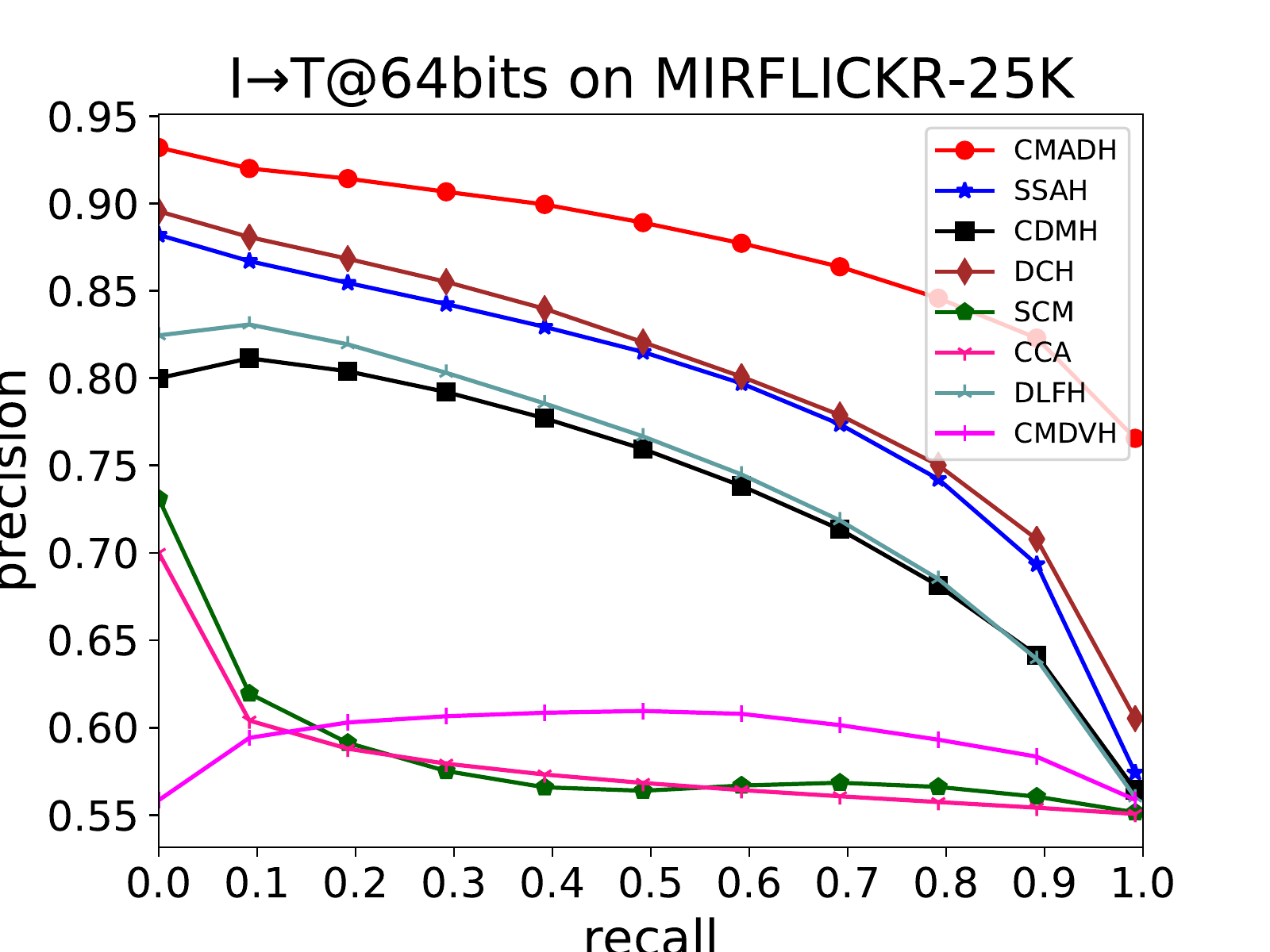}
		\end{minipage}%
	}%
	\caption{Precision-recall curve on MIRFLICKR-25K dataset}
	\label{fig_pr_mir}
\end{figure*}

\begin{figure*}[]
	\centering
	\subfigure[]{
		\begin{minipage}[t]{0.25\textwidth}
			\centering
			\includegraphics[width=\linewidth]{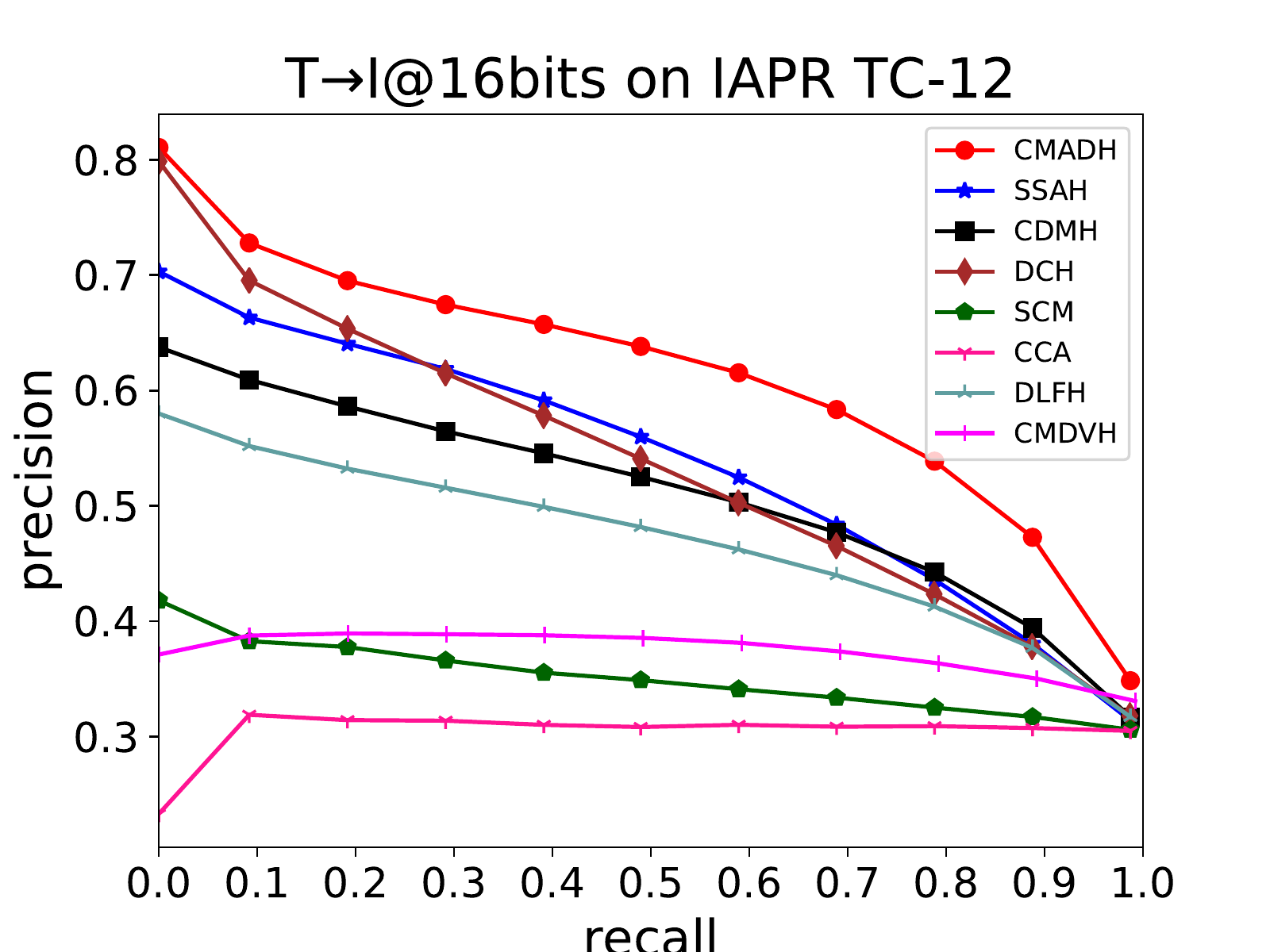}
		\end{minipage}%
	}%
	\subfigure[]{
		\begin{minipage}[t]{0.25\textwidth}
			\centering
			\includegraphics[width=\linewidth]{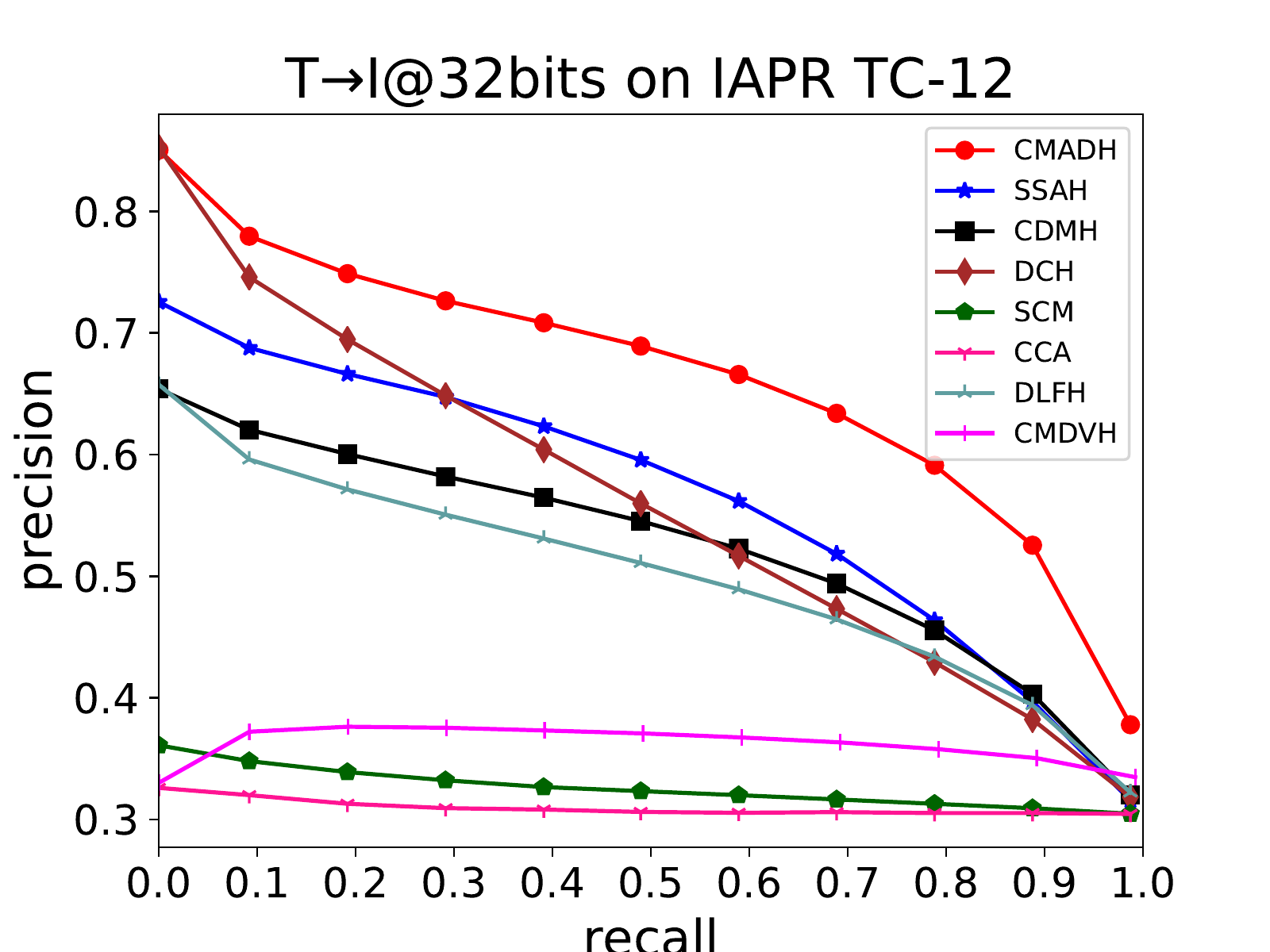}
		\end{minipage}%
	}%
	\subfigure[]{
		\begin{minipage}[t]{0.25\textwidth}
			\centering
			\includegraphics[width=\linewidth]{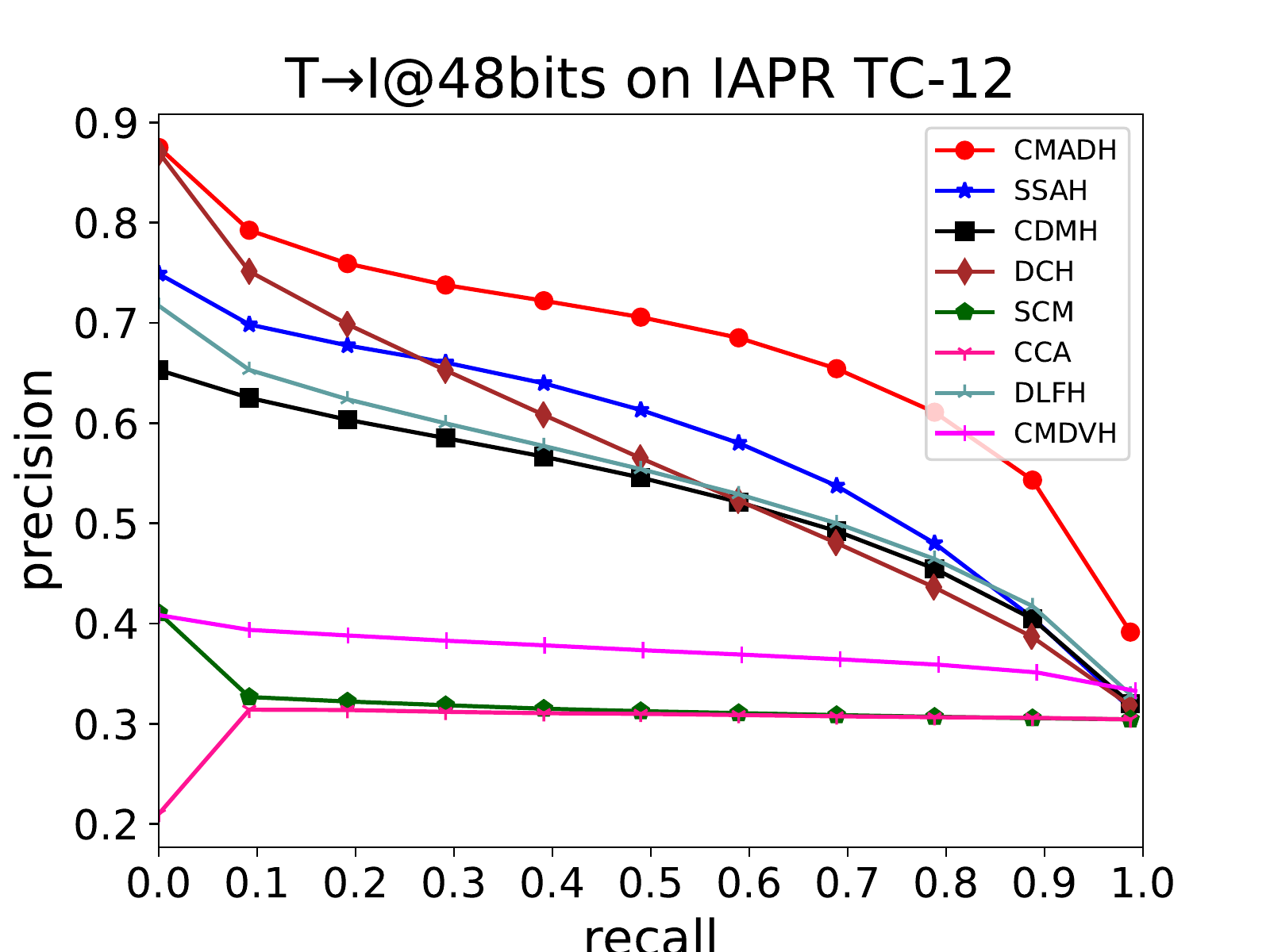}
		\end{minipage}%
	}%
	\subfigure[]{
		\begin{minipage}[t]{0.25\textwidth}
			\centering
			\includegraphics[width=\linewidth]{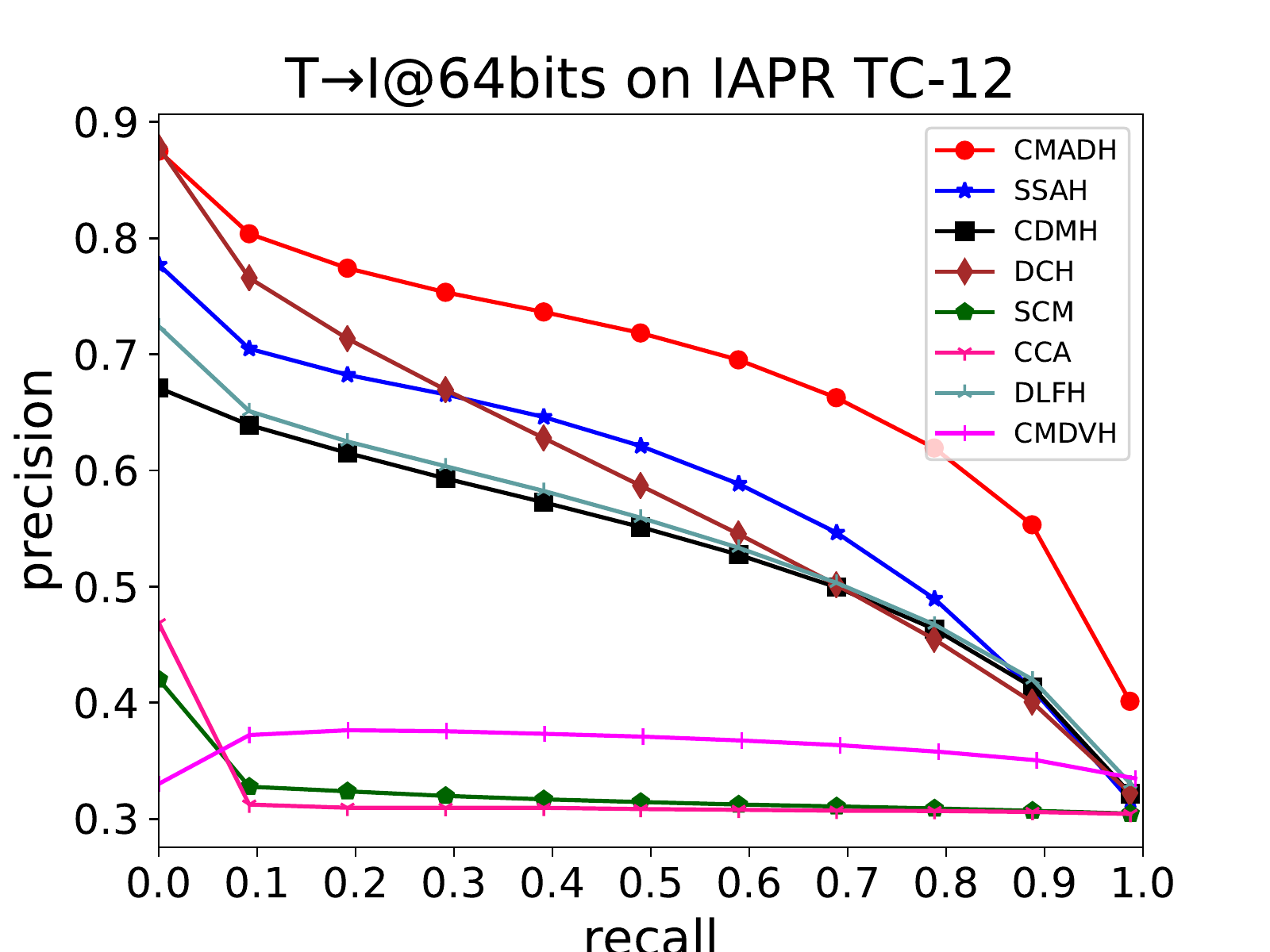}
		\end{minipage}%
	}%
	\quad
	\subfigure[]{
		\begin{minipage}[t]{0.25\textwidth}
			\centering
			\includegraphics[width=\linewidth]{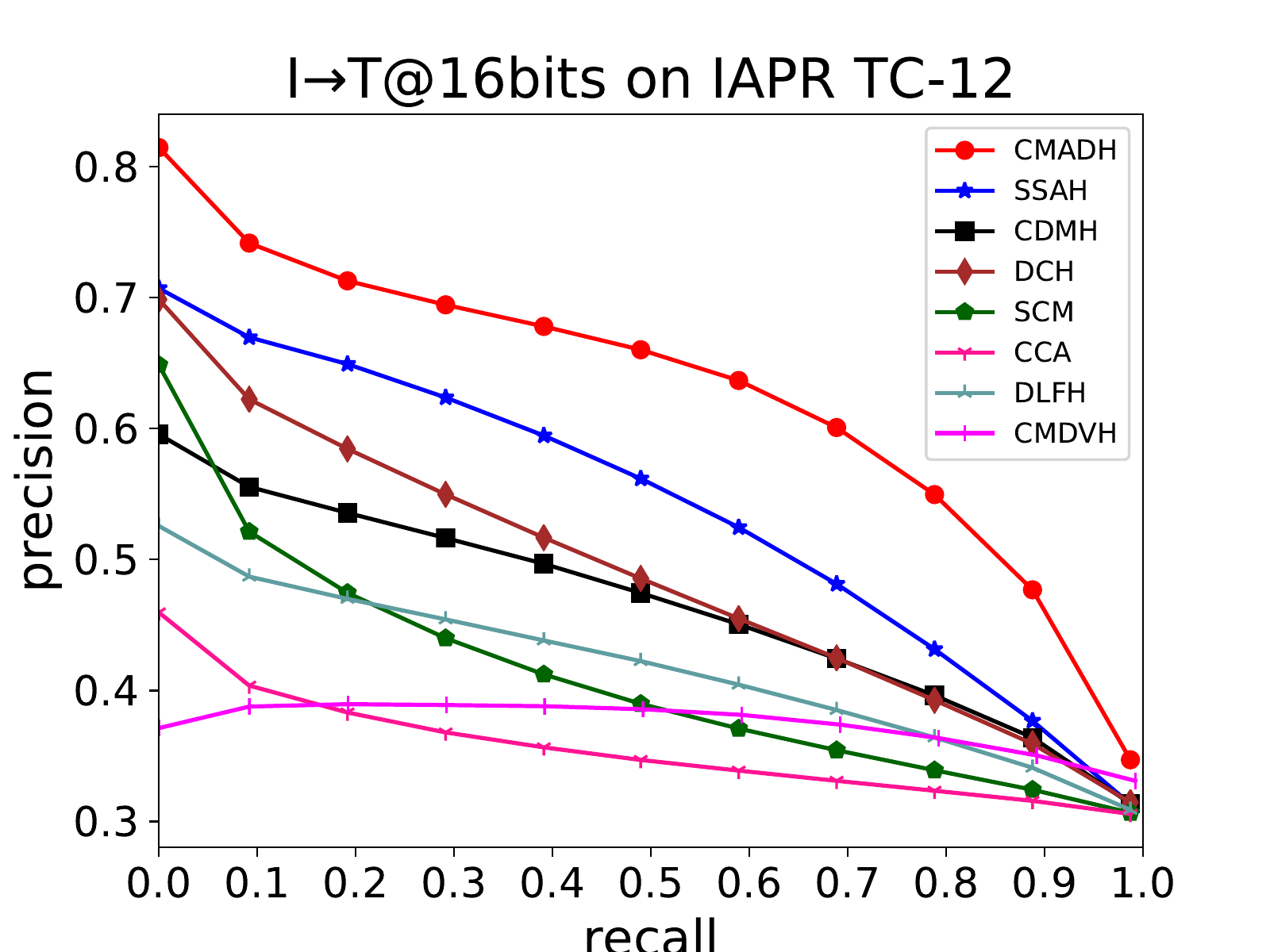}
		\end{minipage}%
	}%
	\subfigure[]{
		\begin{minipage}[t]{0.25\textwidth}
			\centering
			\includegraphics[width=\linewidth]{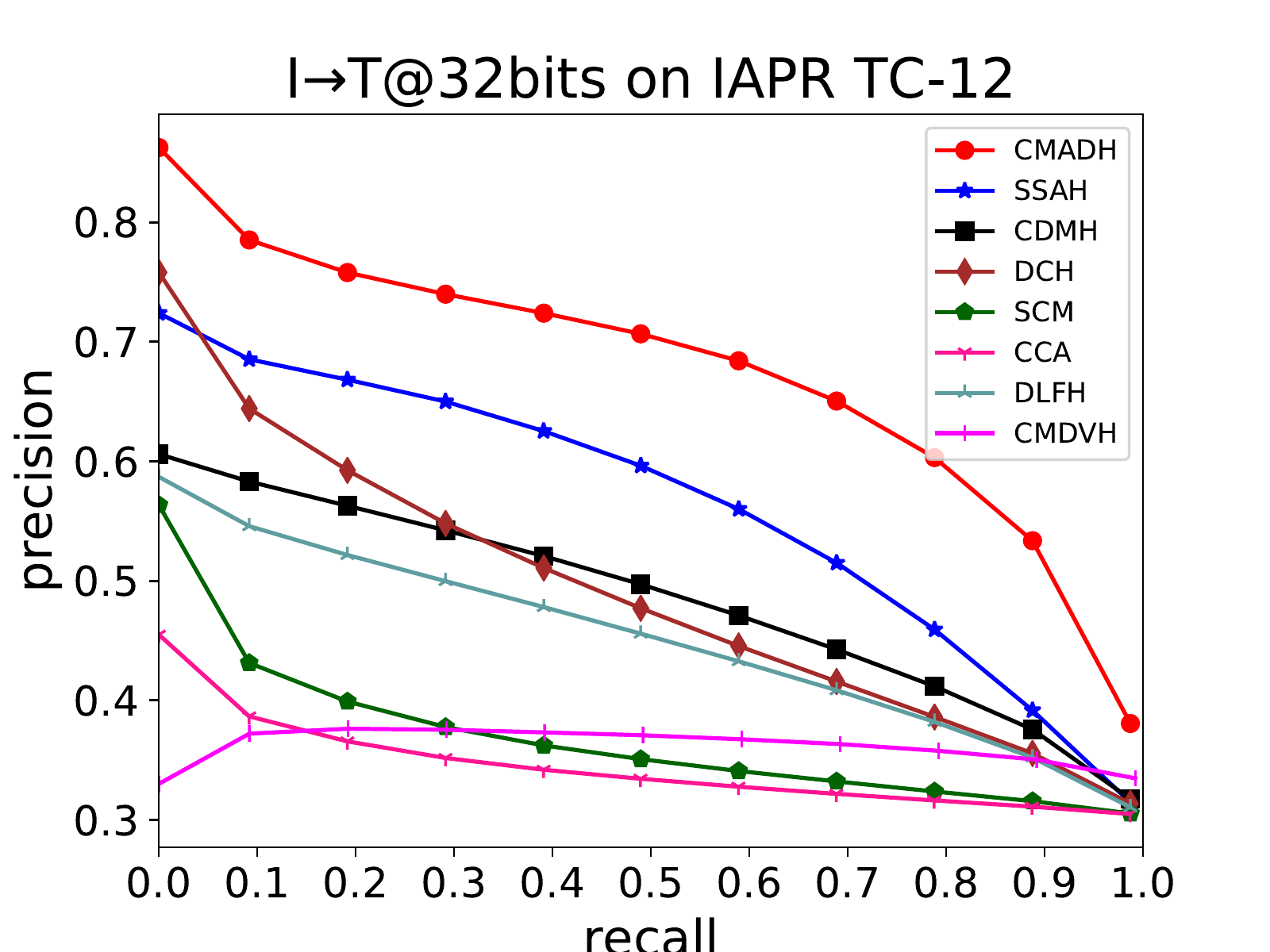}
		\end{minipage}%
	}%
	\subfigure[]{
		\begin{minipage}[t]{0.25\textwidth}
			\centering
			\includegraphics[width=\linewidth]{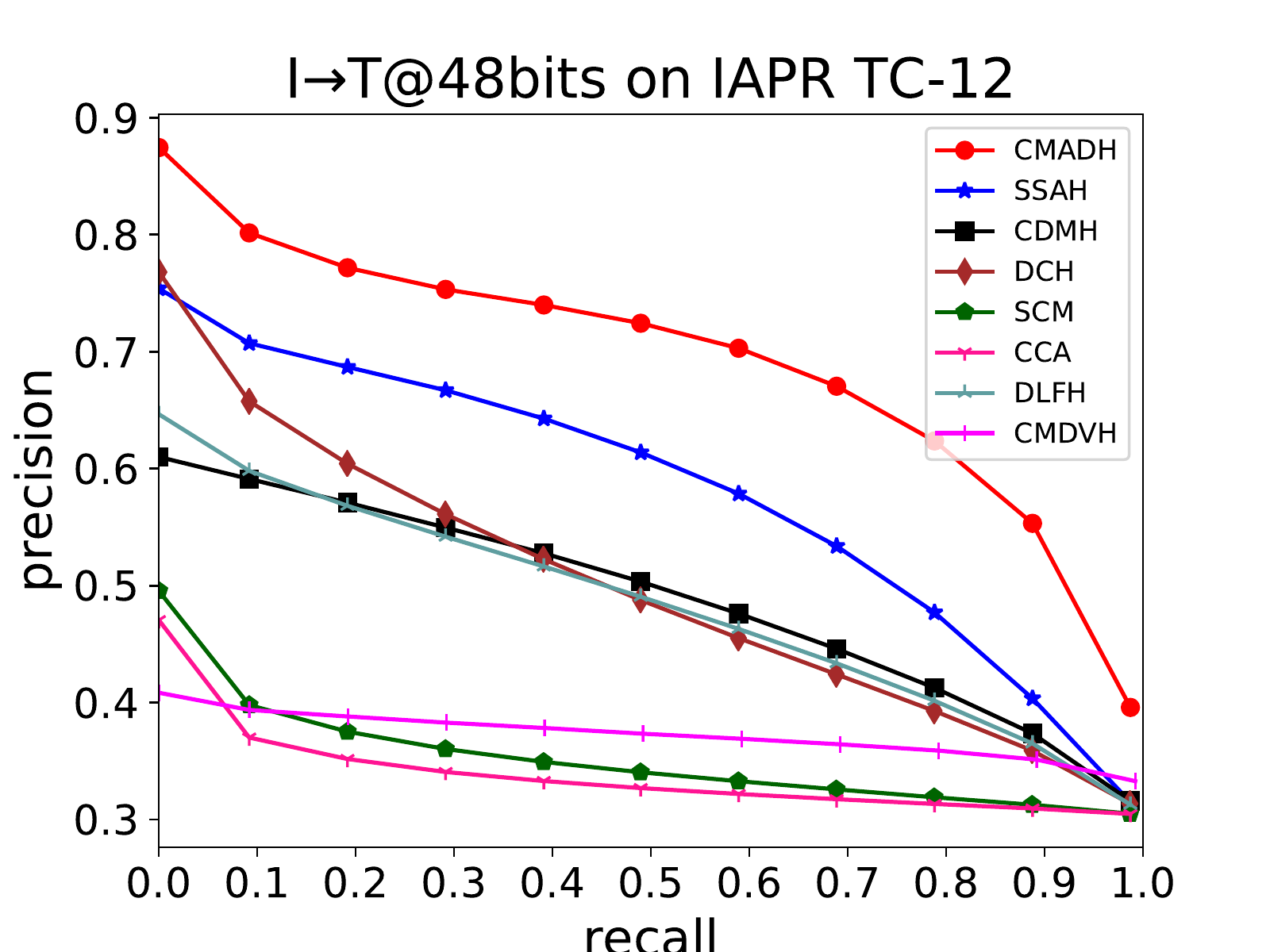}
		\end{minipage}%
	}%
	\subfigure[]{
		\begin{minipage}[t]{0.25\textwidth}
			\centering
			\includegraphics[width=\linewidth]{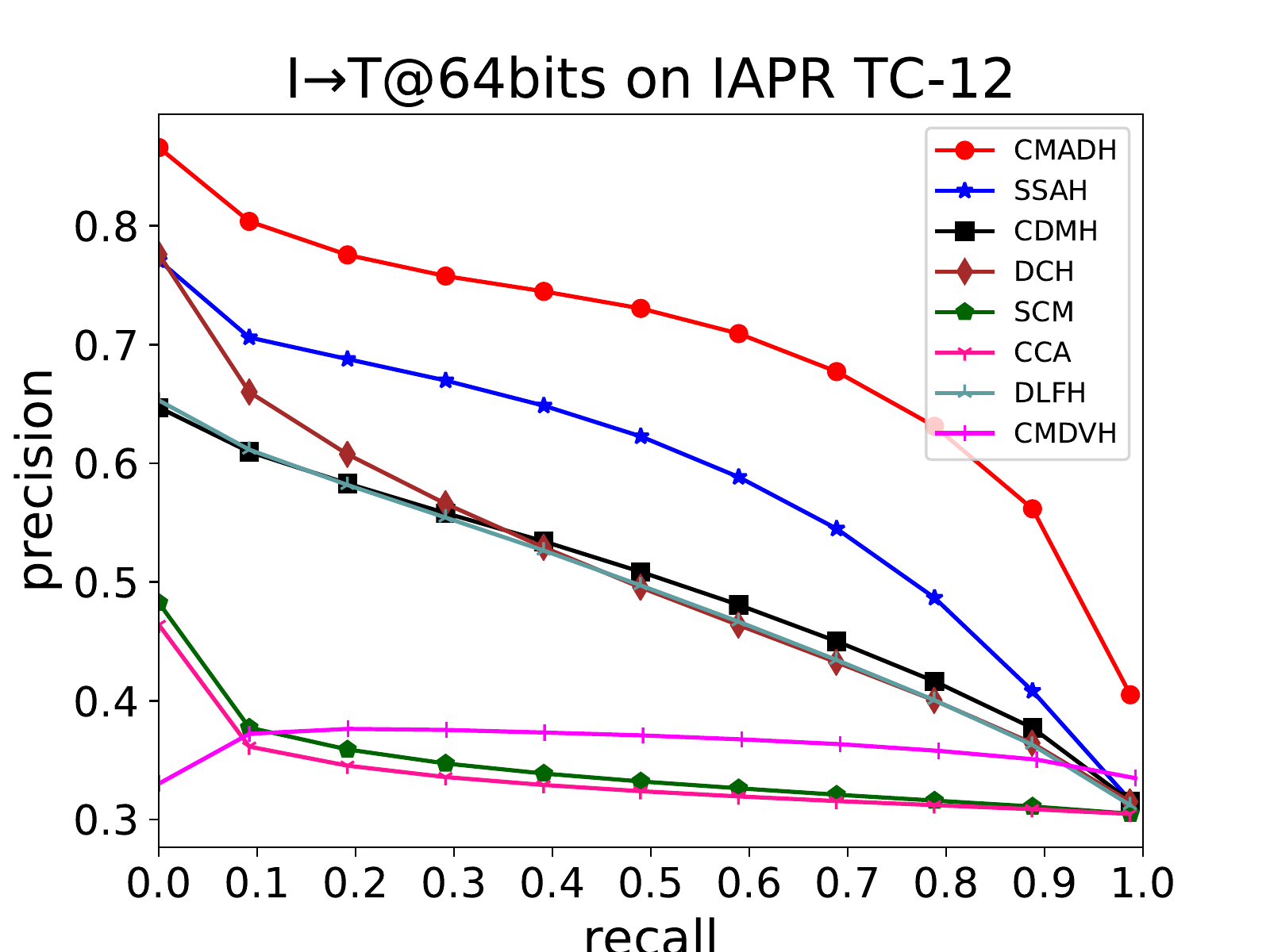}
		\end{minipage}%
	}%
	\caption{Precision-recall curve on IAPR TC-12 dataset}
	\label{fig_pr_iaprtc}
\end{figure*}

\begin{figure*}[]
	\centering
	\subfigure[]{
		\begin{minipage}[t]{0.25\textwidth}
			\centering
			\includegraphics[width=\linewidth]{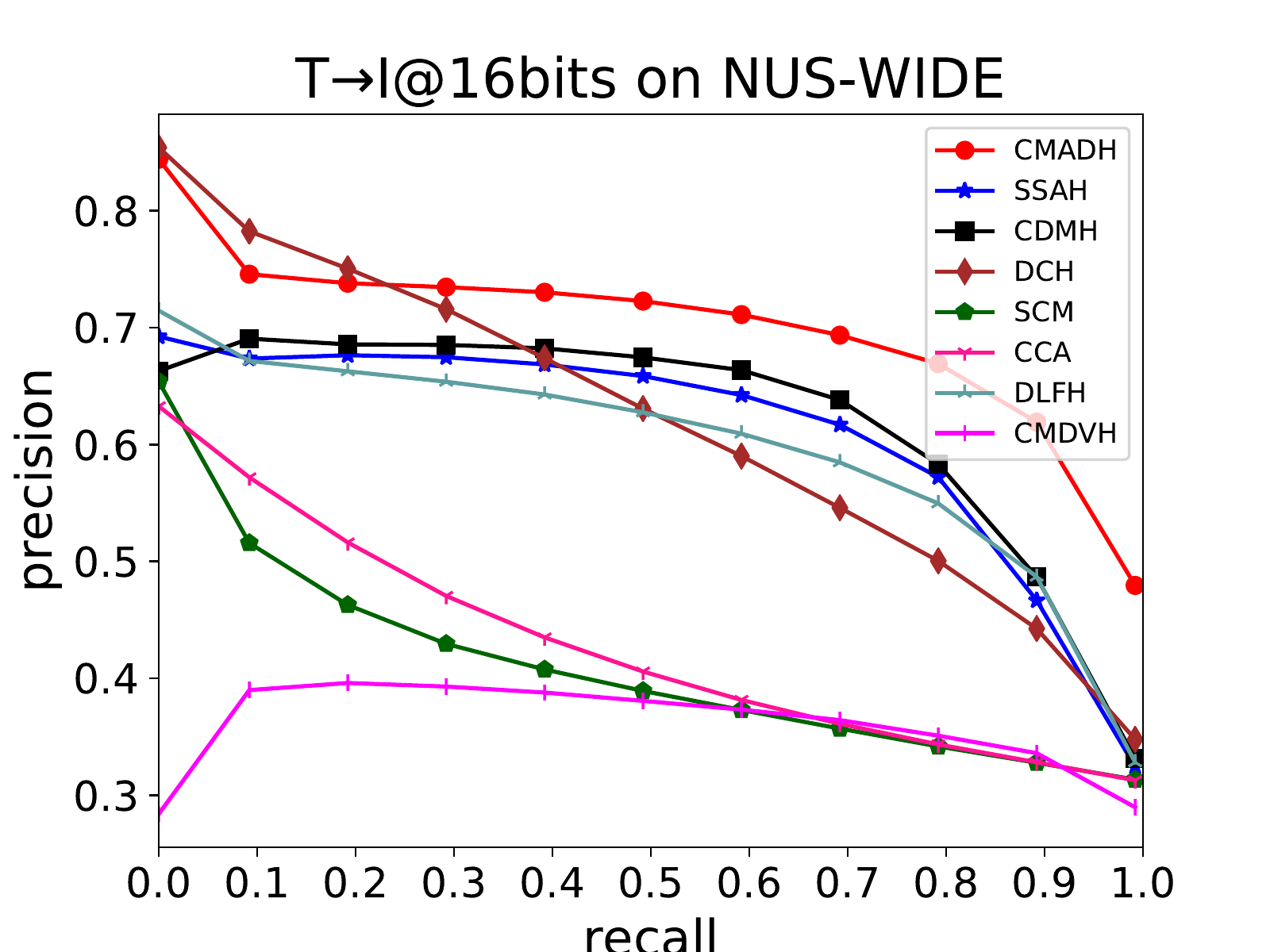}
		\end{minipage}%
	}%
	\subfigure[]{
		\begin{minipage}[t]{0.25\textwidth}
			\centering
			\includegraphics[width=\linewidth]{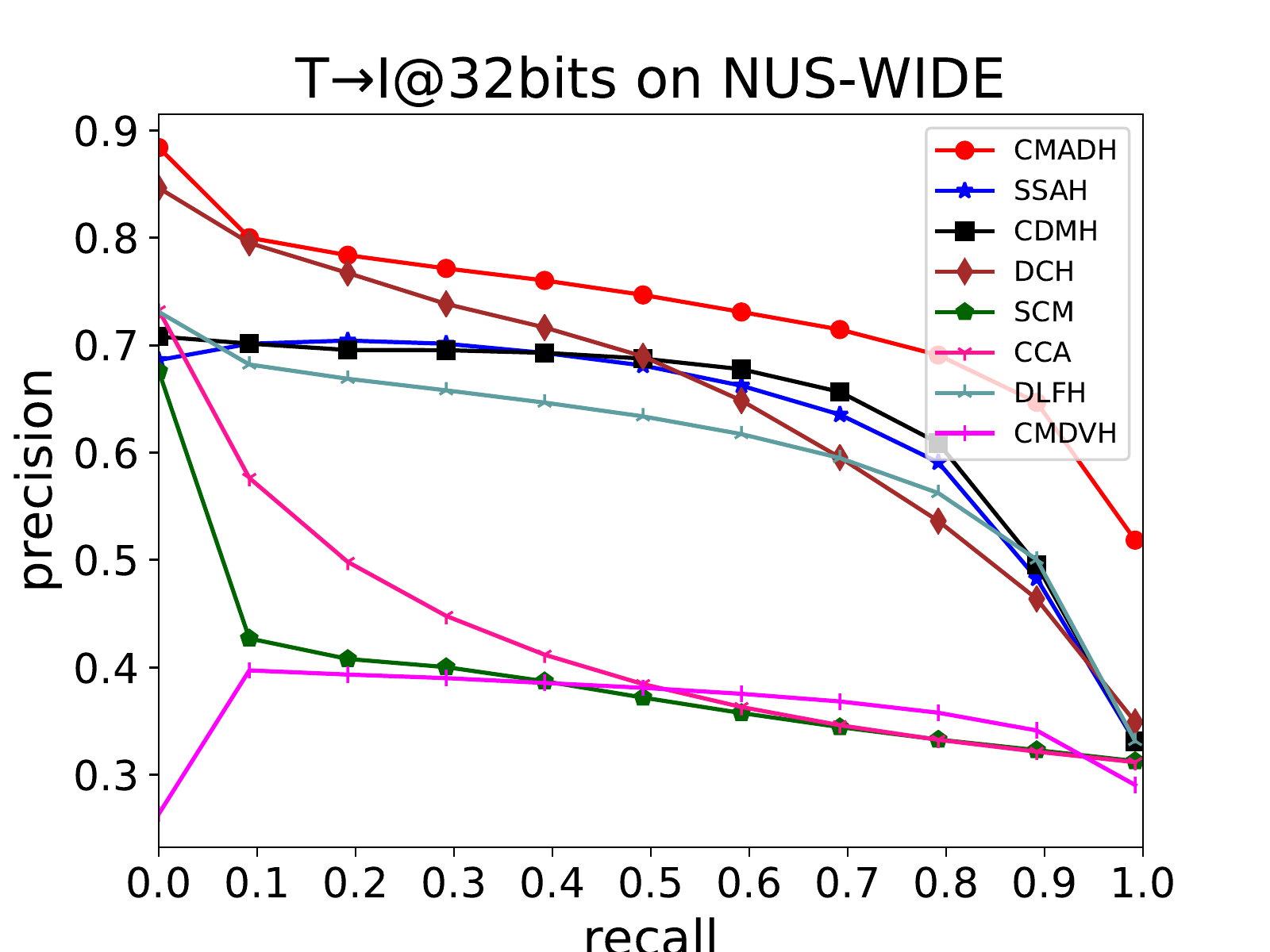}
		\end{minipage}%
	}%
	\subfigure[]{
		\begin{minipage}[t]{0.25\textwidth}
			\centering
			\includegraphics[width=\linewidth]{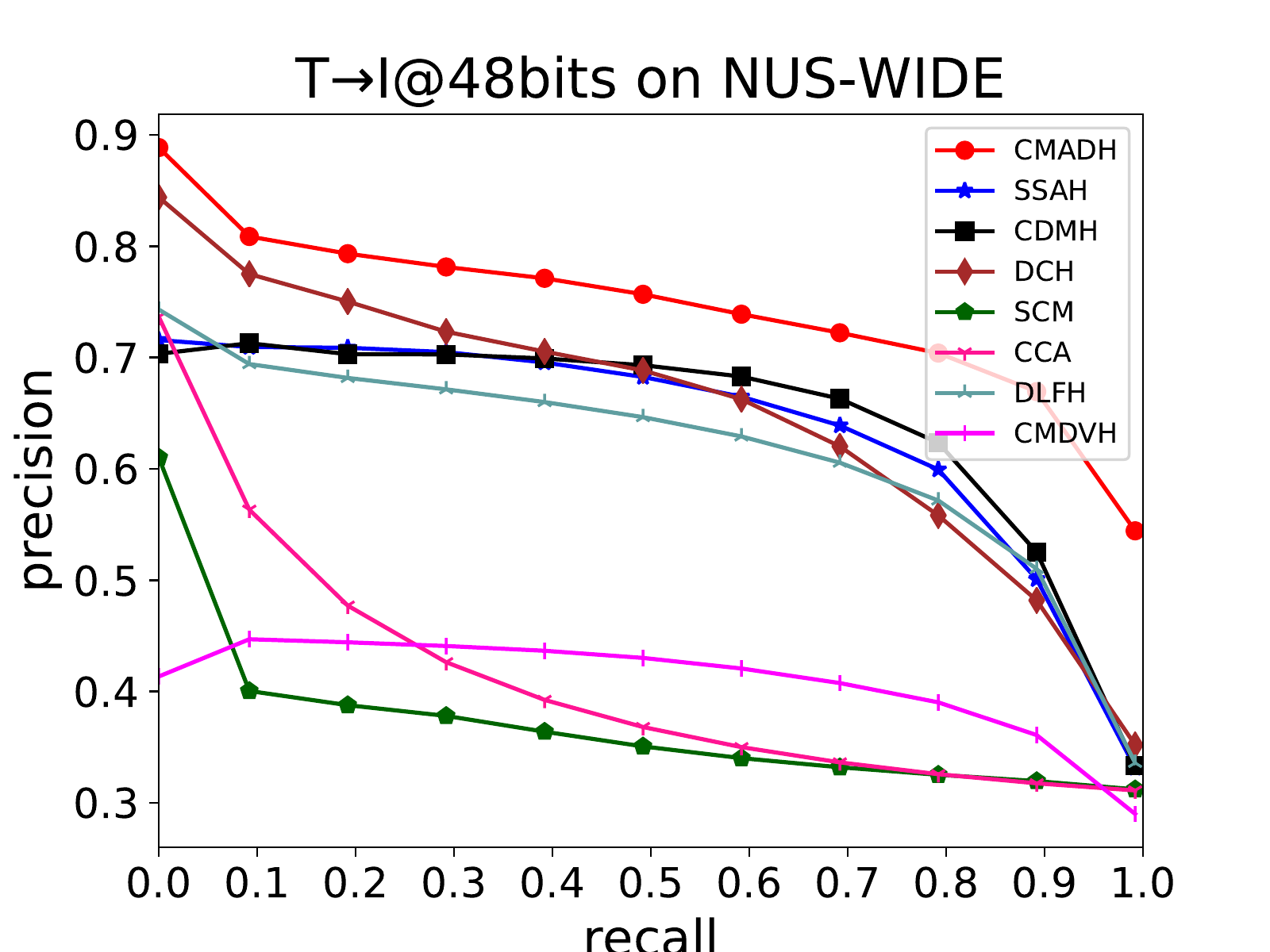}
		\end{minipage}%
	}%
	\subfigure[]{
		\begin{minipage}[t]{0.25\textwidth}
			\centering
			\includegraphics[width=\linewidth]{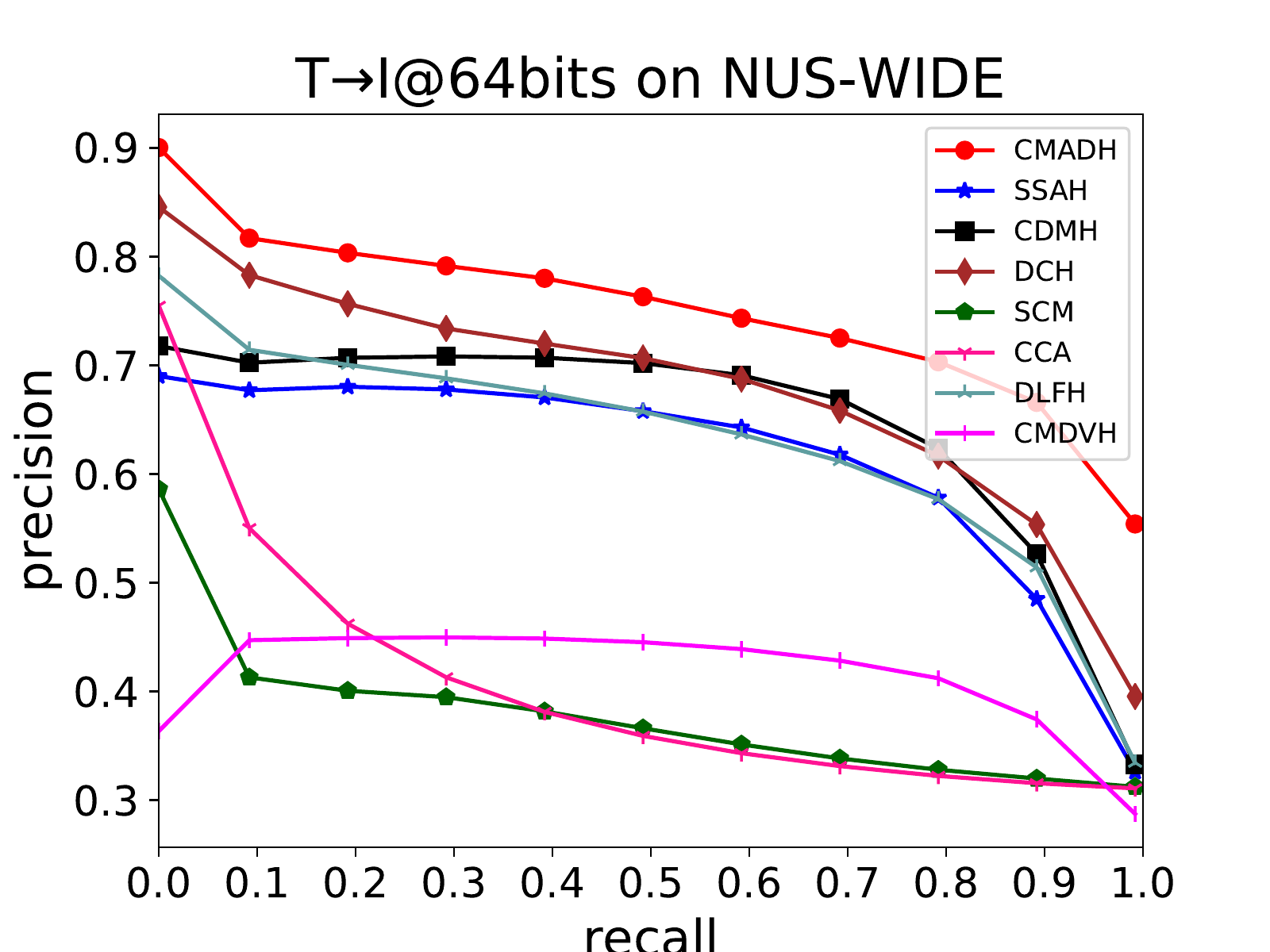}
		\end{minipage}%
	}%
	\quad
	\subfigure[]{
		\begin{minipage}[t]{0.25\textwidth}
			\centering
			\includegraphics[width=\linewidth]{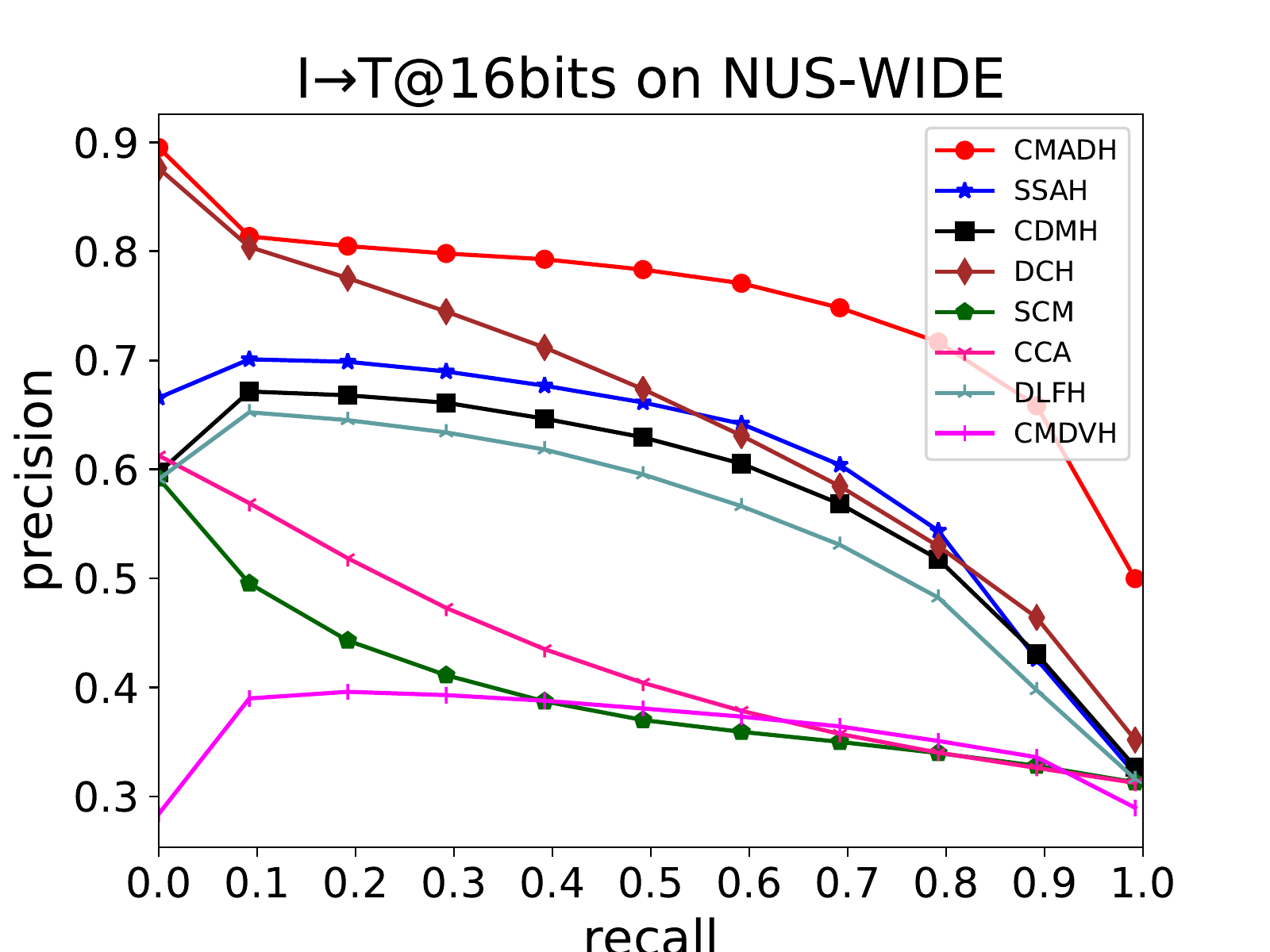}
		\end{minipage}%
	}%
	\subfigure[]{
		\begin{minipage}[t]{0.25\textwidth}
			\centering
			\includegraphics[width=\linewidth]{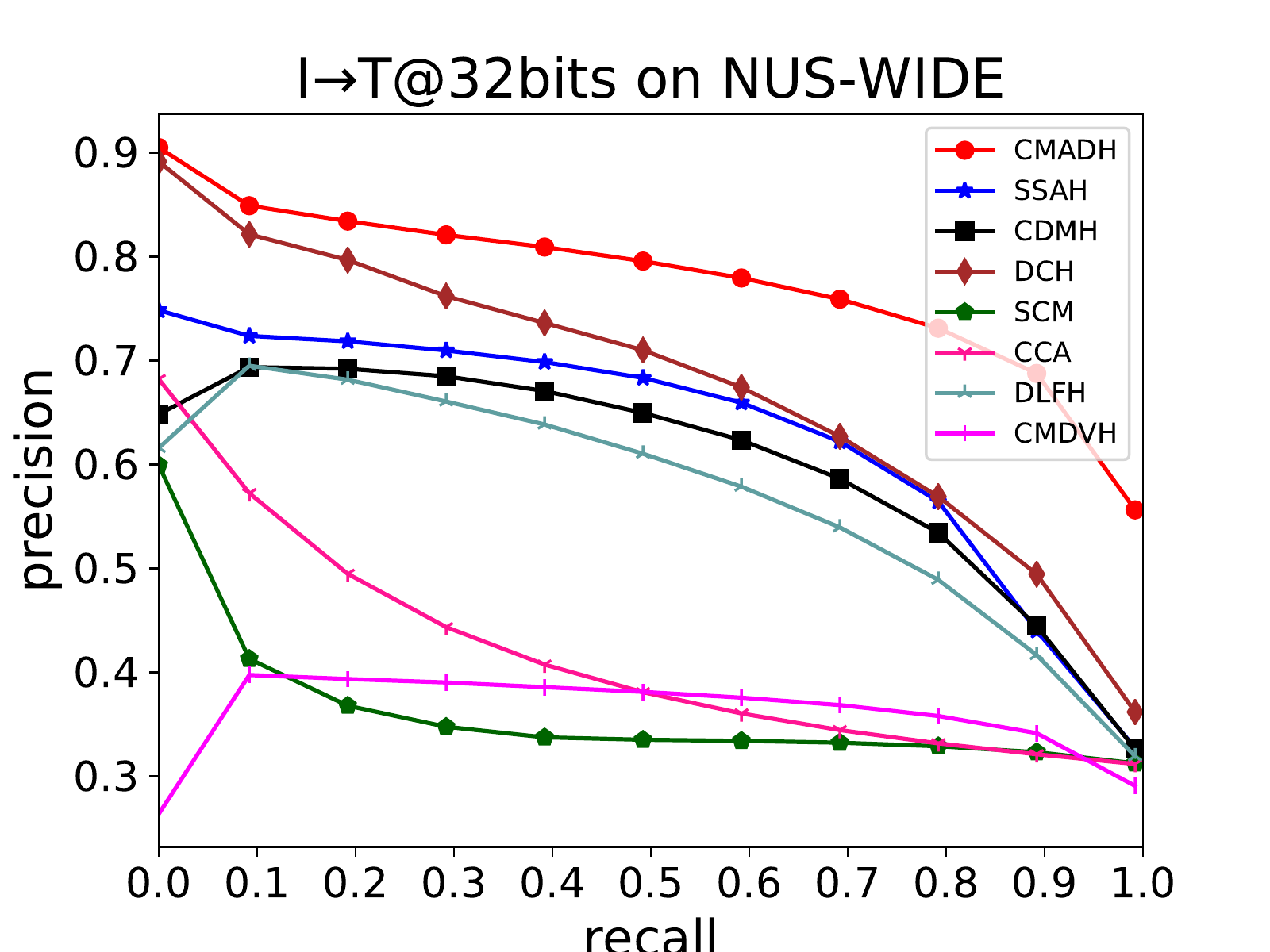}
		\end{minipage}%
	}%
	\subfigure[]{
		\begin{minipage}[t]{0.25\textwidth}
			\centering
			\includegraphics[width=\linewidth]{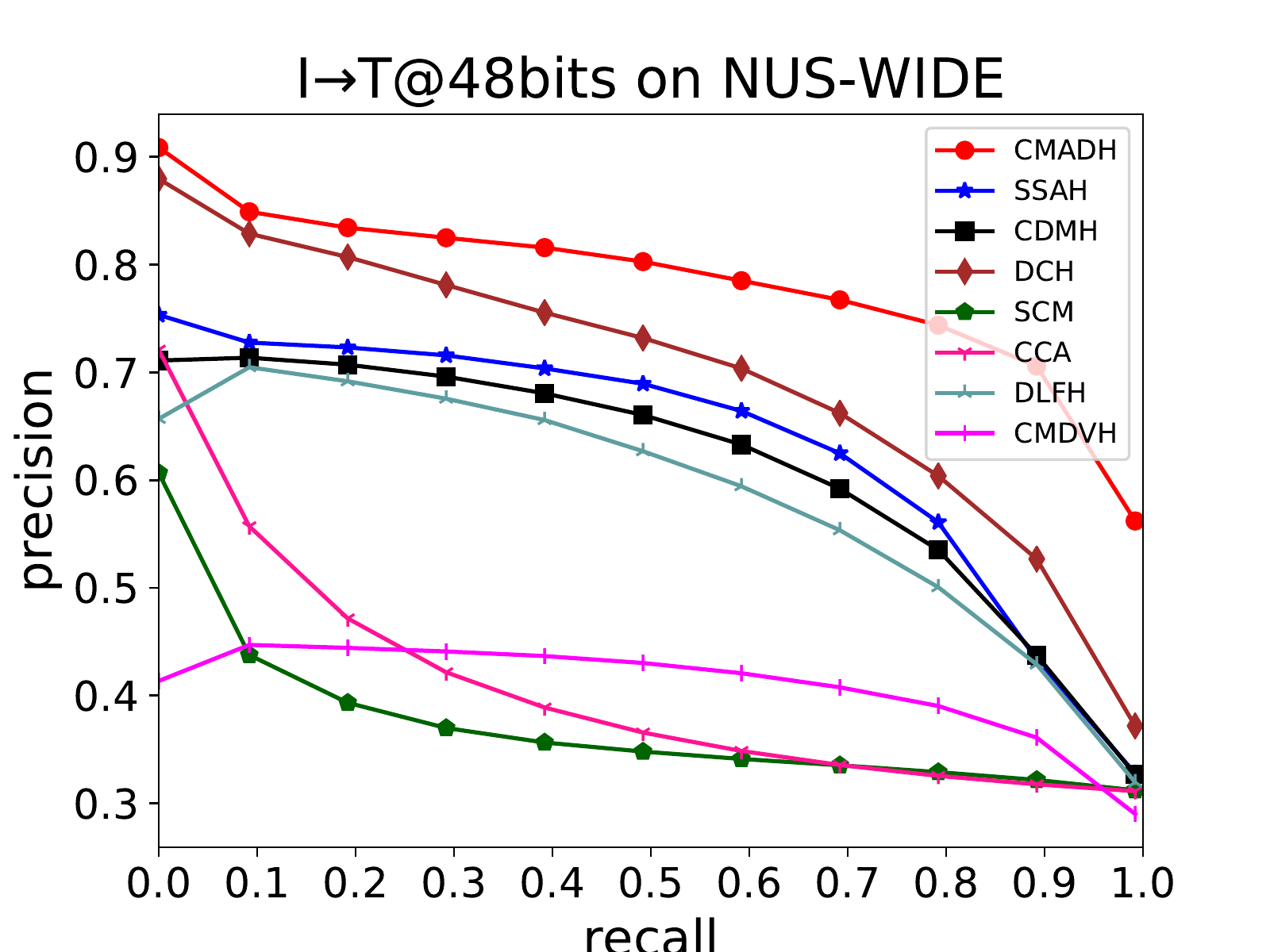}
		\end{minipage}%
	}%
	\subfigure[]{
		\begin{minipage}[t]{0.25\textwidth}
			\centering
			\includegraphics[width=\linewidth]{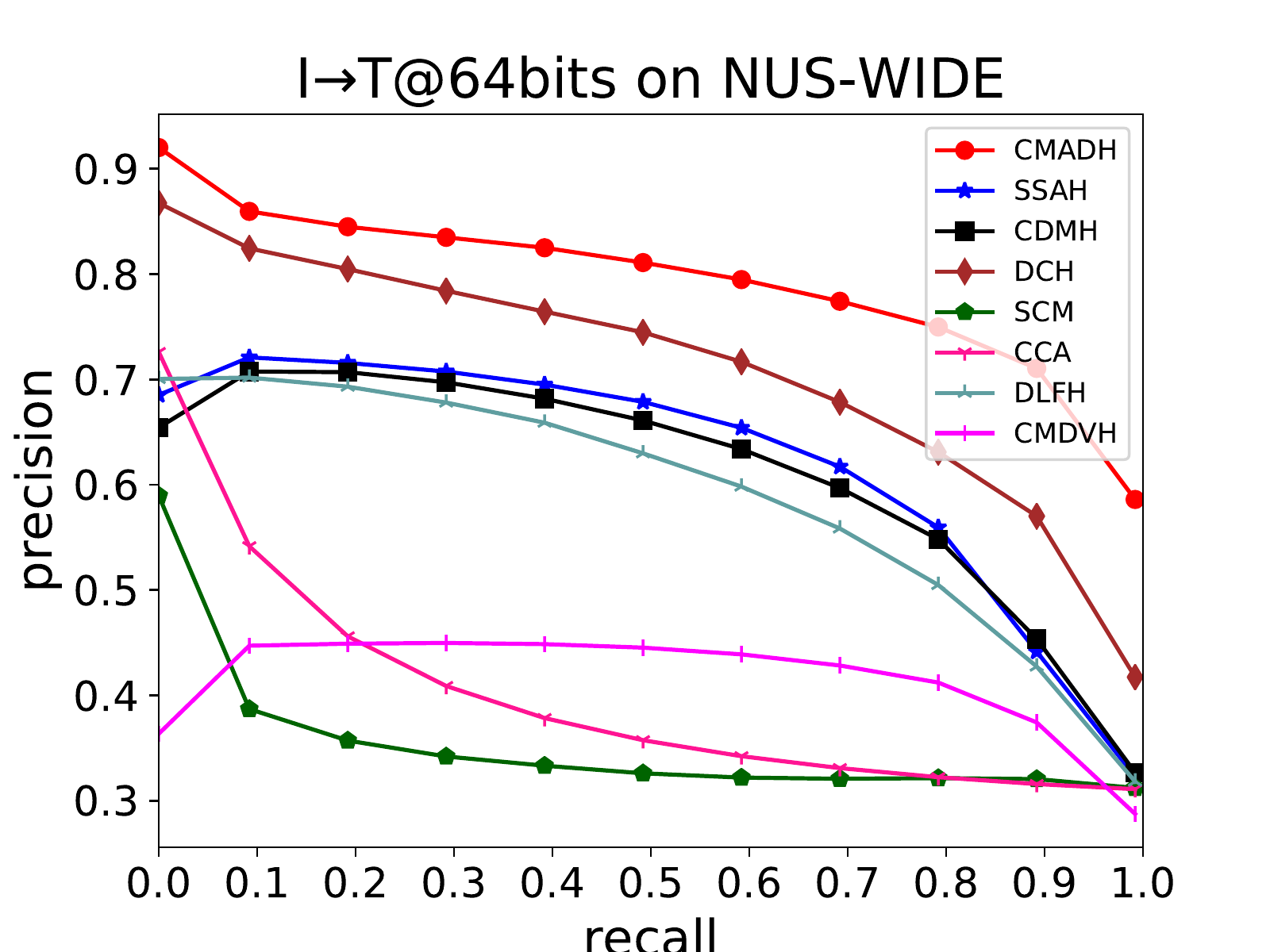}
		\end{minipage}%
	}%
	\caption{Precision-recall curves on NUS-WIDE dataset}
	\label{fig_pr_nus}
\end{figure*}

\begin{figure}[b]
	\centering
	\subfigure[Obj. value@64 bits]{
		\begin{minipage}[t]{0.24\textwidth}
			\centering
			\includegraphics[width=1.1\linewidth, height=0.7\linewidth]{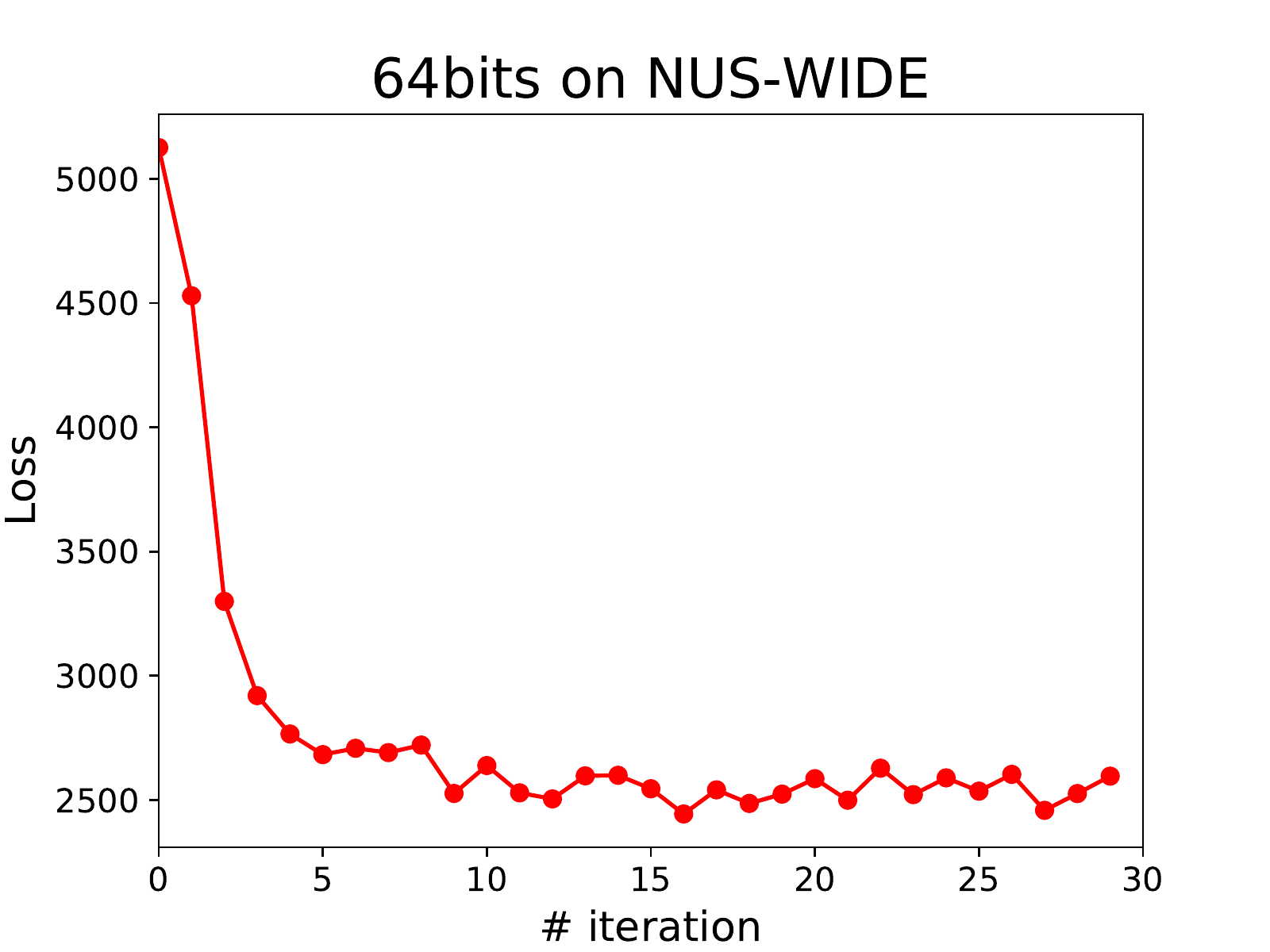}
			\label{loss64}
		\end{minipage}%
	}%
	\subfigure[MAP@64 bits]{
		\begin{minipage}[t]{0.24\textwidth}
			\centering
			\includegraphics[width=1.1\linewidth,height=0.7\linewidth]{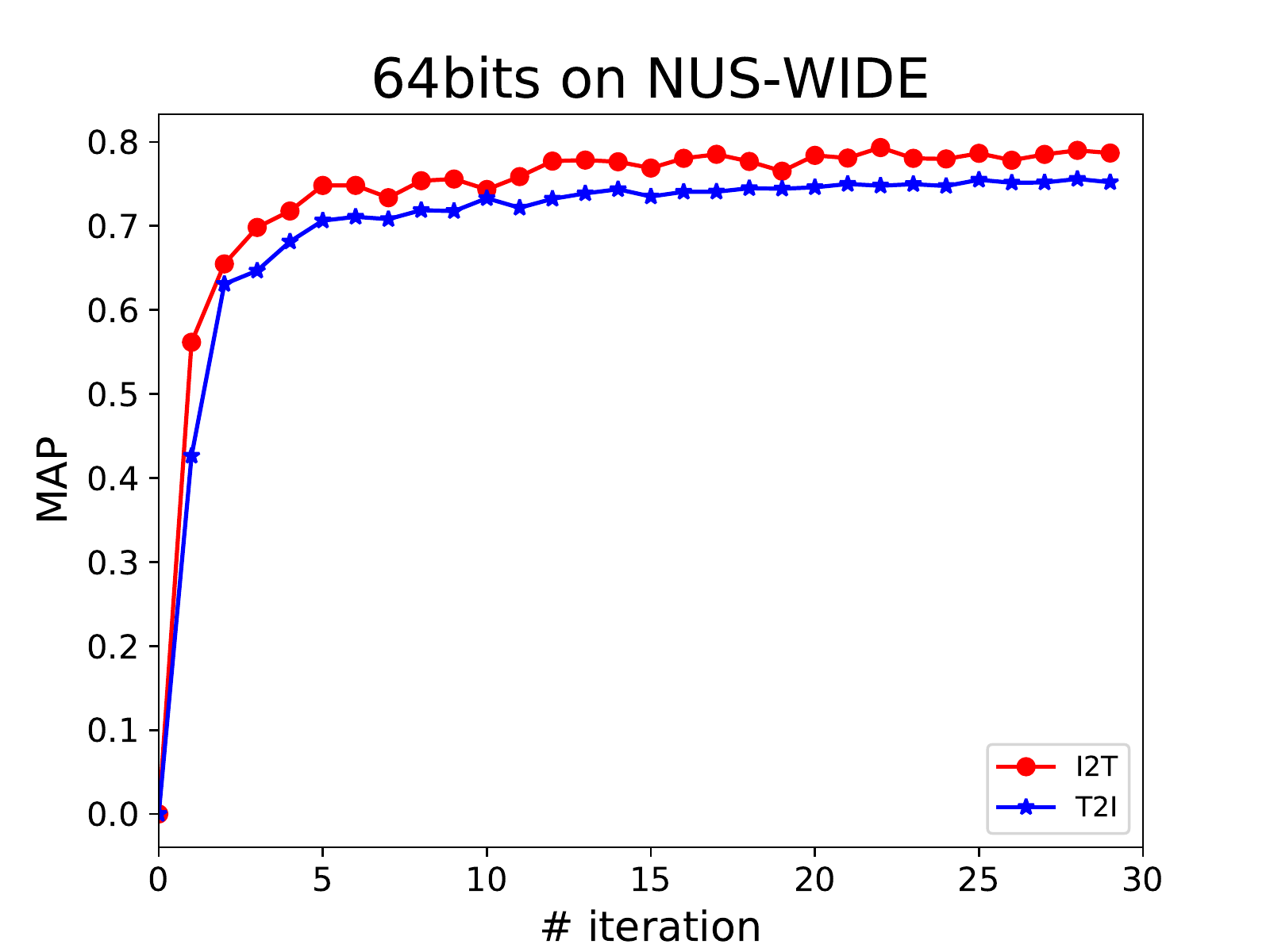}
			\label{map64}
		\end{minipage}%
	}%
	\caption{Objective function value and MAP of DCHUC over NUS-WIDE on 64 bits.}
	\label{loss_map}
\end{figure}
\subsection{Baselines and Implementation Details}
We compare our DCHUC with seven state-of-the-art methods, including four shallow cross-modal hashing methods, i.e., DLFH \cite{jiang2019discrete}, SCM \cite{zhang2014large}, CCA-ITQ \cite{gong2013iterative}  and DCH \cite{xu2017learning}, and three deep cross-modal hashing methods, i.e., DCMH \cite{jiang2017deep}, CMDVH \cite{erin2017cross} and SSAH \cite{li2018self}. The source codes of all baselines but CMDVH and DCH are kindly provided by the authors. We carefully tuned their parameters according to the scheme suggested by the authors. For CMDVH and DCH, we implement it carefully by ourselves. In order to make a fair comparison, we utilize Alexnet \cite{krizhevsky2012imagenet}, which has been pretrained on the ImageNet dataset \cite{russakovsky2015imagenet} to extract deep features as the image inputs of all shallow cross-modal baselines, and the input for image modality hashing network of each  deep cross-modal baseline is the $224 \times 224$ raw pixels.

For the proposed method, we initialize the first seven layers neural network in image feature learning part with the pre-trained Alexnet \cite{krizhevsky2012imagenet} model on ImageNet \cite{russakovsky2015imagenet}. All the parameters of the text modal hashing network and the hashing layer of image hashing network are initialized by Xavier initialization \cite{glorot2010understanding}. The unified binary code B is initialized randomly and zero-centered. The image input is the $224 \times 224$ raw pixels, and the text inputs are the BoW vectors. The hyper-parameters $\alpha, \gamma, \beta, \mu, \eta$ in DCHUC are empirically set as 50, 200, 1, 50, 50, respectively, and they will be discussed in Section \ref{sp}. We set $t_{out} = 30$, $t_{in} = 3$, $|\Phi| = 2000$ by using a validation strategy for all datasets. We adopt SGD with a mini-batch size of 64 as our optimization algorithm. The learning rate is initialized as $0.0001$ for image hashing network and $0.004$ for text modal hashing network. To avoid effect caused by class-imbalance problem between positive and negative similarity information, we empirically set the weight of the element $"-1"$ in S as the ratio between the number of element $"1"$ and the number of element $"-1"$ in S.

The source codes of CMDVH, DCH and our proposed method will be available at: \url{https://github.com/Academic-Hammer}
\subsection{Evaluation Protocol}
For hashing-based cross-modal retrieval task, Hamming ranking and hash lookup are two widely used retrieval protocols to evaluate the performance of hashing methods. In our experiments, we use three evaluation criterions: the mean average precision (MAP), the precision at n (P@n) and the precision-recall (PR) curve. MAP is the widely used metric to measure the accuracy of the Hamming ranking protocol, which is defined as the mean of average precision for all queries. PR curve is used to evaluate the accuracy of the hash lookup protocol, and P@n is used to evaluate precision by considering only the number of top returned points.

\subsection{Experimental results}
All experiments are run 3 times to reduce randomness, then the average accuracy is reported.

\subsubsection{Hamming Ranking Task}
Table \ref{table_map} and Table \ref{table_precision} present the MAP and Precision@1000 on MIRFLICKR-25K, IAPR TC-12  and NUS-WIDE datasets, respectively. $"I \rightarrow T"$ denotes retrieving texts with image queries, and $"T \rightarrow I"$ denotes retrieving images with text queries. In general, from Table \ref{table_map} and Table \ref{table_precision}, we have three observations: ($\romannumeral1$) Our proposed method can outperforms the other cross-modal hashing methods for different length of hash code. For example, on MIRFLICKR-25K, comparing with the best competitor SSAH on 16-bits, the results of DCHUC for $"I\rightarrow T"$ have a relative increase of 12.7\% on MAP and 9.2\% on Precision@1000; the results of DCHUC for $"T\rightarrow I"$ have a relative increase of 8.6\% on MAP and 8.7\% on Precision@1000. On IAPR TC-12, comparing with the competitor SSAH on 64-bits, the results of DCHUC for $"I \rightarrow T"$ have a relative increase of 19.4\% on MAP and 15.5\% on Precision@1000; the results of DCHUC for $"T\rightarrow I"$ have a relative increase of 18.3\% on MAP and 15.5\% on Precision@1000. On NUS-WIDE, comparing with the best competitor DCH on 64-bits, the results of DCHUC for $"I \rightarrow T"$ have a relative increase of 12.3\% on MAP and 7.8\% on Precision@1000; ($\romannumeral2$) Integrating the feature learning of data-points and hashing function learning into an end-to-end network can get the better performance. For example, our proposed method can get a better performance than DCH which also can jointly learning unified hashing codes for instances in the database and modal-specific hashing functions for unseen data-points but the feature extraction procedure is independent of the hash codes learning procedure. ($\romannumeral3$) Jointly learning unified hashing codes for database instances and modality-specific hashing functions for unseen data-points can greatly increase the retrieval performance. For instance, DCHUC can get better performance on MAP and Precision@1000 over three benchmark datasets than CMDVH. Note that, the results of CMDVH is not as good as the results of the original article. It maybe the reason that we used more classes of label to carry out our experimental, which is hard to train the svm used in CMDVH. Furthermore, although DCH is a shallow hashing method, its retrieval performances on MIRFLICKR-25K and IAPR TC-12 datasets are similar to the best deep baseline SSAH, and its retrieval performances on NUS-WIDE dataset is batter than SSAH.
\begin{figure*}[]
	\centering
	\subfigure[I2T@64 bits]{
		\begin{minipage}[t]{0.33\textwidth}
			\centering
			\includegraphics[width=\linewidth]{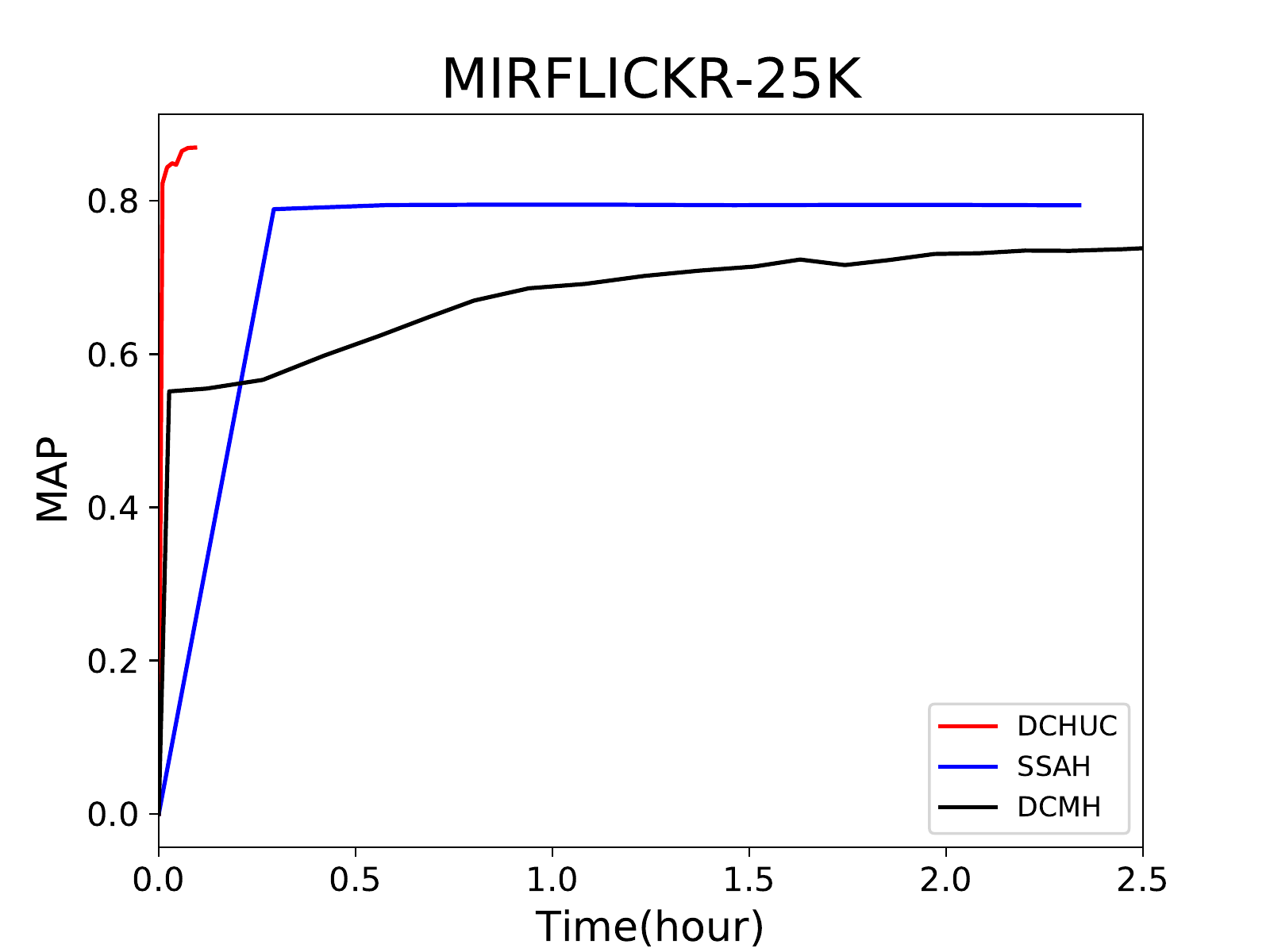}
		\end{minipage}%
	}%
	\subfigure[I2T@64 bits]{
		\begin{minipage}[t]{0.33\textwidth}
			\centering
			\includegraphics[width=\linewidth]{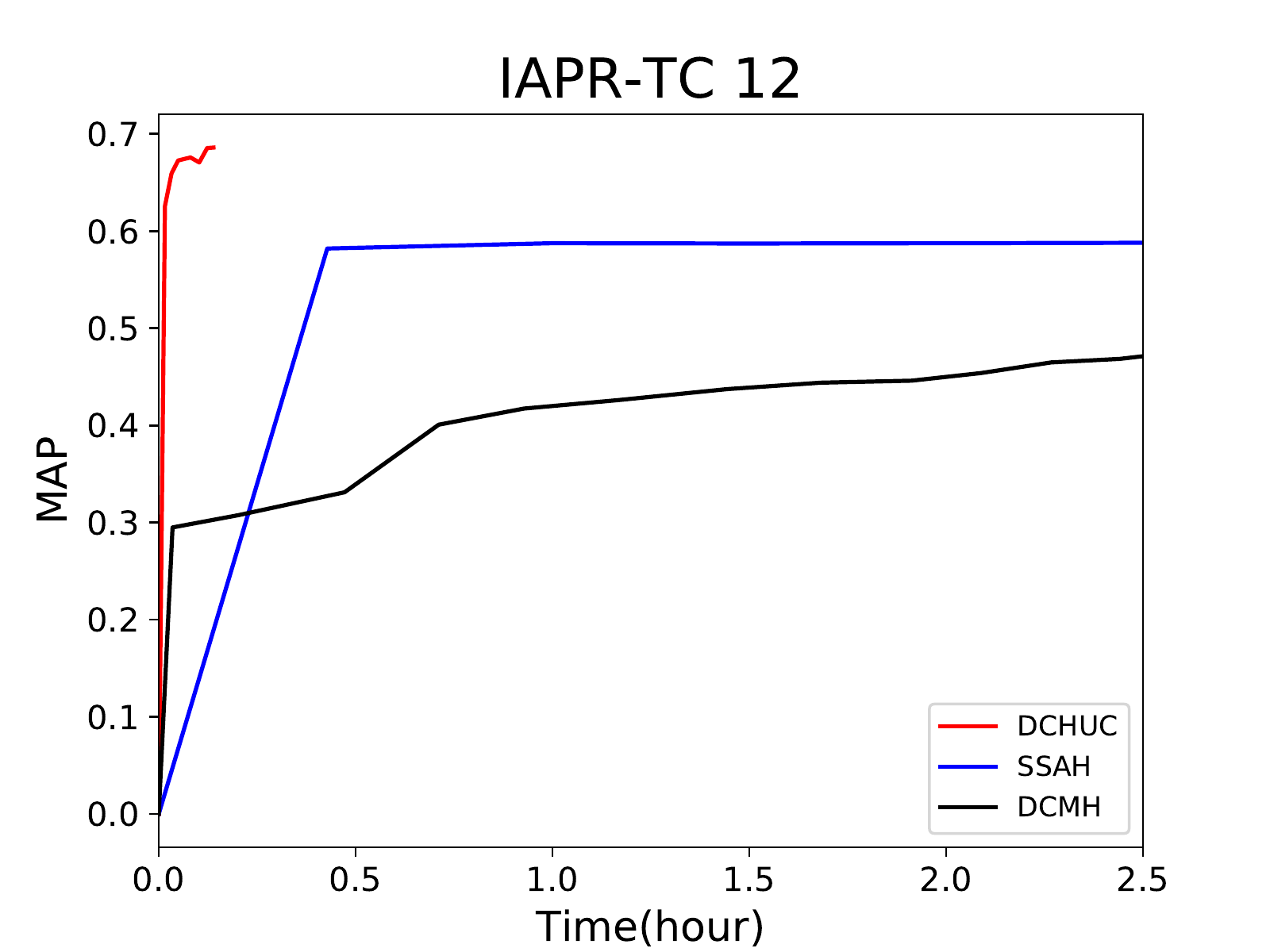}
		\end{minipage}%
	}%
	\subfigure[I2T@64 bits]{
		\begin{minipage}[t]{0.33\textwidth}
			\centering
			\includegraphics[width=\linewidth]{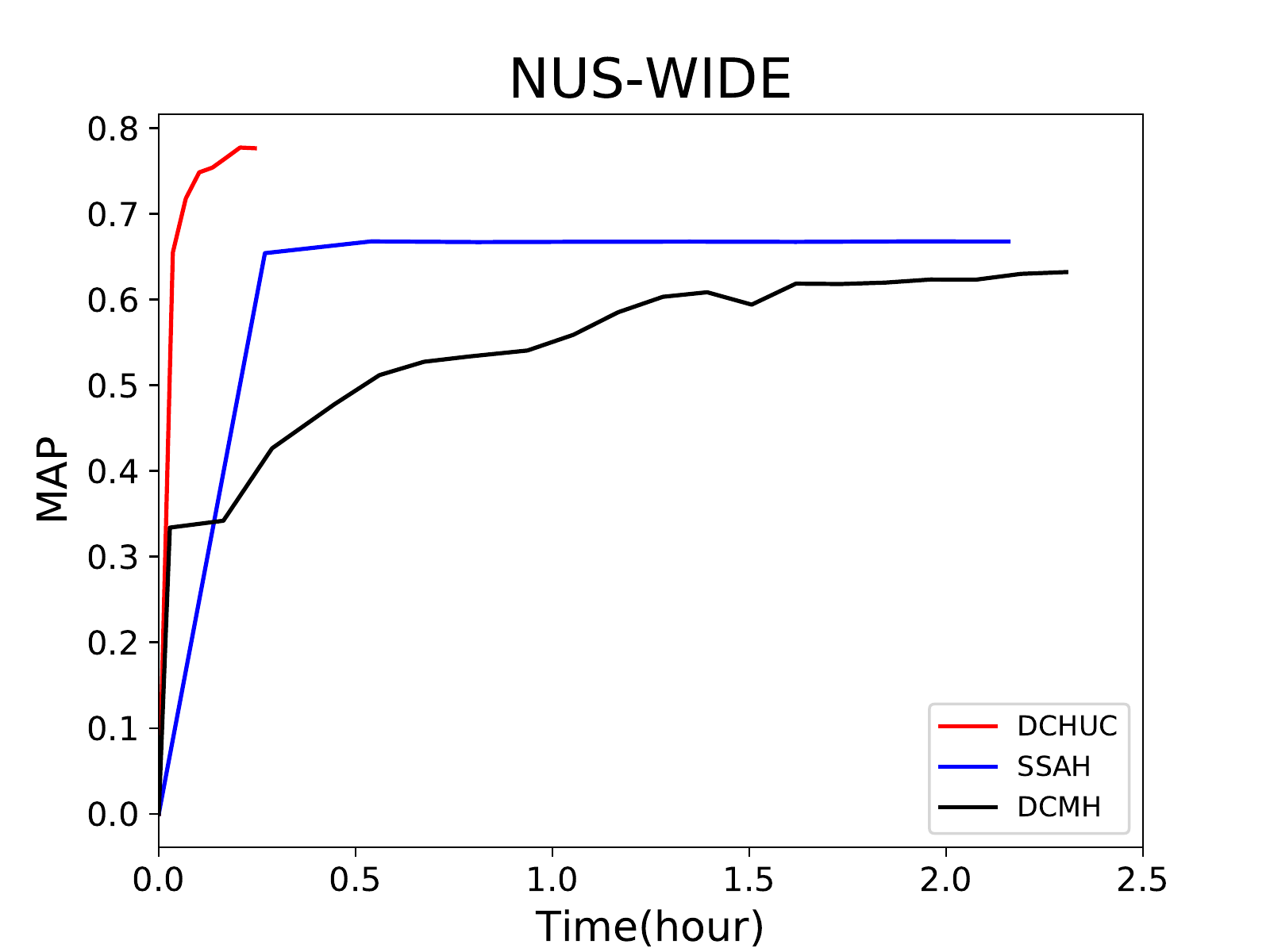}
		\end{minipage}%
	}%
	\quad
	\subfigure[T2I@64 bits]{
		\begin{minipage}[t]{0.33\textwidth}
			\centering
			\includegraphics[width=\linewidth]{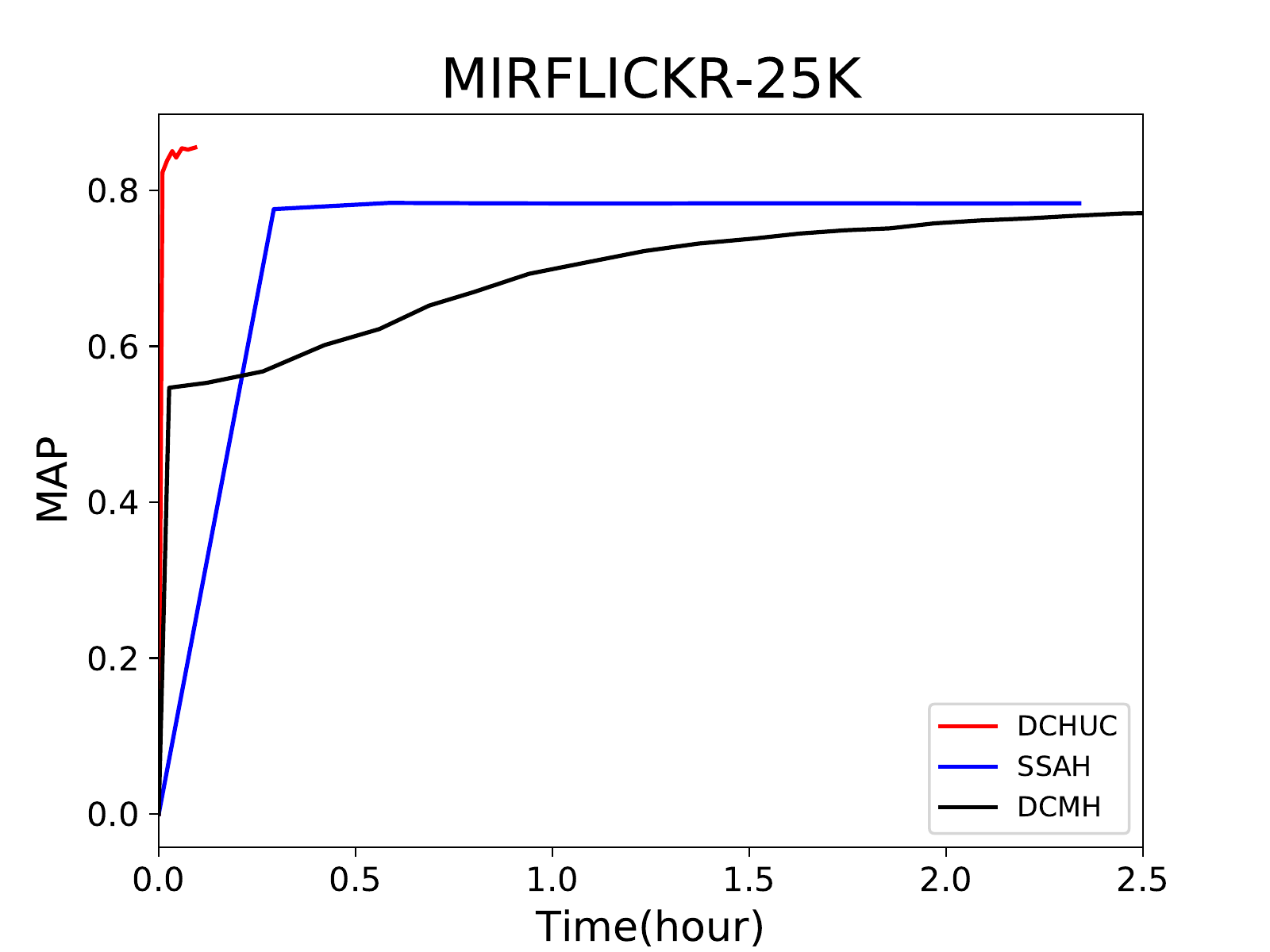}
		\end{minipage}%
	}%
	\subfigure[T2I@64 bits]{
		\begin{minipage}[t]{0.33\textwidth}
			\centering
			\includegraphics[width=\linewidth]{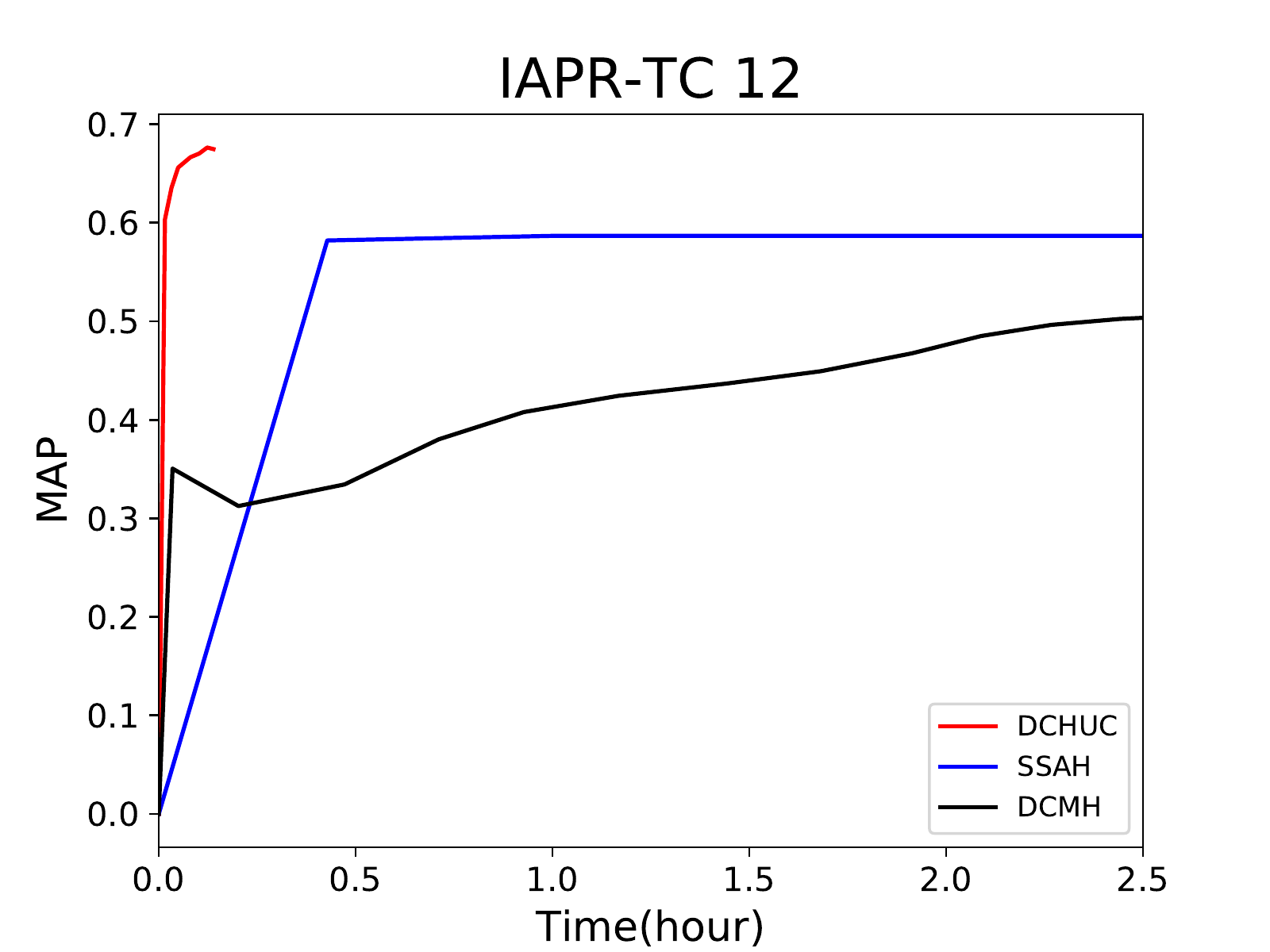}
		\end{minipage}%
	}%
	\subfigure[T2I@64 bits]{
		\begin{minipage}[t]{0.33\textwidth}
			\centering
			\includegraphics[width=\linewidth]{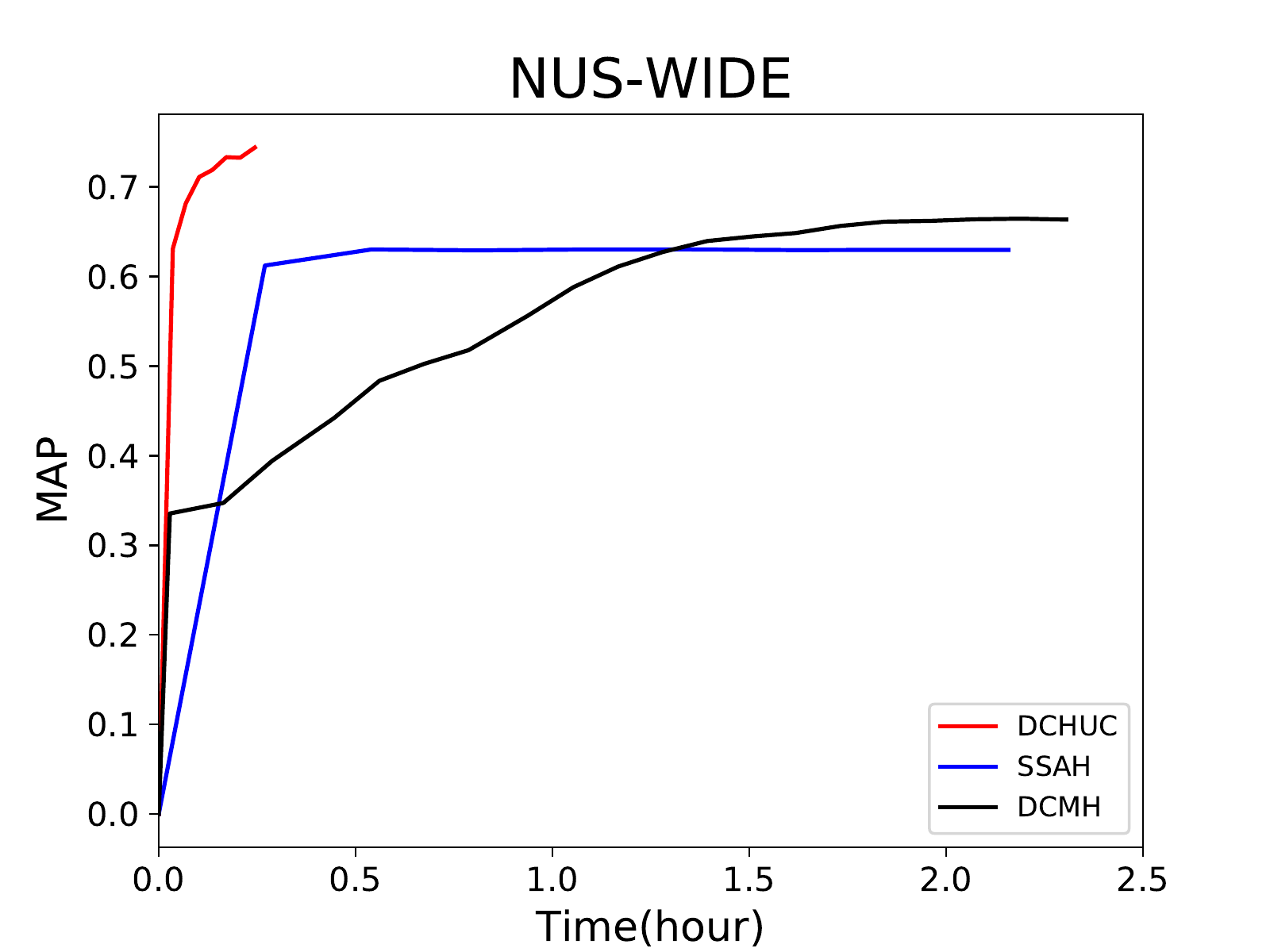}
		\end{minipage}%
	}%
	\caption{Training Efficiency of DCHUC, SSAH and DCMH on Three Datasets.}
	\label{fig_time}
\end{figure*}

\begin{figure*}[]
	\centering
	\subfigure[]{
		\begin{minipage}[t]{0.2\textwidth}
			\centering
			\includegraphics[width=\linewidth]{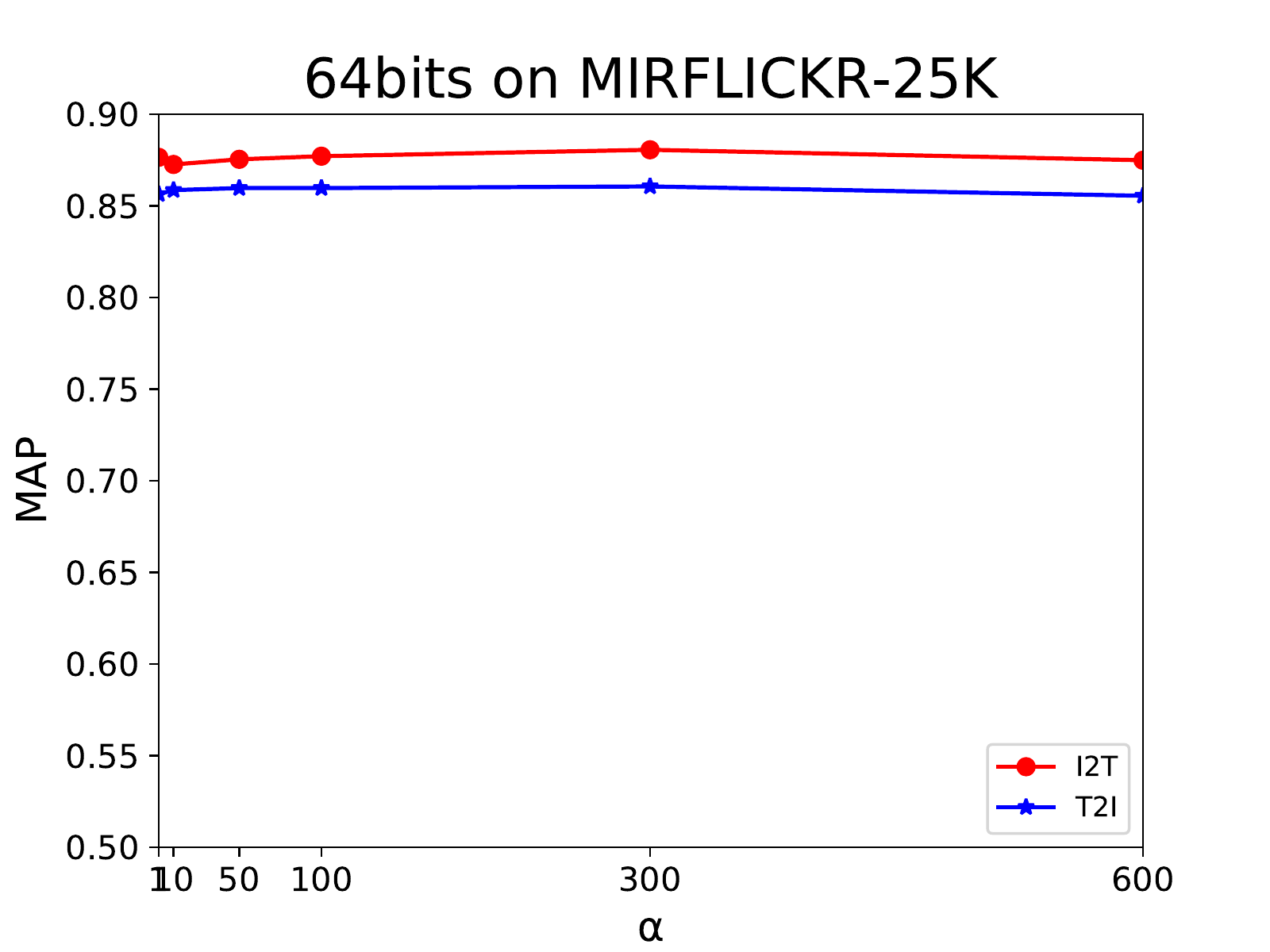}
			\label{mir_alpha}
		\end{minipage}%
	}%
	\subfigure[]{
		\begin{minipage}[t]{0.2\textwidth}
			\centering
			\includegraphics[width=\linewidth]{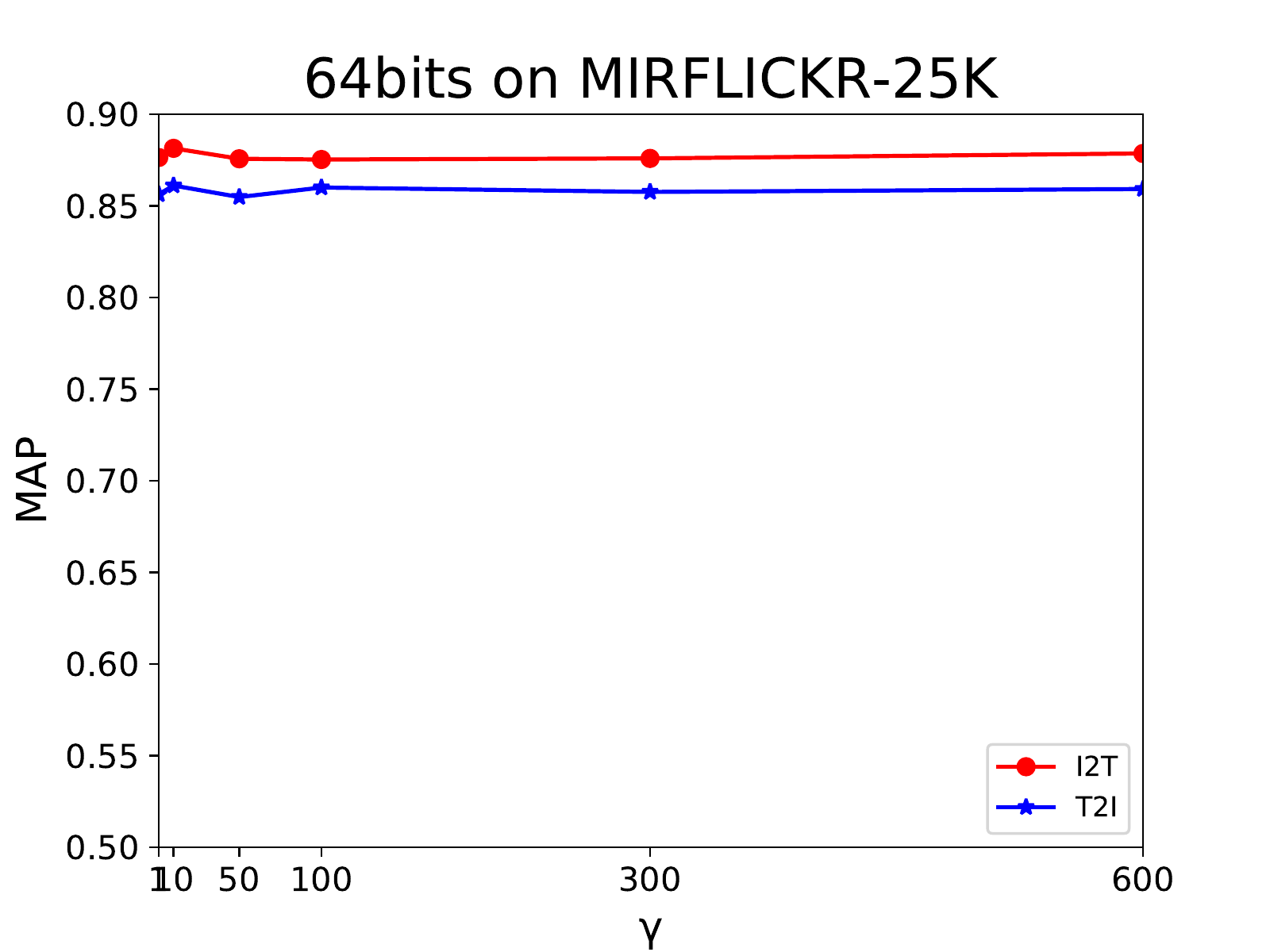}
			\label{mir_gamma}
		\end{minipage}%
	}%
	\subfigure[]{
		\begin{minipage}[t]{0.2\textwidth}
			\centering
			\includegraphics[width=\linewidth]{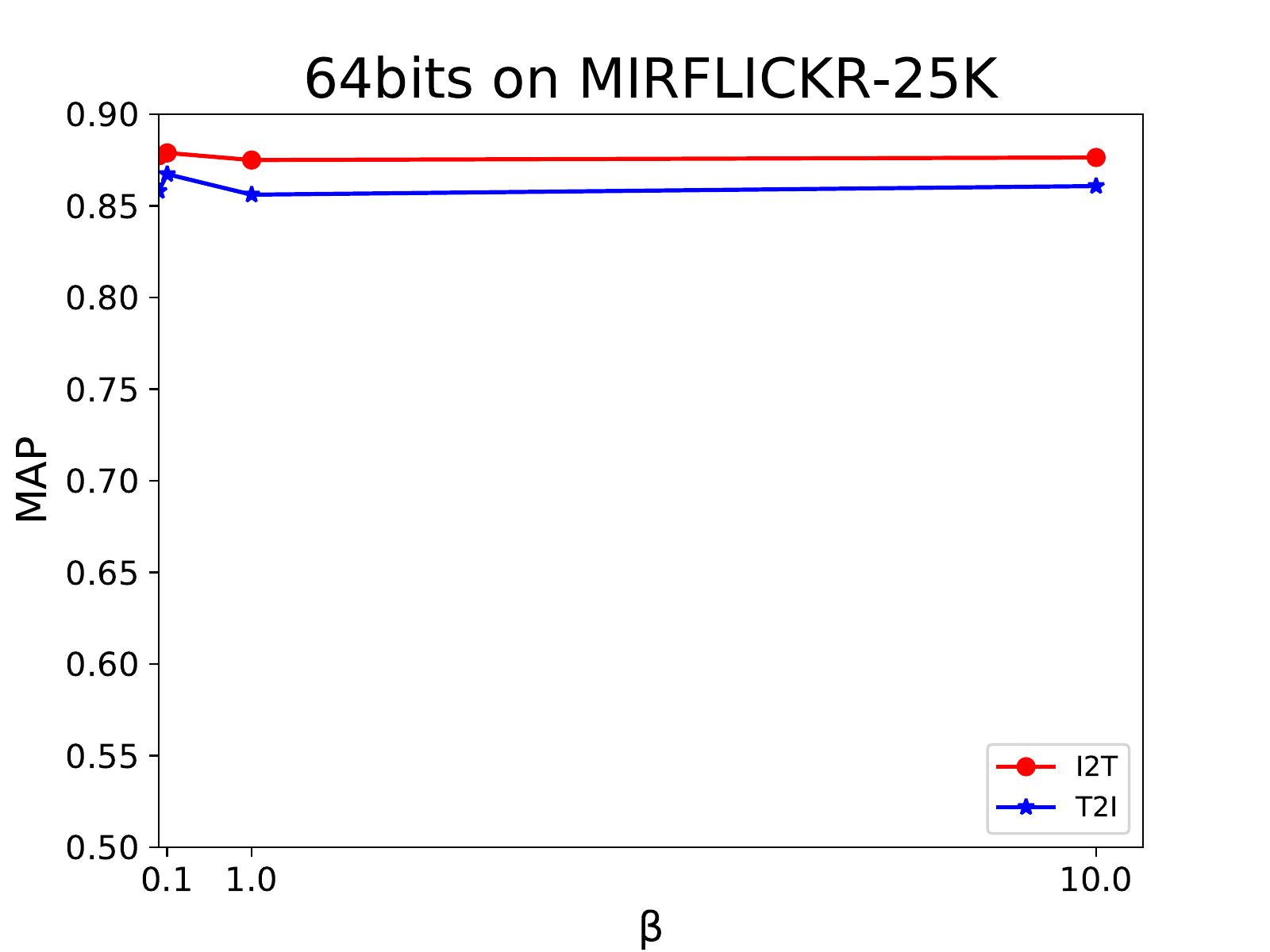}
			\label{mir_beta}
		\end{minipage}%
	}%
	\subfigure[]{
		\begin{minipage}[t]{0.2\textwidth}
			\centering
			\includegraphics[width=\linewidth]{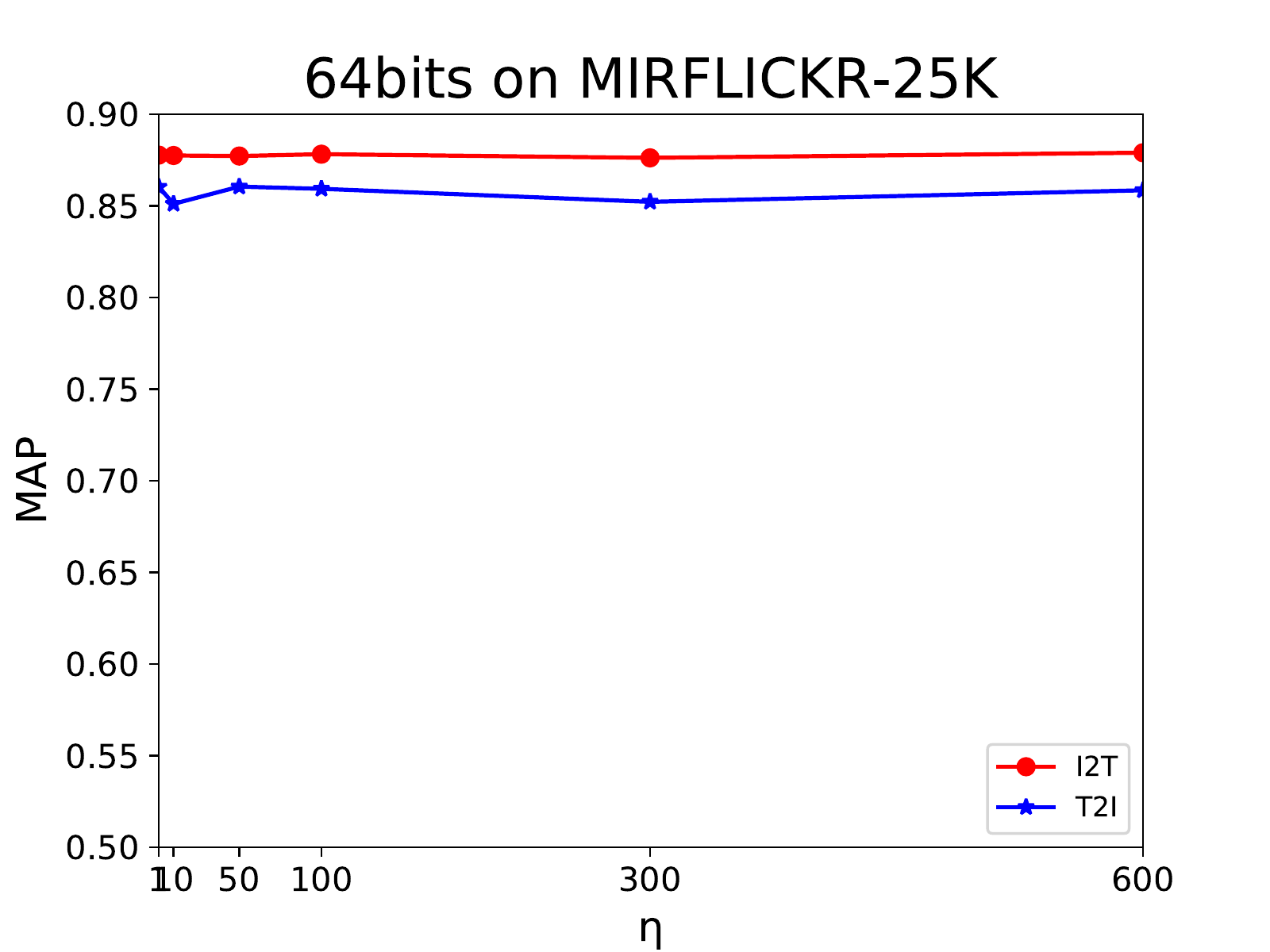}
			\label{mir_yita}
		\end{minipage}%
	}%
	\subfigure[]{
		\begin{minipage}[t]{0.2\textwidth}
			\centering
			\includegraphics[width=\linewidth]{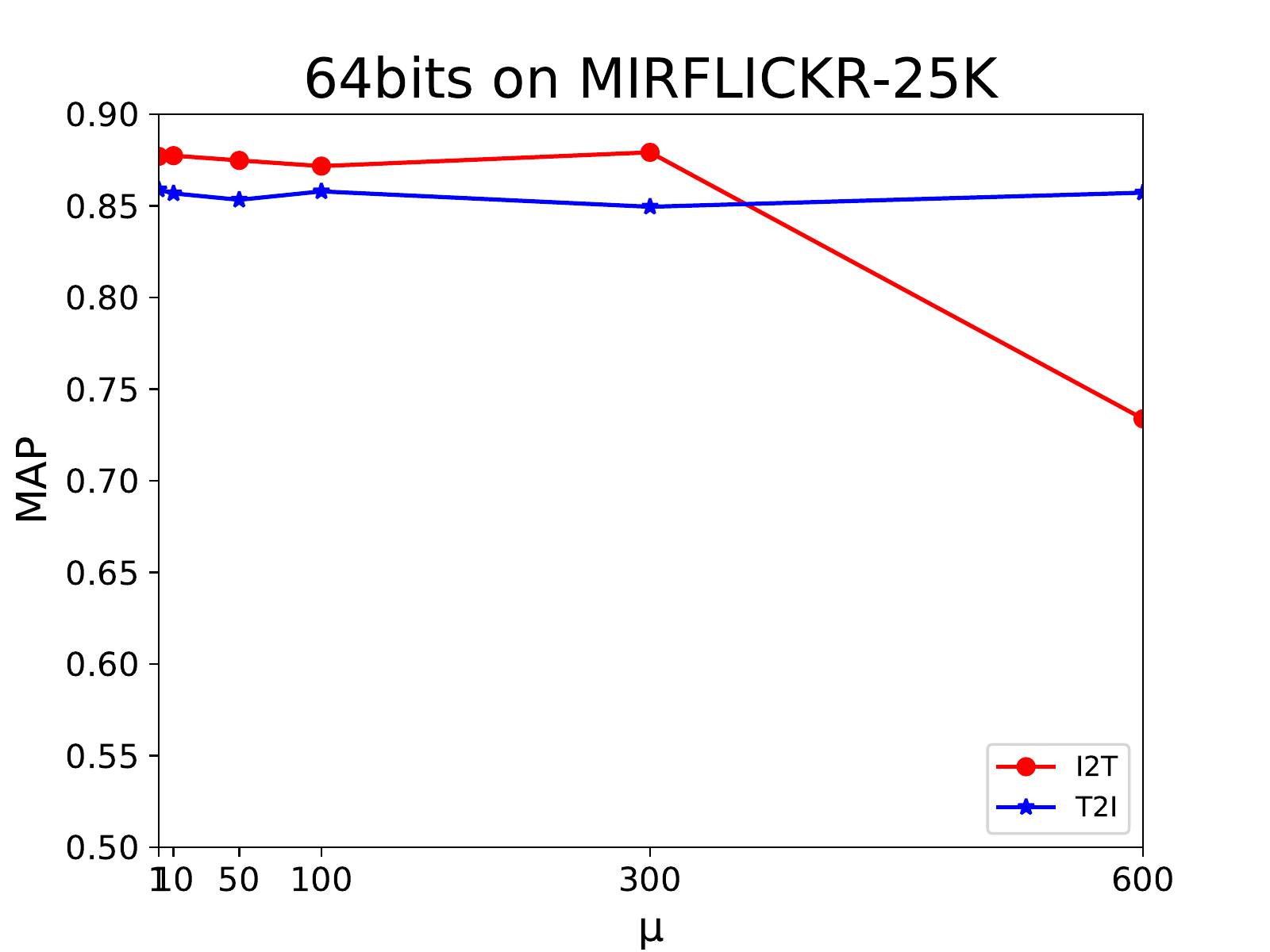}
			\label{mir_mu}
		\end{minipage}%
	}%
	\quad
	\subfigure[]{
		\begin{minipage}[t]{0.2\textwidth}
			\centering
			\includegraphics[width=\linewidth]{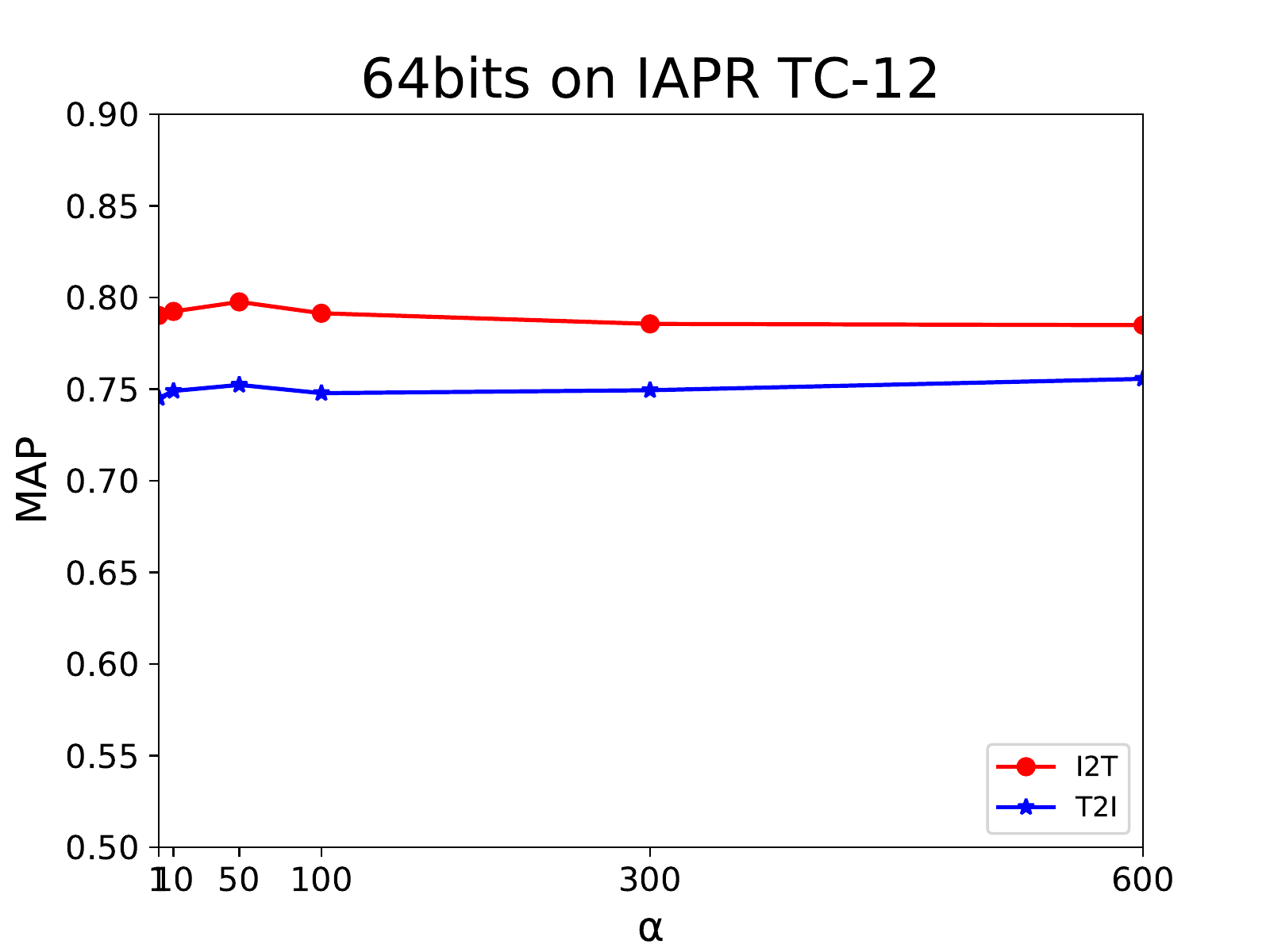}
			\label{iaprtc_alpha}
		\end{minipage}%
	}%
	\subfigure[]{
		\begin{minipage}[t]{0.2\textwidth}
			\centering
			\includegraphics[width=\linewidth]{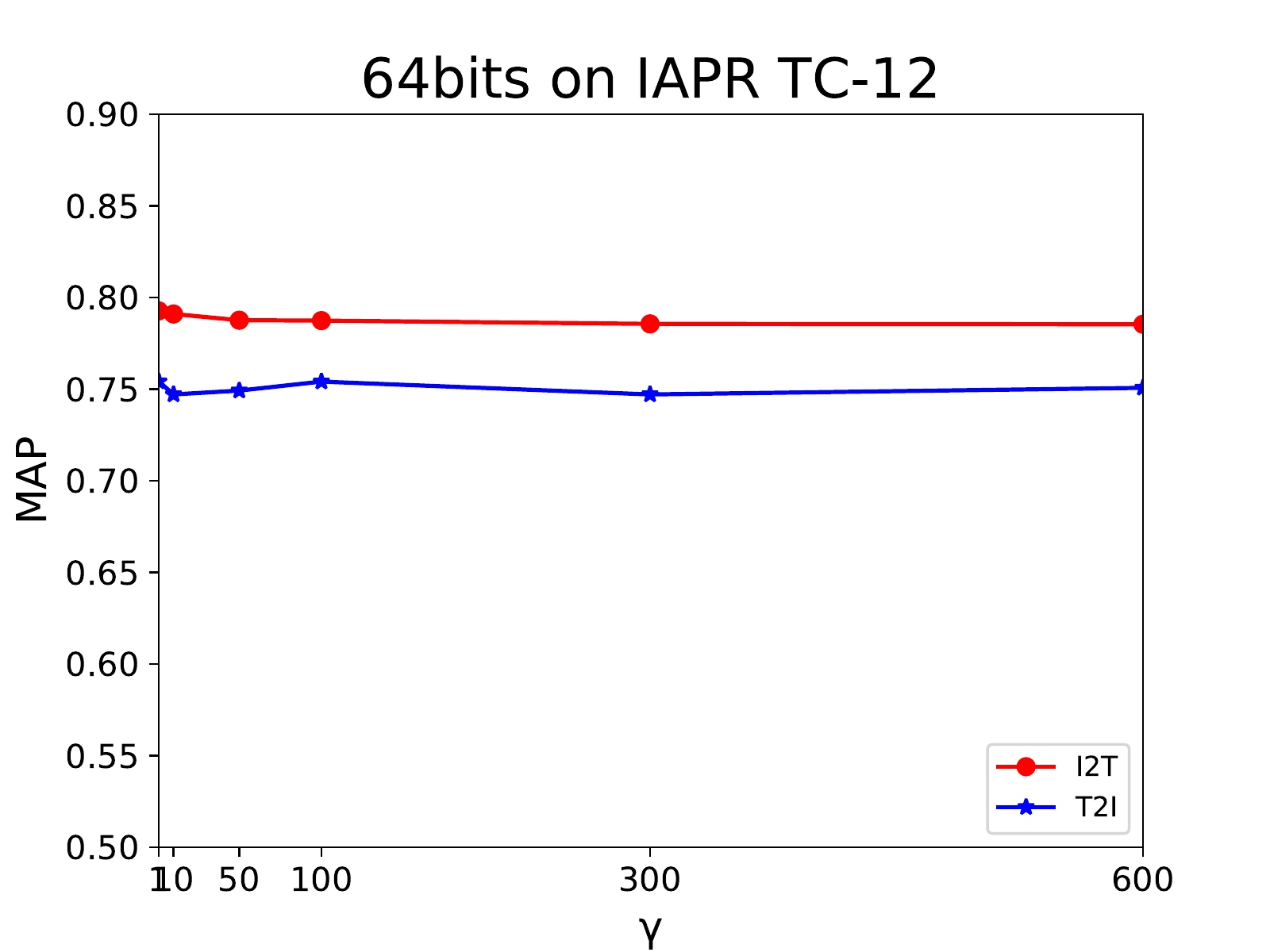}
			\label{iaprtc_gamma}
		\end{minipage}%
	}%
	\subfigure[]{
		\begin{minipage}[t]{0.2\textwidth}
			\centering
			\includegraphics[width=\linewidth]{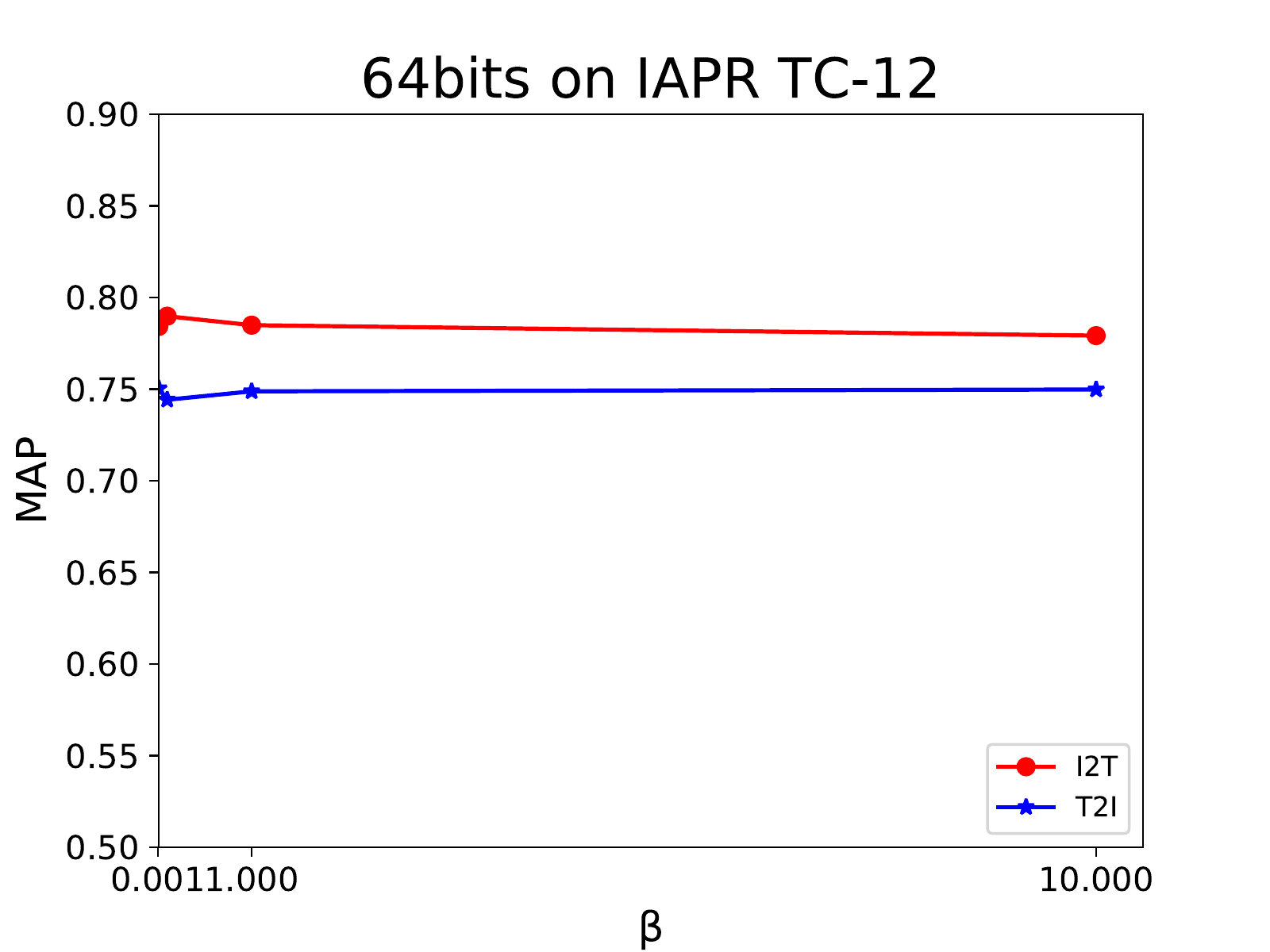}
			\label{iaprtc_beta}
		\end{minipage}%
	}%
	\subfigure[]{
		\begin{minipage}[t]{0.2\textwidth}
			\centering
			\includegraphics[width=\linewidth]{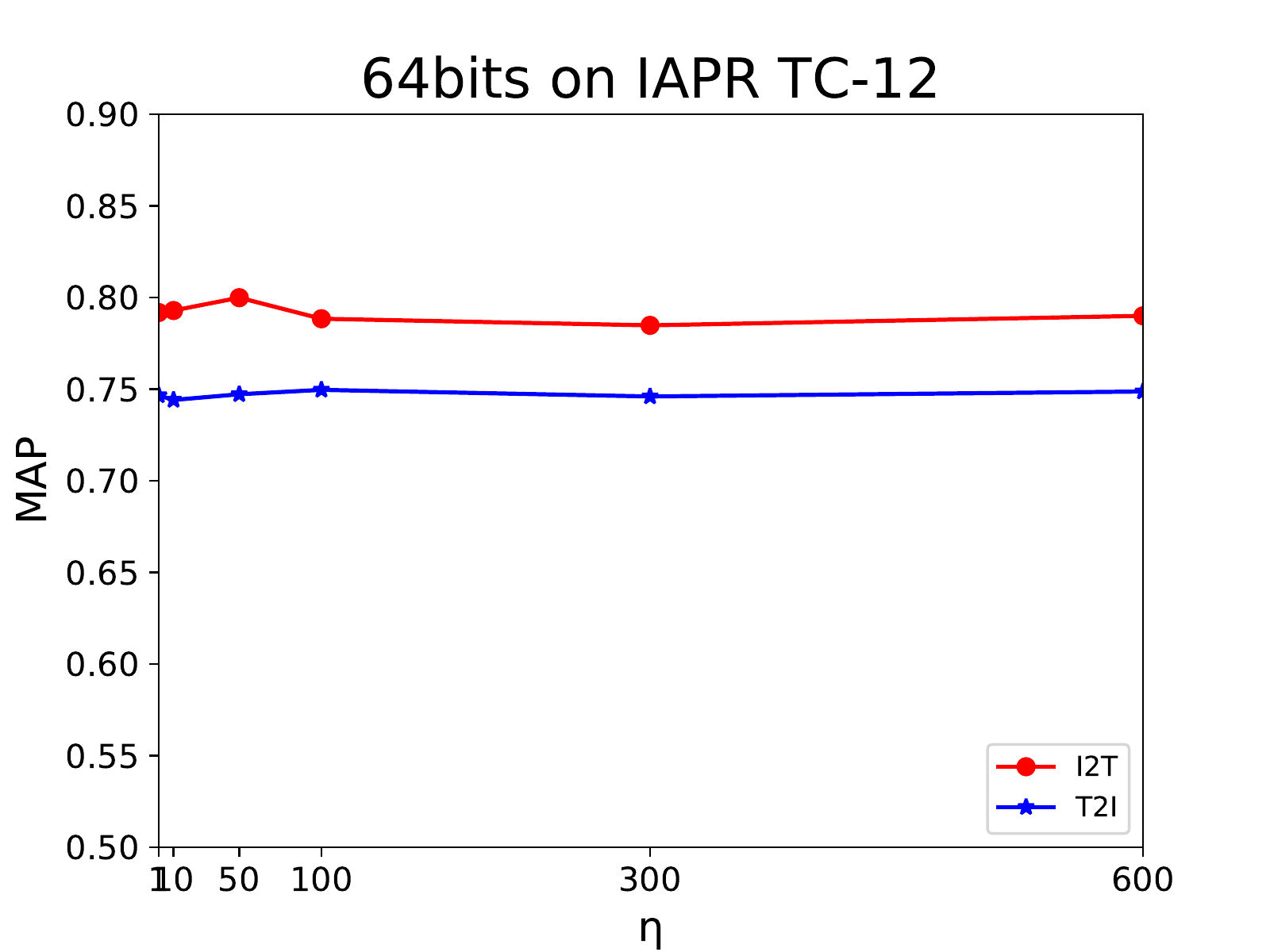}
			\label{iaprtc_yita}
		\end{minipage}%
	}%
	\subfigure[]{
		\begin{minipage}[t]{0.2\textwidth}
			\centering
			\includegraphics[width=\linewidth]{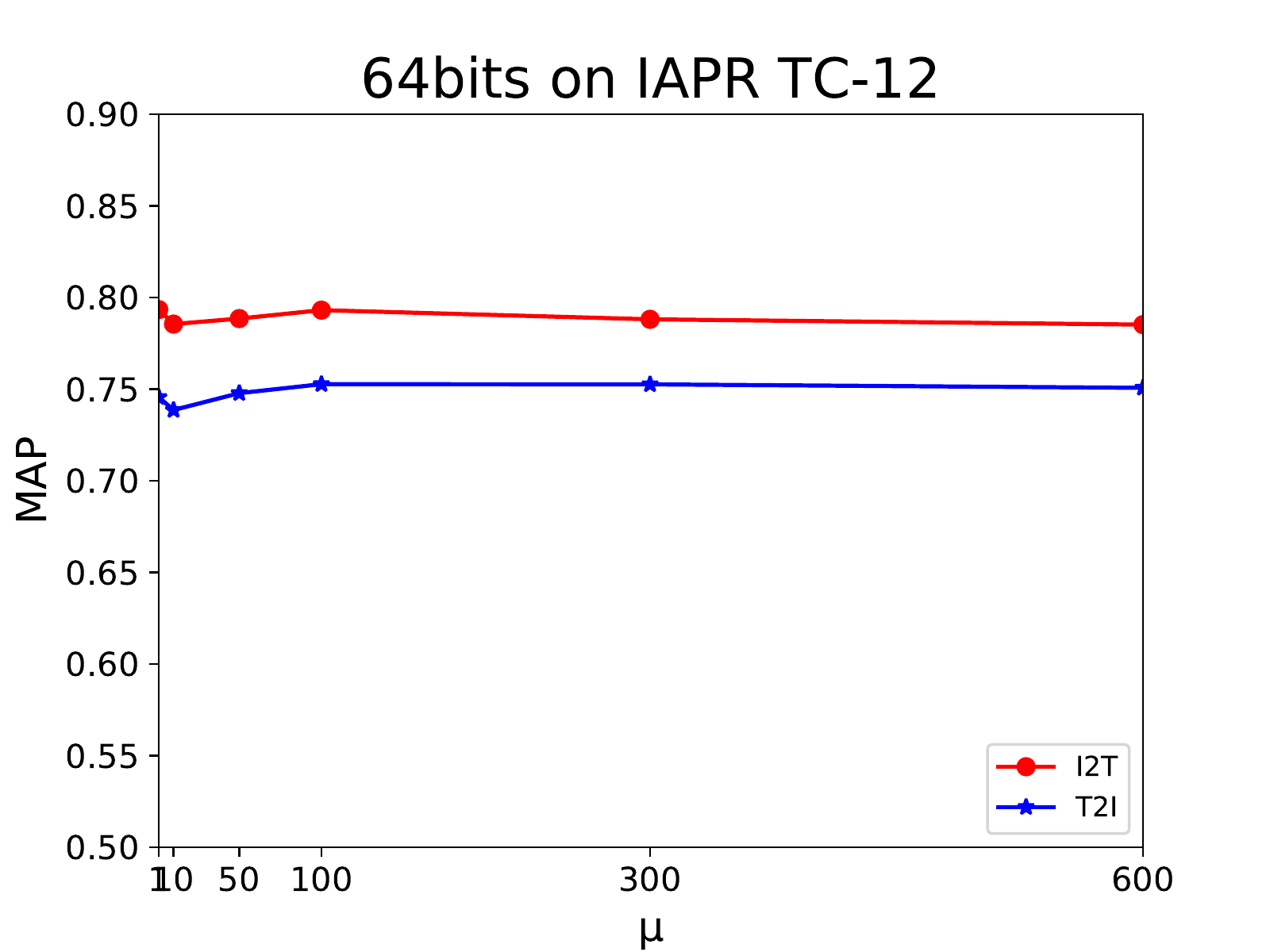}
			\label{iaprtc_mu}
		\end{minipage}%
	}%
	\quad
	\subfigure[]{
		\begin{minipage}[t]{0.2\textwidth}
			\centering
			\includegraphics[width=\linewidth]{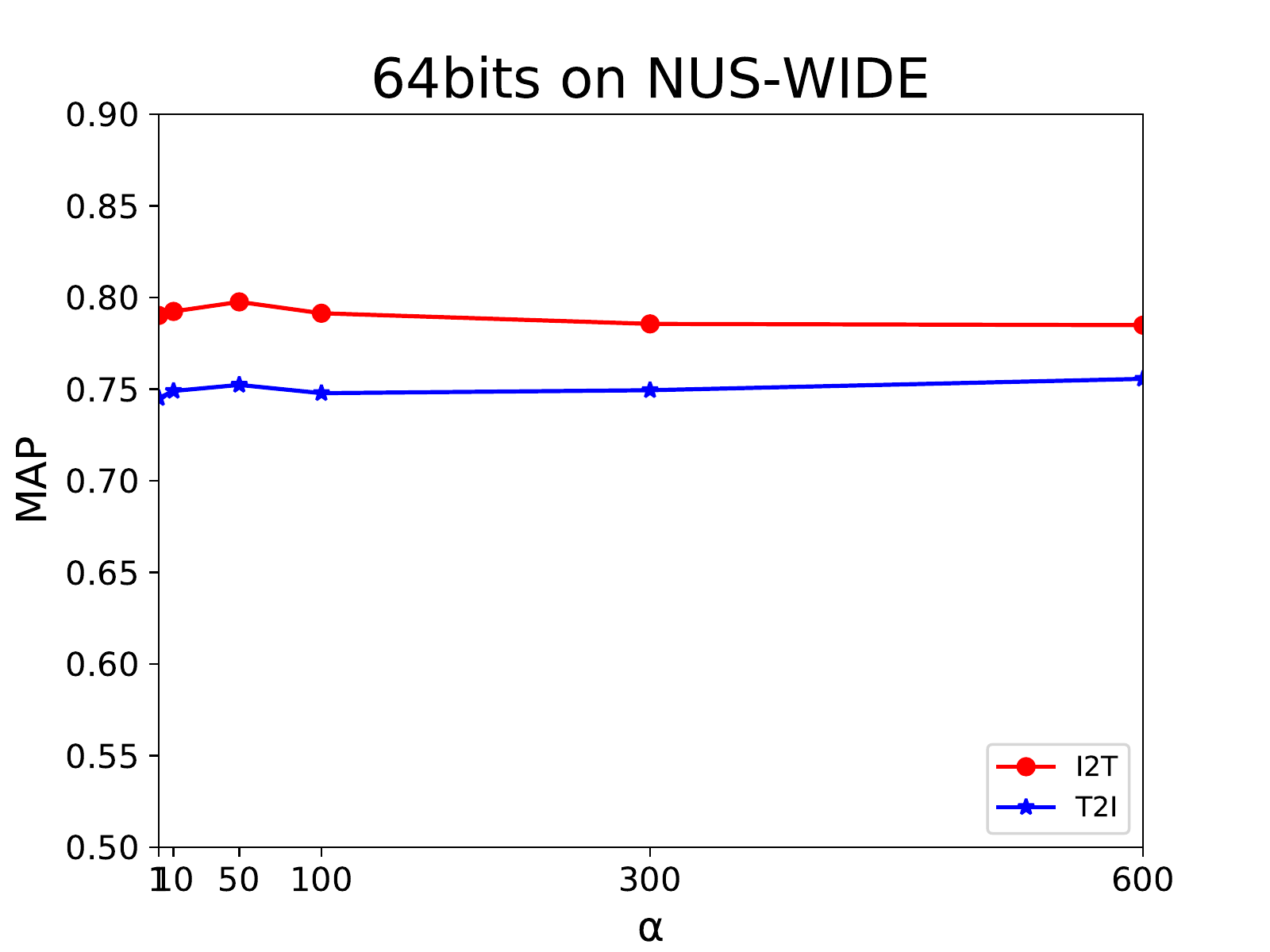}
			\label{nus_alpha}
		\end{minipage}%
	}%
	\subfigure[]{
		\begin{minipage}[t]{0.2\textwidth}
			\centering
			\includegraphics[width=\linewidth]{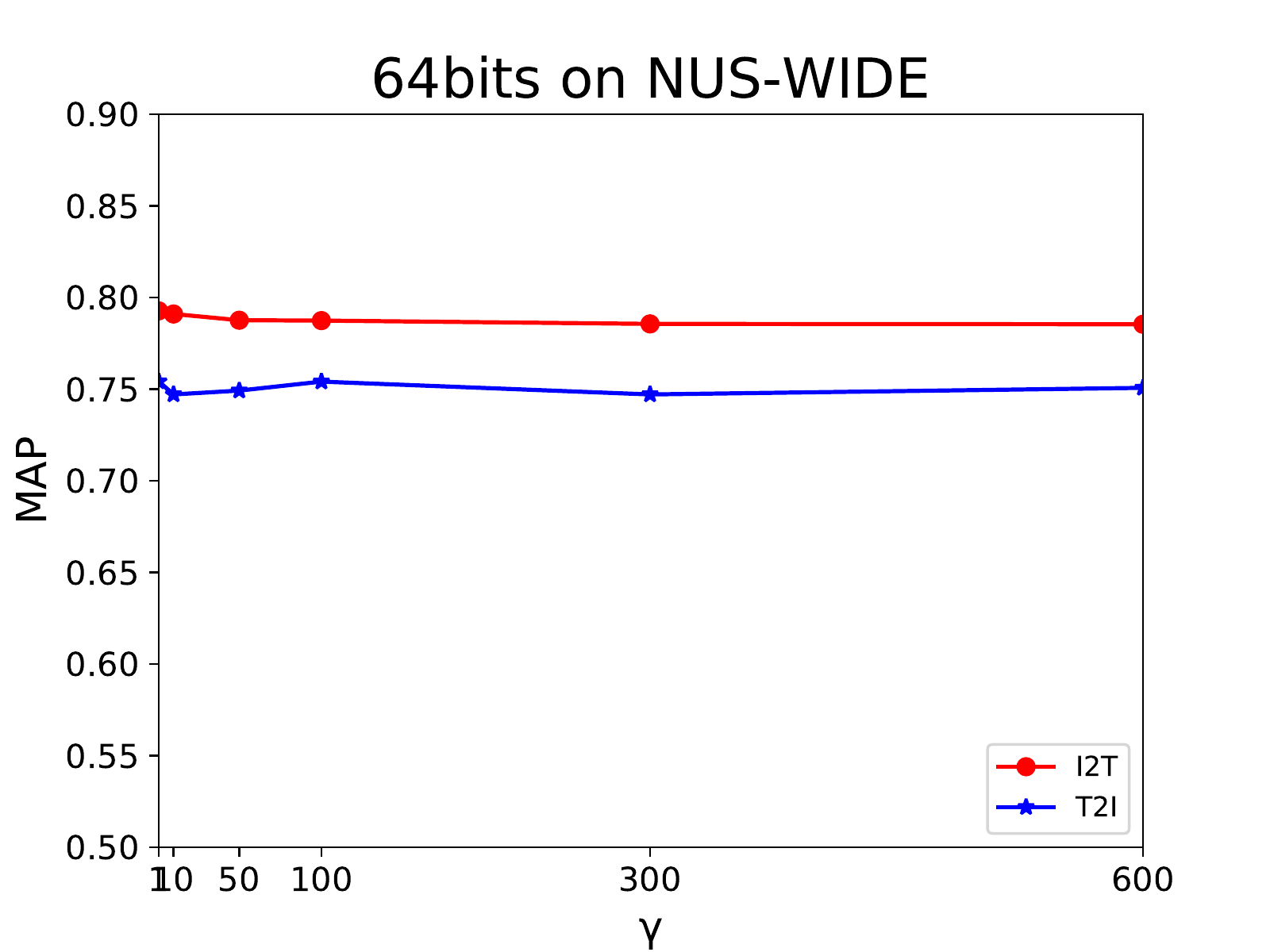}
			\label{nus_gamma}
		\end{minipage}%
	}%
	\subfigure[]{
		\begin{minipage}[t]{0.2\textwidth}
			\centering
			\includegraphics[width=\linewidth]{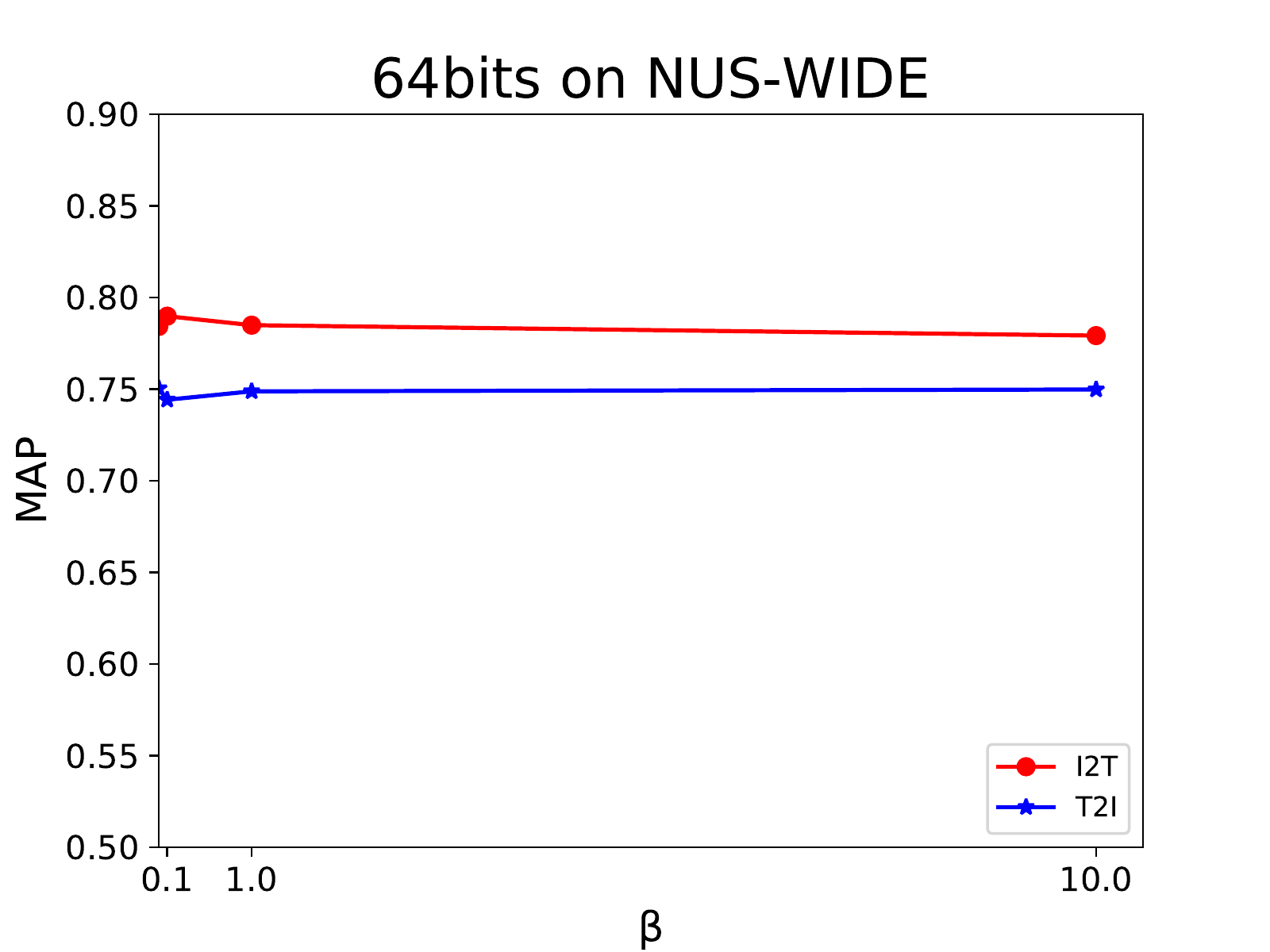}
			\label{nus_beta}
		\end{minipage}%
	}%
	\subfigure[]{
		\begin{minipage}[t]{0.2\textwidth}
			\centering
			\includegraphics[width=\linewidth]{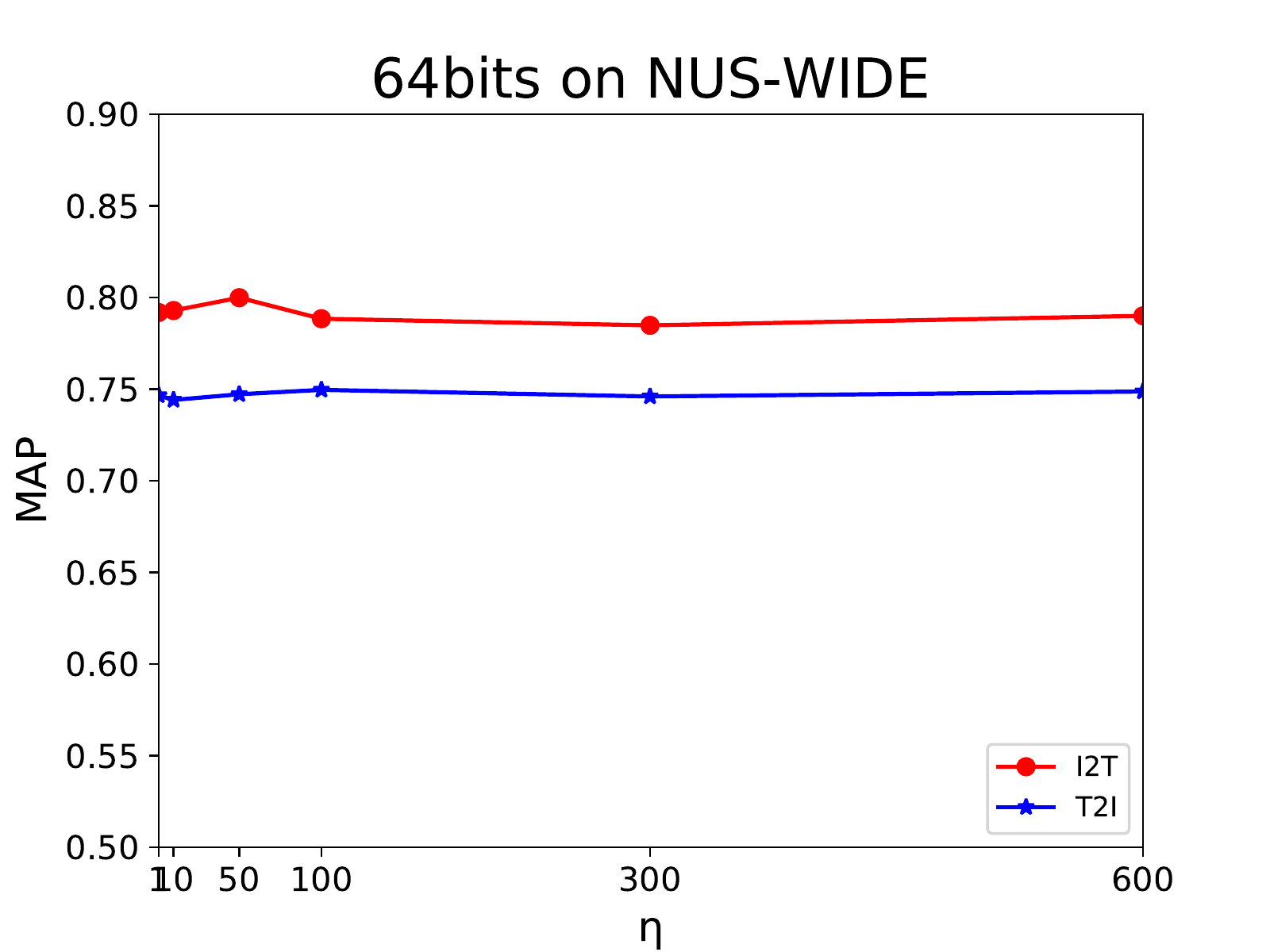}
			\label{nus_yita}
		\end{minipage}%
	}%
	\subfigure[]{
		\begin{minipage}[t]{0.2\textwidth}
			\centering
			\includegraphics[width=\linewidth]{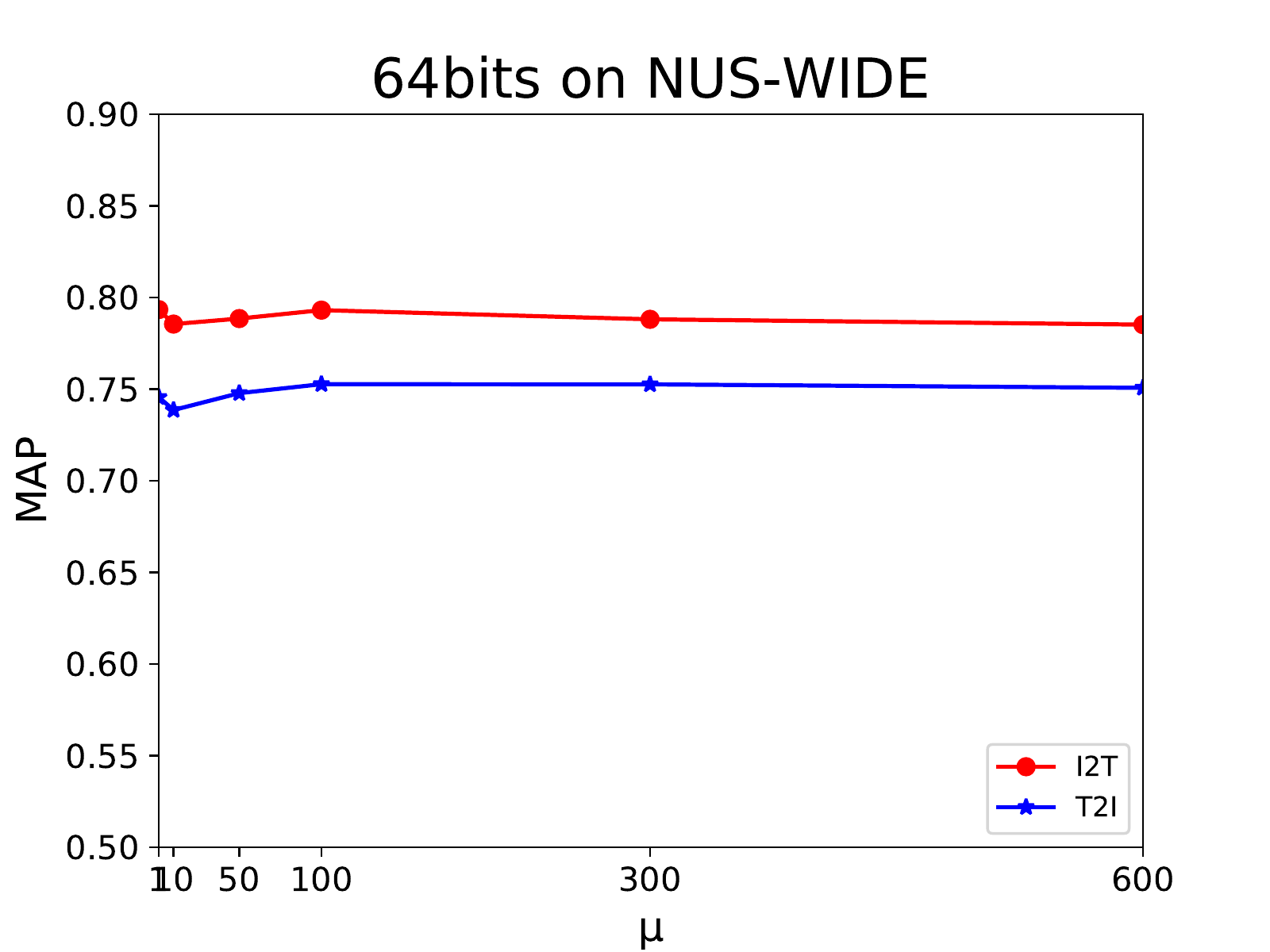}
			\label{nus_mu}
		\end{minipage}%
	}%
	\caption{MAP values with different parameters on three datasets.}
	\label{fig_par_nus}
\end{figure*}

\subsubsection{Hash Lookup Task}
When considering the lookup protocol, we compute the precision and recall (PR) curve for the returned points given any Hamming radius. The PR curve can be obtained by varying the Hamming radius from 0 to $k$ with a step-size of 1. Fig. \ref{fig_pr_mir}, Fig. \ref{fig_pr_iaprtc} and Fig. \ref{fig_pr_nus} show the precision-recall curve on MIRFLICKR-25K, IAPR TC-12 and NUS-WIDE datasets, respectively. It is easy to find that DCHUC can dramatically outperform the state-of-the-art baselines, which means our DCHUC generates hash codes for similar points in a small Hamming radius. For example, compared with baselines, the precision value of DCHUC decreases more slowly with the recall value increasing, and DCHUC can get a high precision value even though the recall value increasing to 0.9 on MIRFLICKR-25K and NUS-WIDE datasets.

\subsection{Convergence Analysis}
To verify the convergence property of DCHUC, we conduct an experiment over NUS-WIDE dataset with the code length being 64. Fig. \ref{loss_map} shows the convergence of objective function value and MAP. As shown in Fig. \ref{loss_map} (a), the objective function value can convergence after only 10 iterations. In Fig. \ref{loss_map} (b),  $"I \rightarrow T"$ denotes retrieving texts with image queries, and $"T \rightarrow I"$ denotes retrieving images with text queries. We can find the MAP values of both the two retrieval task can convergence. Furthermore, combining the two subfigure Fig. \ref{loss_map} (a) and (b), we can find both the two map values can increase with the objective function value decrease and eventually converge. 

\subsection{Training efficiency}
To evaluate the training speed of DCHUC, we conduct experiments between the deep cross-modal baselines except CMDVH on three datasets. Fig. \ref{fig_time} shows the variation between MAP and training time on the three datasets for DCHUC, SSAH and DCMH. It can be find that DCHUC can not only training faster than the two deep cross-modal baselines, but also get a better performance on retrieval tasks than them. For the CMDVH baseline, it is a two step method. Then it is unfair to compare MAP-Time curve. In here, we calculate the the whole training time of CMDVH. The cost times of training phase on IAPR TC-12, MIRFLICKR-25K and NUS-WIDE datasets with 32-bits are 16.3s, 21.2s and 39.2s for CMDVH, and are 11.8s, 12.9s and 28.2s for DCHUC, respectively. We can find that DCHUC is also the faster one.

\subsection{Sensitivity to Parameters} 
\label{sp}
We study the influence of the hyper-parameters $\alpha,\gamma,\beta,\eta$ and $\mu$ on IAPR TC-12, MIRFLICKR-25K and NUS-WIDE datasets with the code length being 64-bits. More specially,Fig. \ref{fig_par_nus} (a), (f) and (k) show the affect of the hyper-parameter $\alpha$ over the three datasets with the value between $1$ and $600$. Fig. \ref{fig_par_nus} (b), (g) and (i) show the affect of the hyper-parameter $\gamma$ over the three datasets with the value between $1$ and $600$. Fig. \ref{fig_par_nus} (c), (h) and (m) show the affect of the hyper-parameter $\beta$ over the three datasets with the value between $10^{-3}$ and $10$. Fig. \ref{fig_par_nus} (d), (i) and (n) show the affect of the hyper-parameter $\eta$ over the three datasets with the value between $1$ and $600$. Fig. \ref{fig_par_nus} (e), (j) and (o) show the affect of the hyper-parameter $\mu$ over the three datasets with the value between $1$ and $600$. It can be found that DCHUC is not sensitive to $\alpha,\gamma,\beta,\eta$ and $\mu$. For instance, DCHUC can achieve good performance on all the three datasets in the range of $1$ to $600$ for the hyper-parameters $\alpha,\gamma$ and $\eta$, and also can achieve good performance on all the three datasets with $1 \leq \beta \leq 300$. Furthermore, DCHUC can get  the high MAP values with different $\beta$ from the range of $10^{-3}$ to $10$.

\section{Conclution}
In this paper, we have proposed a novel cross-modal deep hashing method for cross-modal data, called DCHUC. To the best of our knowledge, DCHUC is the first deep method to jointly learn unified hash codes for database instances and hashing functions for unseen query points in an end-to-end framework. Extensive experiments on three real-world public datasets have shown that the proposed DCHUC method outperforms the state-of-the-art cross-modal hashing methods.
\section*{Acknowledgment}
The work is supported by SFSMBRP(2018YFB1005100), BIGKE(No. 20160754021), NSFC (No. 61772076 and 61751201), NSFB (No. Z181100008918002), Major Project of Zhijiang Lab (No. 2019DH0ZX01), CETC (No. w-2018018) and OPBKLICDD (NO. ICDD201901).
\bibliographystyle{IEEEtran}
\bibliography{my}


\ifCLASSOPTIONcaptionsoff
  \newpage
\fi



%
%

\begin{IEEEbiography}[{\includegraphics[width=1in,height=1.25in,clip,keepaspectratio]{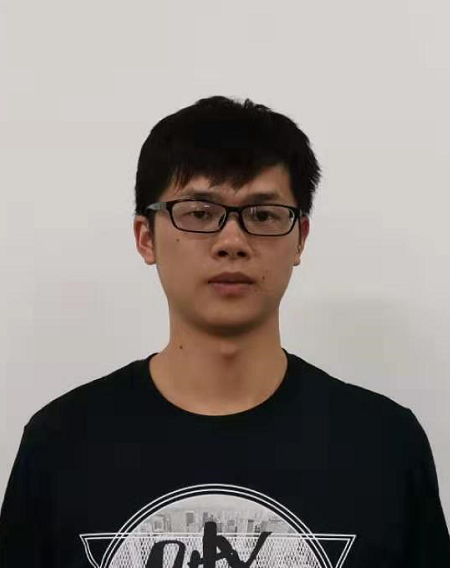}}]{Rong-Cheng Tu}
received the bachelor's degree from Beijing Institute of Technology, Chain, in 2018. He is currently working toward the master's degree in the Department of Computer Science and Technology, Beijing Institute of Technology, Chain. His research interests are in deep learning and learning to hash.
\end{IEEEbiography}

\begin{IEEEbiography}[{\includegraphics[width=1in,height=1.25in,clip,keepaspectratio]{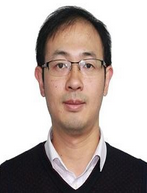}}]{Xian-Ling Mao}
received the Ph.D. science degree from Peking University, Chain, in 2012. He is currently an Associate Professor with the Department of Computer Science and Technology, Beijing Institute of Technology, Chain. His major research interests include deep learning, machine learning, information retrieval, natural language processing, artificial intelligence and network data mining.
\end{IEEEbiography}



\begin{IEEEbiography}[{\includegraphics[width=1in,height=1.25in,clip,keepaspectratio]{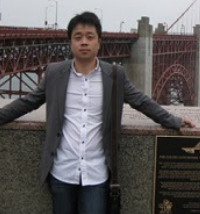}}]{Wei wei}
received the Ph.D. degree from the Huazhong University of Science and Technology, Chain, in 2012. He is currently an Associate Professor with School of Computer Science and Technology and the Director of Cognitive Computing and Intelligent Information Processing (CCIIP) Laboratory in Huazhong University of Science and Technology, Chain. His major research interests include information retrieval, natural language processing, artificial intelligence, data mining (text mining), statistics machine learning, social media analysis and mining recommender system.
\end{IEEEbiography}

\begin{IEEEbiography}[{\includegraphics[width=1in,height=1.25in,clip,keepaspectratio]{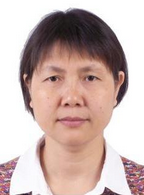}}]{Heyan Huang}
	received the bachelor's degree from Wuhan University of Surveying and Mapping, Chain, in 1983, the master's degree from  National University of Defense Technology, Chain, in 1986, and the Ph.D. degree from the Institute of Computing Technology, Chinese Academy of Sciences, Chain, in 1989. She is currently a professor and the Dean with the Department of Computer Science and Technology, Beijing Institute of Technology, Chain. Her major research interests include natural language processing, information content security, intelligent application system.
\end{IEEEbiography}



\end{document}